\documentclass{aastex61}
\usepackage{multirow}

\received{XXX}
\revised{YYY}
\accepted{ZZZ}
\submitjournal{ApJ}

\begin{document}

\title{A systematic TMRT observational study of Galactic $^{12}$C/$^{13}$C ratios from Formaldehyde}

\correspondingauthor{JiangShui Zhang }
\email{ jszhang@gzhu.edu.cn }

\author{Y. T. Yan}
\affil{Center for Astrophysics, Guangzhou University, 510006 Guangzhou, PR China}
\author{J. S. Zhang}
\affiliation{Center for Astrophysics, Guangzhou University, 510006 Guangzhou, PR China}

\author{  C. Henkel }
\affiliation{Max-Planck-Institut f\"{u}r Radioastronomie, Auf dem H\"{u}gel 69, 53121 Bonn, Germany}
\affiliation{Astronomy Department, King Abdulaziz University, P.O. Box 80203, 21589 Jeddah, Saudi Arabia}
\author{  T. Mufakharov }
\affiliation{Shanghai Astronomical Observatory, Chinese Academy of Sciences, 200030 Shanghai, PR China }
\affiliation{Kazan Federal University, 18 Kremlyovskaya St., 420008 Kazan, Russia }
\author{L. W. Jia}
\affiliation{Center for Astrophysics, Guangzhou University, 510006 Guangzhou, PR China}
\author{X. D. Tang}
\affiliation{Max-Planck-Institut f\"{u}r Radioastronomie, Auf dem H\"{u}gel 69, 53121 Bonn, Germany}
\affiliation{Xinjiang Astronomical Observatory, Chinese Academy of Sciences, 830011 Urumqi, PR China}
\affiliation{Key Laboratory of Radio Astronomy, Chinese Academy of Sciences, PR China}
\author{Y. J. Wu}
\affiliation{Shanghai Astronomical Observatory, Chinese Academy of Sciences, 200030 Shanghai, PR China }
\affiliation{Key Laboratory of Radio Astronomy, Chinese Academy of Sciences, PR China}
\author{J. Li}
\affiliation{Shanghai Astronomical Observatory, Chinese Academy of Sciences, 200030 Shanghai, PR China }
\affiliation{Key Laboratory of Radio Astronomy, Chinese Academy of Sciences, PR China}
\author{Z. A. Zeng}
\affiliation{Center for Astrophysics, Guangzhou University, 510006 Guangzhou, PR China}
\author{Y. X. Wang}
\affiliation{Center for Astrophysics, Guangzhou University, 510006 Guangzhou, PR China}
\author{Y. Q. Li}
\affiliation{Center for Astrophysics, Guangzhou University, 510006 Guangzhou, PR China}
\author{J. Huang}
\affiliation{Center for Astrophysics, Guangzhou University, 510006 Guangzhou, PR China}
\author{J. M. Jian}
\affiliation{Center for Astrophysics, Guangzhou University, 510006 Guangzhou, PR China}


\begin{abstract}
We present observations of the C-band $1_{10}-1_{11}$ (4.8 GHz) and Ku-band $2_{11}-2_{12}$ (14.5 GHz) K-doublet lines of H$_2$CO and the C-band $1_{10}-1_{11}$ (4.6 GHz) line of H$_2$$^{13}$CO toward a large sample of Galactic molecular clouds, through the Shanghai Tianma 65-m radio telescope (TMRT). Our sample with 112 sources includes strong H$_2$CO sources from the TMRT molecular line survey at C-band and other known H$_2$CO sources. All three lines are detected toward 38 objects (43 radial velocity components) yielding a detection rate of 34\%. Complementary observations of their continuum emission at both C- and Ku-bands were performed. Combining spectral line parameters and continuum data, we calculate the column densities, the optical depths and the isotope ratio H$_2$$^{12}$CO/H$_2$$^{13}$CO for each source. To evaluate photon trapping caused by sometimes significant opacities in the main isotopologue's rotational mm-wave lines connecting our measured K-doublets, and to obtain $^{12}$C/$^{13}$C abundance ratios, we used the RADEX non-LTE model accounting for radiative transfer effects. This implied the use of the new collision rates from \citet{Wiesenfeld2013}. Also implementing distance values from trigonometric parallax measurements for our sources, we obtain a linear fit of $^{12}$C/$^{13}$C = (5.08$\pm$1.10)D$_{GC}$ + (11.86$\pm$6.60), with a correlation coefficient of 0.58. D$_{GC}$ refers to Galactocentric distances. Our $^{12}$C/$^{13}$C ratios agree very well with the ones deduced from CN and C$^{18}$O but are lower than those previously reported on the basis of H$_2$CO, tending to suggest that the bulk of the H$_2$CO in our sources was formed on dust grain mantles and not in the gas phase.

\end{abstract}


\section{Introduction} \label{sec:intro}

Isotope abundance ratios provide a powerful tool to trace stellar nucleosynthesis, to evaluate the enrichment of the interstellar medium (ISM) by stellar ejecta and to constrain the chemical evolution of the Milky Way (\citealt{Wilson1994}). In particular, the $^{12}$C/$^{13}$C ratio is one of the most useful tracers of the relative degree of primary to secondary processing. $^{12}$C is known to be produced primarily via Helium burning in massive stars, on rapid time scales, whereas $^{13}$C is thought to be formed primarily through CNO processing of $^{12}$C seeds from earlier stellar generations. This occurs on a slower timescale during the red giant phase in low and intermediate-mass stars or novae (\citealt{1994LectureNFhys...439..72}, \citealt{Meyer1994}; \citealt{Wilson1994}). Thus the $^{12}$C/$^{13}$C ratio is expected to decrease with time and a $^{12}$C/$^{13}$C gradient should exist with Galactocentric distance, assuming an inside-out formation scenario of the Milky way (\citealt{Pilkington}).

Under certain conditions, isotope abundance ratios can be effectively determined from the strengths of corresponding molecular lines (e.g., H$_2$CO, CO and CN). Measurements of absorption lines of H$_2$CO and H$_2$$^{13}$CO toward strong continuum sources are believed to be one of the best ways to determine the isotope ratio between $^{12}$C and $^{13}$C. Previous observations support the existence of a gradient of $^{12}$C/$^{13}$C, i.e., the ratio increases as a function of Galactocentric distance (Wilson et al. \citeyear{1976A&A....51..303W}; Gardner \& Whiteoak \citeyear{1979MNRAS...188..331}; Henkel et al. \citeyear{1980A&A...82..41},\citeyear{1982A&A...109..344},\citeyear{1983A&A...127..388},\citeyear{1985A&A...143..148H}). A $^{12}$C/$^{13}$C ratio cannot be directly deduced from $^{12}$CO (hereafter CO) and $^{13}$CO, since CO (and sometimes $^{13}$CO) is optically thick. Therefore, their isotopologues C$^{18}$O and $^{13}$C$^{18}$O, exhibiting optically thin lines, are sometimes used to determine $^{12}$C/$^{13}$C (\citealt{1990ApJ...357..477}). However, this is time-consuming, since thermally excited emission lines with low optical depths can have extremely low intensities (\citealt{Wilson1994}). CN is also a tracer of the $^{12}$C/$^{13}$C ratio. Although CN has a distinct hyperfine structure, which can be used to directly evaluate opacities (\citealt{2002ApJ...578..211}; \citealt{2005ApJ...634...1126}), it is still difficult to obtain a true $^{12}$C/$^{13}$C value from $^{12}$CN/$^{13}$CN because of sometimes large opacities of $^{12}$CN. Milam et al. (\citeyear{2005ApJ...634...1126}) proposed a gradient of $^{12}$C/$^{13}$C with Galactocentric distance, from a combination of all previously published measurements of CN, C$^{18}$O and H$_2$CO. The goal of our study is to critically review their results using new distances and H$_2$CO collision rates as well as a larger sample of sources and to examine in how far previous results have to be modified.

Thus we have performed a more systematic study of H$_2$CO and H$_2$$^{13}$CO toward a big sample of Galactic molecular clouds to determine $^{12}$C/$^{13}$C isotope ratios more accurately. Based on a TMRT formaldehyde survey at C-band ($\sim 5$ GHz; Li et al. 2019, in preparation), we chose sources with strong H$_2$CO signals as our targets for deep integration. Some previously studied strong H$_2$CO sources with absolute flux densities larger than 0.5 Jy (Wilson et al. \citeyear{1976A&A....51..303W}; Henkel et al. \citeyear{1982A&A...109..344},\citeyear{1983A&A...127..388},\citeyear{1985A&A...143..148H}; Araya et al. \citeyear{2007ApJS...170..152}) are also included in our sample. H$_2$CO collision rates were taken from \citet{Wiesenfeld2013} who consider H$_2$CO collisions with H$_2$ based on the high accuracy potential energy surface introduced by Troscompt et al. (\citeyear{2009aA&A...493..687}). These rates differ significantly from the scaled rates of H$_2$CO in collision with He (\citealt{1991ApJS...76..979}) used in previous analyses. At the same time we also took improved distances for our sources to evaluate more accurately the previously reported gradient of the carbon isotope ratio as a function of Galactocentric distance. In Section \ref{sec:obser}, the observations of our large sample are described, encompassing Galactocentric radii from 0 to 10.5 kpc. In Section \ref{sec:DataReduction}, we first perform the analysis of spectra and the radio continuum, followed by modelling work based on radiative transfer calculations. Section \ref{sec:discussions} provides discussions on the isotope ratio gradient, derived from our observational data alone as well as also including previously obtained data. Our main results are summarized in Section \ref{sec:summary}.

\section{Observations} \label{sec:obser}

The $1_{10}-1_{11}$ and $2_{11}-2_{12}$ transitions were observed toward our sources in March and October 2016 as well as in May and July 2017 with the Shanghai Tianma Radio Telescope (TMRT). We used two cryogenically cooled receivers covering the frequency ranges of 4.0-8.0 GHz (C-band) and 12.0-18.2 GHz (Ku-band). The rest frequencies of the $1_{10}-1_{11}$ transitions of H$_2$$^{12}$CO and H$_2$$^{13}$CO were set to be 4.829660 and 4.593089 GHz, respectively (\citealt{1971ApJ...169..429T}; \citealt{1976A&A....51..303W}). 14.48848 GHz was adopted for the $2_{11}-2_{12}$ transition of H$_2$$^{12}$CO. The beam sizes are $\sim$4 arcmin at 5 GHz and $\sim$1.3 arcmin at 14.5 GHz, respectively. A Digital Backend System (DIBAS) was used for data recording (see \citealt{2016ApJ...824...136}). The DIBAS mode 22 was adopted for our observations, with eight spectral windows, to cover the H$_2$$^{12}$CO and H$_2$$^{13}$CO lines simultaneously, each with 16384 channels and a bandwidth of 23.4 MHz, supplying channel widths of 0.09 km s$^{-1}$ and 0.03 km s$^{-1}$ at 5 GHz and 14.5 GHz, respectively. The system temperature was 20-30 and 30-80 K on a T$_A^*$ scale for the $1_{10}-1_{11}$ and $2_{11}-2_{12}$ transition observations, respectively. The active surface system of the primary dish and a subreflector were used to improve the aperture efficiency during our Ku-band observations. We measured an aperture efficiency of about 55 per cent (\citealt{2017MNRAS...466..4364}; \citealt{2018ExpAstron..45...397}) at both the C-band and Ku-band. The spectra were obtained in a position switching mode with total on source integration times of 0.3 - 3 hours and 2 - 40 mins for the $1_{10}-1_{11}$ and $2_{11}-2_{12}$ transition observations, respectively. Observational parameters of our sample are listed in Appendix \ref{A}.

Among our 112 targets, 84 sources were successfully detected in the $1_{10}-1_{11}$ line of H$_2$CO, and 38 sources of these 84 were also successfully detected in the $1_{10}-1_{11}$ line of H$_2$$^{13}$CO. Subsequently we observed the $2_{11}-2_{12}$ lines of H$_2$CO for these 38 sources and all were detected. 5 sources are showing two velocity components. Thus, 43 radial velocity components were detected towards these 38 sources. The Signal-to-Noise (S/N) ratios of the $1_{10}-1_{11}$ line of H$_2$CO are greater than 25 for these 38 sources. For the $1_{10}-1_{11}$ line of H$_2$$^{13}$CO and the $2_{11}-2_{12}$ line of H$_2$CO, the corresponding S/N ratios are greater than 5. To determine apparent optical depths, complementary observations for their continuum at both C- and Ku-bands were performed toward those 38 sources, from 2017 October 26 to November 4 and during 2018 February 4 to 8. We used the C- and Ku-band receivers, centered at 4.829660 GHz and 14.48848 GHz with 500 MHz bandwidth, respectively. A cross-scan mode was used for observations, where the antenna beam pattern was driven repeatedly in right ascension (RA) and declination (Dec) over the source position. The data were acquired with the DIBAS backend system. Before observing each target source, we chose one calibration source (e.g., 3C147, 3C295, 3C358, 3C401, 3C383 or NGC7027) near the target source to observe it for pointing correction of the telescope. We took one and two scans for calibration and target sources, respectively, with scan lengths of three minutes each. C- and Ku-band receiver features were observed with left and right circular polarization and were subsequently averaged. The continuum at both C- and Ku-bands was successfully detected toward 32 of the 38 sources. For the other 6 sources, the continuum at C-band was detected, while no Ku-band signal was obtained. 4 sources remained undetected, while 2 sources (Mol12, J205703.98) were not observed at Ku-band. Table \ref{Table1} presents obtained continuum temperatures. The typical rms value of the continuum at Ku-band for the 4 undetected targets was $\sim$ 0.1 K ($\sim$ 0.15 Jy, see below), which is also listed in Table \ref{Table1}. The two unobserved targets are marked by "-". Nevertheless, this does not affect our main results (see details in Sect. \ref{subsec:correct}).

The noise diode signal inserted into the receiver system may differ slightly between the C-band frequencies of H$_2$CO and H$_2$$^{13}$CO. Thus we performed separate flux calibration observations at 4.829 GHz and 4.593 GHz, respectively, in January 24 to 29, 2019. Also for comparisons, we made repeated observations toward several strong sources (W3IRS5, NGC2024, SgrB2, W33) in our sample. The new results related to those strong sources are consistent with our previous ones. Based on the frequency dependencies of the spectral flux densities in \citet{2013ApJS...204..19}, the radio sources 3C123, 3C147, 3C196 and 3C295 were taken as calibrators in our observations. The cross scan mode with a bandwidth of 10 MHz was used for calibrations at 4.829 GHz and 4.593 GHz and the telescope gains are 1.49 Jy K$^{-1}$ and 1.28 Jy K$^{-1}$, respectively. Thus a correction factor of (1.49/1.28) from the different gains at 4.829 GHz and 4.593 GHz should be applied to obtain the accurate value of the isotope abundance ratio (see details in Sect. \ref{subsec:opticaldepth}). For C- ($\sim$ 5 GHz) and Ku-band ($\sim$ 14.5 GHz), the sensitivities are both 1.5 Jy K$^{-1}$. The calibrators were observed at suitable elevations of $40\arcdeg-60\arcdeg$, while the elevation range of all the target sources is $20\arcdeg-80\arcdeg$. Beside noise, the uncertainty of the measured flux density is mainly caused by the elevation-dependent gain curve of the TMRT (\citealt{2018ITAP..66...2044}; see also Sect. \ref{subsec:opticaldepth}). 

%\begin{longrotatetable}
\begin{deluxetable}{l|c|ccc|ccc|ccc}
%% Keep a portrait orientation
%% Over-ride the default font size
%% Use 8pt
\tabletypesize{\scriptsize}
%% This is the title of the table.
\tablecaption{Line Parameters derived from Gaussian-fitting}

%\tablenum{1}

\tablehead{
\colhead{}  & \multicolumn{7}{c}{$1_{10}-1_{11}$ transitions} & \multicolumn{3}{c}{$2_{11}-2_{12}$ transitions} \\
\colhead{} & \colhead{Continuum } & \multicolumn{3}{c}{H$_2$$^{12}$CO Parameters} & \multicolumn{3}{c}{H$_2$$^{13}$CO Parameters} & \multicolumn{3}{c}{H$_2$$^{12}$CO Parameters}\\
\colhead{}  & \colhead{antenna} & \colhead{Line} & \colhead{Full Width} & \colhead{V$_{LSR}$} & \colhead{Line} & \colhead{Full Width} & \colhead{V$_{LSR}$} & \colhead{Continuum } & \colhead{Line} & \colhead{Apparent}\\
\colhead{Source}   & \colhead{temp.} & \colhead{antenna} & \colhead{to} & \colhead{} & \colhead{ antenna} & \colhead{to} & \colhead{} & \multicolumn{2}{c}{antenna} & \colhead{optical}\\
\colhead{ }  & \colhead{} & \colhead{temp.} & \colhead{Half Power} & \colhead{} & \colhead{temp.} & \colhead{Half Power} & \colhead{} & \multicolumn{2}{c}{temperature} & \colhead{depth}\\
\colhead{}   & \colhead{(K)} & \colhead{(K)} & \colhead{(km s$^{-1}$)} & \colhead{(km s$^{-1}$)} & \colhead{(K)} & \colhead{(km s$^{-1}$)} & \colhead{(km s$^{-1}$)} & \colhead{(K)} & \colhead{(K)} & \colhead{}}
\tablecolumns{13}
\colnumbers
\startdata
\label{Table1}
W3IRS5  & 38.8 & -3 $\pm$ 0.048& 3.49 & -39.6 & -0.035 $\pm$ 0.019 & 2.08 & -38.6 & 31.5 & -1.55 & 0.05\\
Mol12 &  0.12 & -0.366$ \pm$ 0.005& 2.37 & 2.46 & -0.012 $\pm$ 0.006 & 2.21 & 2.31  &  -  & -0.47 & - \\
NGC2024 &  26.5 & -4.926 $\pm$ 0.025& 1.59 & 9.39 & -0.062 $\pm$ 0.021 & 2.22 & 9.30 &  1.7 & -2.1 & 1.34\\
M-0.13 &   8.79 & -3.388 $\pm$ 0.041& 33.55 &19.19 & -0.526 $\pm$ 0.032 & 30.16 & 19.75 & 1.3  & -0.29 & 0.14\\
SgrA &   79.3 & -9.317 $\pm$ 0.099 & 38.17 & 39.64 & -1.078 $\pm$ 0.099 & 37.23 & 40.11 &  4.3 & -0.38 & 0.07\\
 Sgr B2&   20.1 & -2.255 $\pm$ 0.048 & 14.33  & 4.71  & -0.061 $\pm$ 0.024 & 8.16  & 2.56 & 54.6 & -0.9 & 0.02 \\
 & &-12.75 $\pm$ 0.048 & 24.79 & 64.91  &  -1.991 $\pm$ 0.024  & 22.96 & 64.93 &   & -16.0 & 0.34 \\
M+1.6 &   0.3 & -1.054 $\pm$ 0.023 & 17.07 & 53.13 & -0.140 $\pm$ 0.017 & 17.12 & 52.65 &  $<0.1$\tablenotemark{*}  & -0.29 & $>0.11$ \\
G10.16-0.35  & 4.55 & -0.167 $\pm$ 0.014 & 5.12 & 0.55 & -0.044 $\pm$ 0.008 & 0.68 & 1.11 & 15.9  & -0.32 & 0.02\\
 &   & -0.325 $\pm$ 0.014 & 11.90 & 10.28 & -0.021 $\pm$ 0.008 & 7.45 & 9.18 &   & -0.24 & 0.01\\
W31  & 3.77 & -0.842 $\pm$ 0.012 & 6.41 & -1.02 & -0.046 $\pm$ 0.010 & 8.09 & -0.81 & 8.7  & -0.69 & 0.07\\
  &    & -0.748 $\pm$ 0.012 & 4.40 & 28.51 & -0.020 $\pm$ 0.010 & 3.08 & 28.37 &   & -0.07 & 0.007\\
G11.93-0.61   & 1.27 & -0.808 $\pm$ 0.014 & 3.78 & 37.67 & -0.026 $\pm$ 0.012 & 4.18 & 37.60 &  2.5 & -0.32 & 0.1\\
W33 &   20.8 & -6.3 $\pm$ 0.034 & 5.44 & 34.15 & -0.251 $\pm$ 0.026 & 5.61 & 34.44 & 4.5  & -12.05 & -  \\
G13.88+0.28  & 3.67 & -0.772 $\pm$ 0.015 & 3.59 & 48.52 & -0.032 $\pm$ 0.013 & 2.55 & 48.54 & 4.8  & -0.17 & 0.03\\
G12.91-0.26   & 2.23 & -1.354 $\pm$ 0.015 & 5.58 & 35.23 & -0.056 $\pm$ 0.010 & 4.2 & 34.88 & 0.8  & -0.25 & 0.16\\
G19.62-0.23  & 4.2 & -0.649 $\pm$ 0.010 & 6.40 & 44.45 & -0.03 $\pm$ 0.009 & 3.85 & 44.24 & 0.76 & -0.26 & 0.15\\
G023.44-00.18   & 5.19 & -0.324 $\pm$ 0.009 & 5.09 & 80.81 & -0.016 $\pm$ 0.011 & 15.33 & 83.03  &  0.5 & -0.02  & 0.01 \\
 &   & -0.917 $\pm$ 0.009  & 6.62  & 98.97  & -0.027  $\pm$ 0.011  & 7.89  & 99.58   &    &  -0.1  & 0.06 \\
G23.43-0.21   & 5.19 & -0.88 $\pm$ 0.006  & 7.47  & 99.54  &  -0.026 $\pm$ 0.006  &  7.51 & 100.31   &  1.9 & -0.05  & 0.02\\
G29.9-0.0  & 7.1  & -0.707 $\pm$ 0.008  & 5.61  &  99.74 & -0.012  $\pm$ 0.010  & 7.91  & 100.5  & 0.45  & -0.12  & 0.08 \\
G31.41+0.31   & 1.02 &  -0.864 $\pm$ 0.013 & 5.99  &  97.65 & -0.045  $\pm$ 0.012  & 7.18  & 98.4  &  1.8 &  -0.42 & 0.19 \\
W43   & 26 & -3.367 $\pm$ 0.029  & 6.85  & 92.39  &  -0.122 $\pm$ 0.025  & 7.78  & 92.09   & $<0.1$\tablenotemark{*}   & -0.27  & $>0.25$ \\
G34.26+0.15   & 3.07 & -1.738 $\pm$ 0.015   &  4.32 & 60.17  &  -0.055 $\pm$ 0.017  & 4.44  & 59.81  & 11.4  & -0.89  & 0.08 \\
G34.3+0.1   & 9.6 &  -2.170 $\pm$ 0.019  & 3.79  & 60.22  &  -0.062 $\pm$ 0.017  & 4.64  & 59.98  & 1.6  & -0.88  & 0.4 \\
J185648.26   & 2.44 &  -1.212 $\pm$  0.016 & 6.14  & 44.47  &  -0.029 $\pm$  0.013 & 6.90  & 44.08  &  0.3 & -0.17  & 0.15 \\
G37.76-0.20   & 3 &  -0.612 $\pm$ 0.010  & 5.33  &  63.68 & -0.017  $\pm$ 0.011  & 2.31  & 63.25   & 1.79   &  -0.1 &0.03  \\
G35.20-1.74   & 13.6 &  -1.406 $\pm$ 0.018  & 3.44  & 43.35  & -0.022  $\pm$ 0.013  & 4.58  & 43.11  & 16.1  &  -0.46 & 0.03 \\
G35.2-1.8  & 11.2 & -1.169 $\pm$ 0.328  &  3.44  &  43.26 &  -0.023 $\pm$ 0.011  & 2.41  & 43.22  & 1.8  & -0.19  & 0.07 \\
G043.16+00.01   & 34.5 & -2 $\pm$ 0.045  & 10.26  & 12.46  & -0.06  $\pm$ 0.020  & 20.43  & 11.06  & 33  &  -0.99 & 0.03 \\
G43.17+0.00   & 37.5 & -2.151 $\pm$ 0.016  & 10.64  &  12.49 & -0.055  $\pm$ 0.017  & 14.2  & 10.93   & 12  & -0.88  & 0.07\\
G43.2+0.0  & 30.4 & -2.368 $\pm$ 0.061  & 9.81  & 12.67  &  -0.031 $\pm$ 0.019  &  9.1 & 12.84  &  3.58 & -1.15  & 0.28 \\
G45.45+0.06   & 7.18 &  -0.907 $\pm$ 0.013  & 4.37  & 59.44  & -0.025  $\pm$ 0.010  & 4.29  & 60.22  & 2.1  &  -0.63 & 0.22 \\
G49.21-0.35   & 9.7 &  -1.976 $\pm$ 0.014  &  3.31 &  65.69 & -0.03  $\pm$ 0.015  &  2.78 &  65.84  & 0.87  & -0.17  & 0.09\\
J192311.17   & 12.7 & -0.548 $\pm$ 0.018  & 4.62  &  50.63 & -0.025  $\pm$ 0.012  &  1.39 & 51.05  & 11  &  -0.22 & 0.02 \\
 &     & -2.924 $\pm$ 0.018  & 4.41  & 63.34  & -0.039  $\pm$ 0.012  & 5.93  & 63.43   &   & -0.27  & 0.022\\
G49.4-0.3   & 13.7 &  -2.804 $\pm$ 0.014  & 4.41  & 63.36  & -0.037  $\pm$ 0.015  & 4.21  &  63.08  &  1.14 & -0.19  & 0.09 \\
G49.5-0.4   & 54.2 &  -6.212 $\pm$ 0.039 & 7.94  & 65.64  & -0.144  $\pm$ 0.051  &  13.45 & 63.89  & 6.18  &  -0.79 & 0.11 \\
J192345.73   & 35.5 &  -1.188 $\pm$ 0.019  & 8.31  &  57.08 & -0.064  $\pm$ 0.026  & 12  &  57.16  &  $<0.1$\tablenotemark{*}  & -0.18  & $>0.15$ \\
 &   & -4.014 $\pm$ 0.019  & 6.55  & 66.39  & -0.087  $\pm$ 0.026  & 6.16  & 66.77   &    & -0.15  & $>0.13$ \\
J203901.04  & 18.1 & -3.84 $\pm$ 0.022  & 3.75  & -2.608  &  -0.079 $\pm$ 0.033  & 2.06  &  -2.24  & 25.5  & -1.94  & 0.08\\
DR21   & 15.2 & -4.208 $\pm$ 0.018  & 3.74   & -2.61  & -0.0786  $\pm$ 0.019  & 2.06  & -2.42   & 3.5  & -2.05  & 0.6\\
J205658.56   & 0.13 & -0.71 $\pm$ 0.015   & 3.29  & 0.84  & -0.037  $\pm$ 0.013  & 1.81  & 1.13  &  $<0.1$\tablenotemark{*}  & -0.25  & $>0.09$  \\
J205703.98   & 0.06 & -0.804 $\pm$ 0.014   & 3.05  & 1.15  & -0.041  $\pm$ 0.012  & 1.28  & 1.13   & -   & -0.15  & -  \\
\enddata

%% Include any \tablenotetext{key}{text}, \tablerefs{ref list},
%% or \tablecomments{text} between the \enddata and
%% \end{deluxetable} commands
%% No \tablecomments indicated
\tablenotetext{*}{The typical rms values of the continuum at Ku-band for undetected sources was $\sim$ 0.1 K.}
\tablecomments{ Column (1): source name; Column (2): continuum temperatures at 4.8 GHz; Columns (3), (4) and (5): line temperatures, FWHP line widths and H$_2$CO $1_{10}-1_{11}$ line center velocities, respectively; Columns (6), (7) and (8): H$_2$$^{13}$CO $1_{10}-1_{11}$ line temperatures, FWHP line widths, and line center velocities; Column (9): 14.5 GHz continuum temperatures; Column (10): H$_2$$^{12}$CO $2_{11}-2_{12}$ line temperatures; Column (11): apparent optical depths of the H$_2$$^{12}$CO $2_{11}-2_{12}$ lines. }

\end{deluxetable}
%\end{longrotatetable}

\section{Data reduction and Results} \label{sec:DataReduction}

\subsection{Optical Depth and Column Density} \label{subsec:opticaldepth}

We used the GILDAS/CLASS package to reduce the spectral line data. A first order polynomial was subtracted from each spectrum for baseline removal. Then we obtained the line parameters via Gaussian fitting for those 38 sources, which are presented in Table \ref{Table1}. The spectra of the $2_{11}-2_{12}$ lines of H$_2$$^{12}$CO, as well as the $1_{10}-1_{11}$ lines of H$_2$$^{12}$CO and H$_2$$^{13}$CO, after subtracting baselines and applying Hanning smoothing, are shown in Figure \ref{fig1}. For those 46 sources with only a detection of the H$_2$CO $1_{10}-1_{11}$ line, spectra and line parameters derived from Gaussian fitting are presented in Appendices \ref{B} and \ref{C}, respectively.

\begin{figure*}[h]
\center
  \includegraphics[width=233pt]{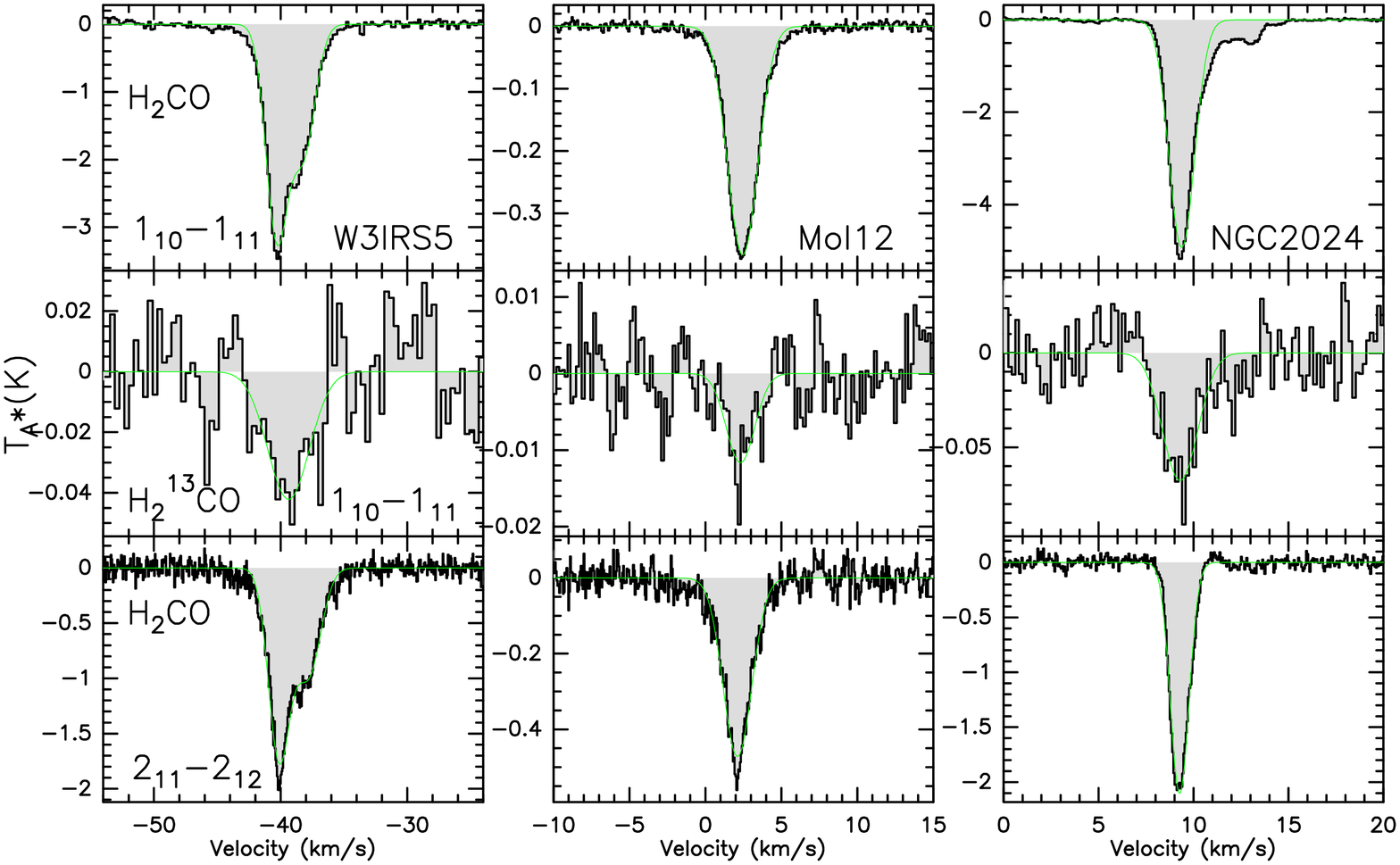}
  \includegraphics[width=233pt]{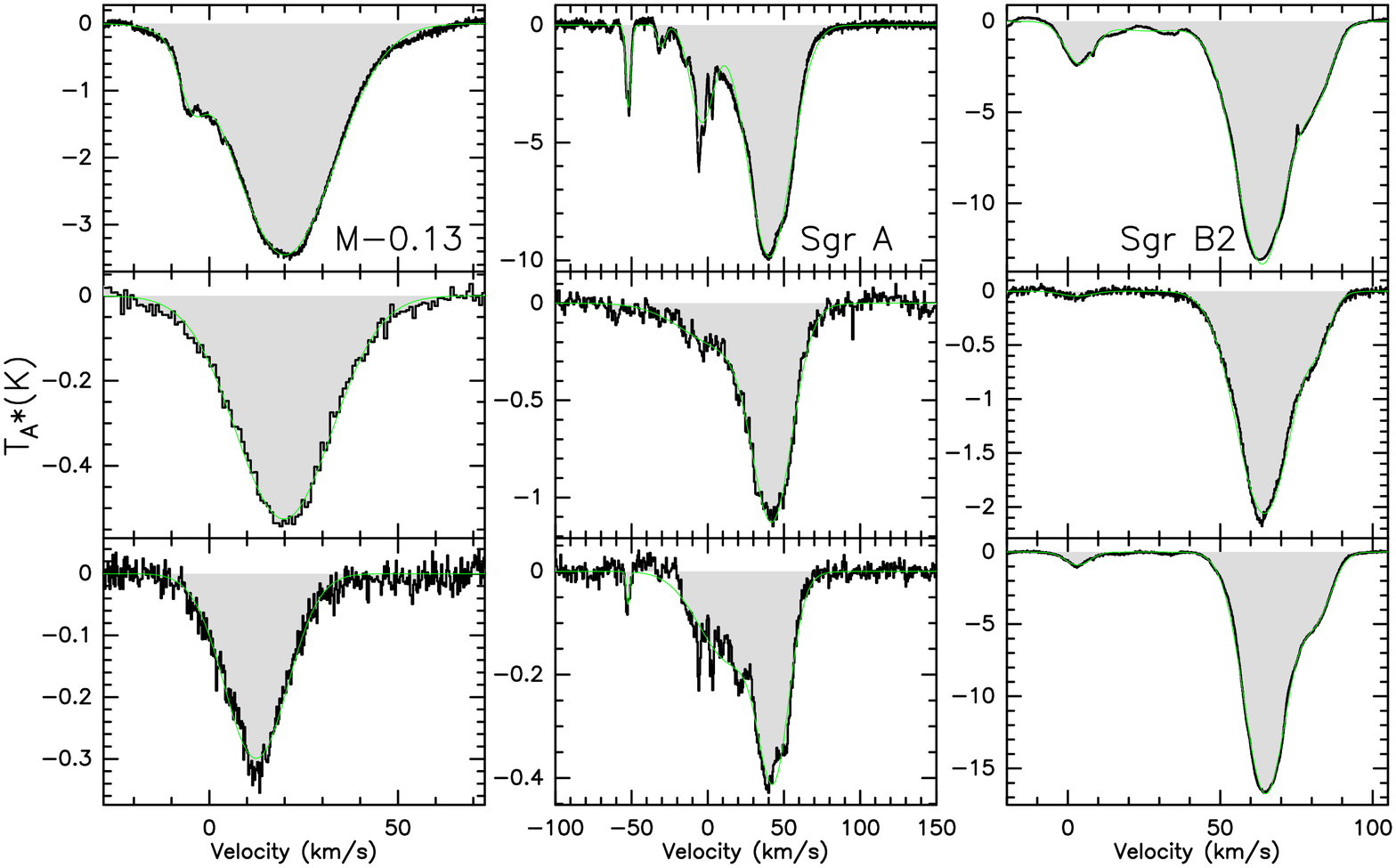}
    \includegraphics[width=233pt]{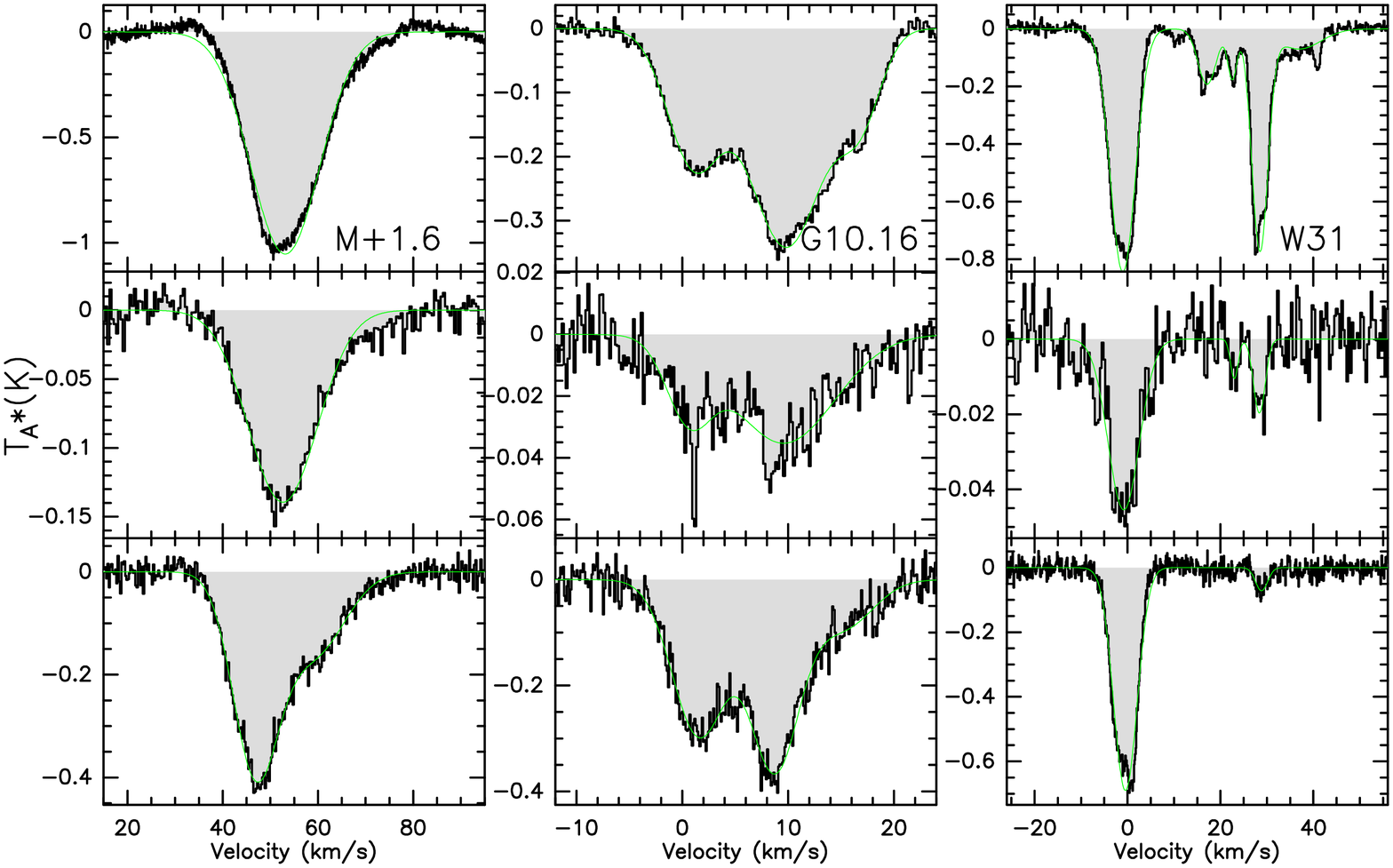}
  \includegraphics[width=233pt]{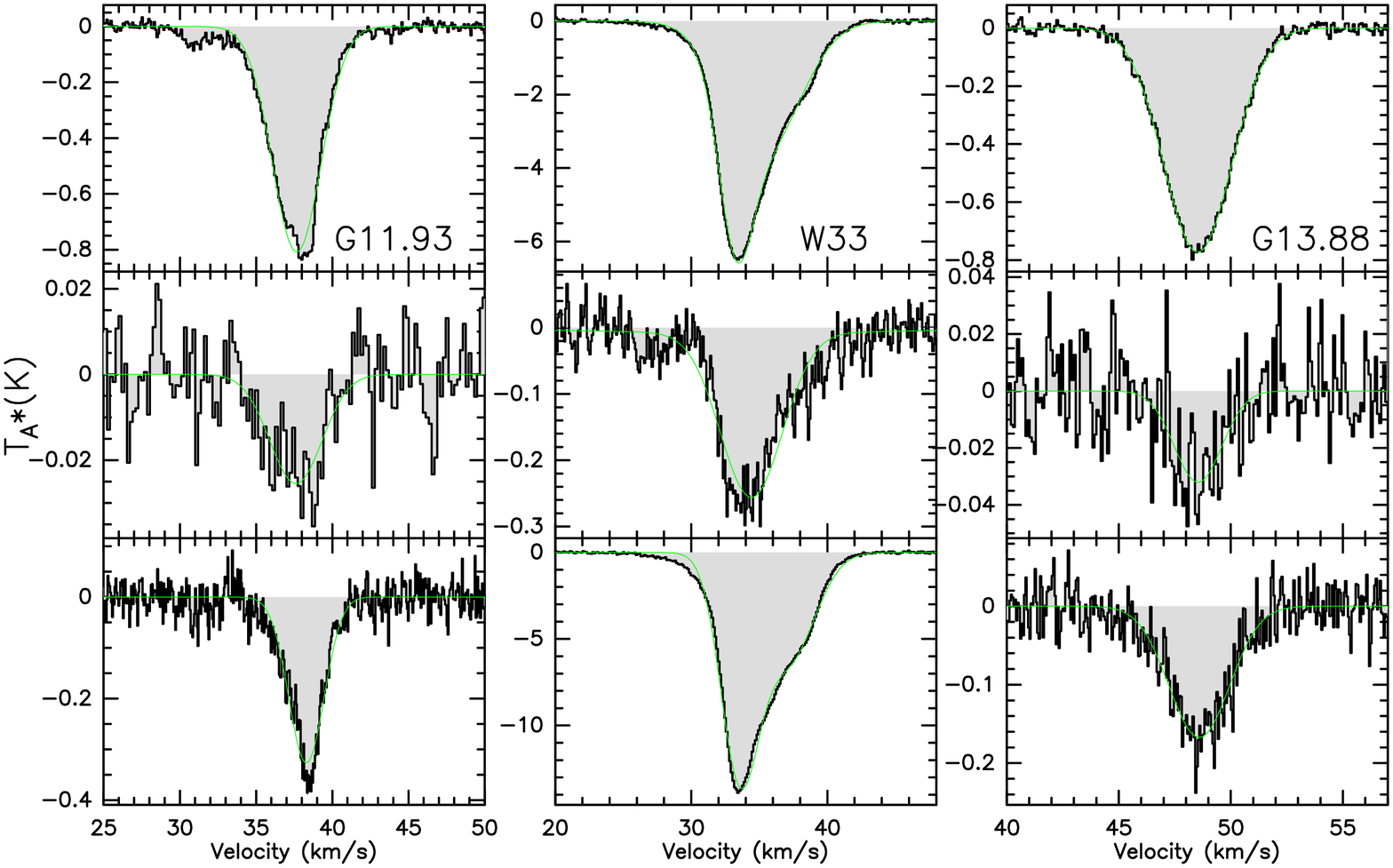}
    \includegraphics[width=233pt]{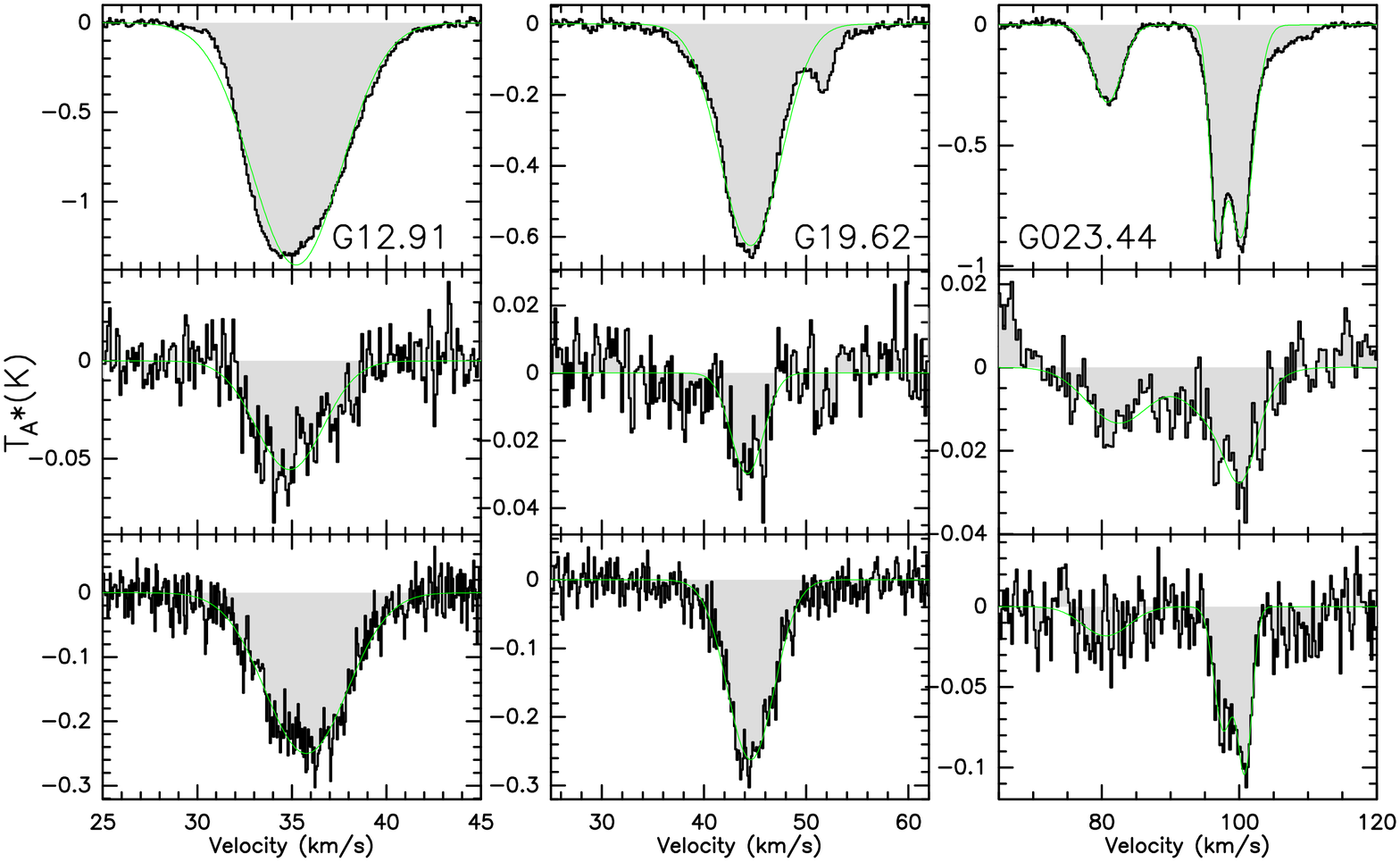}
  \includegraphics[width=233pt]{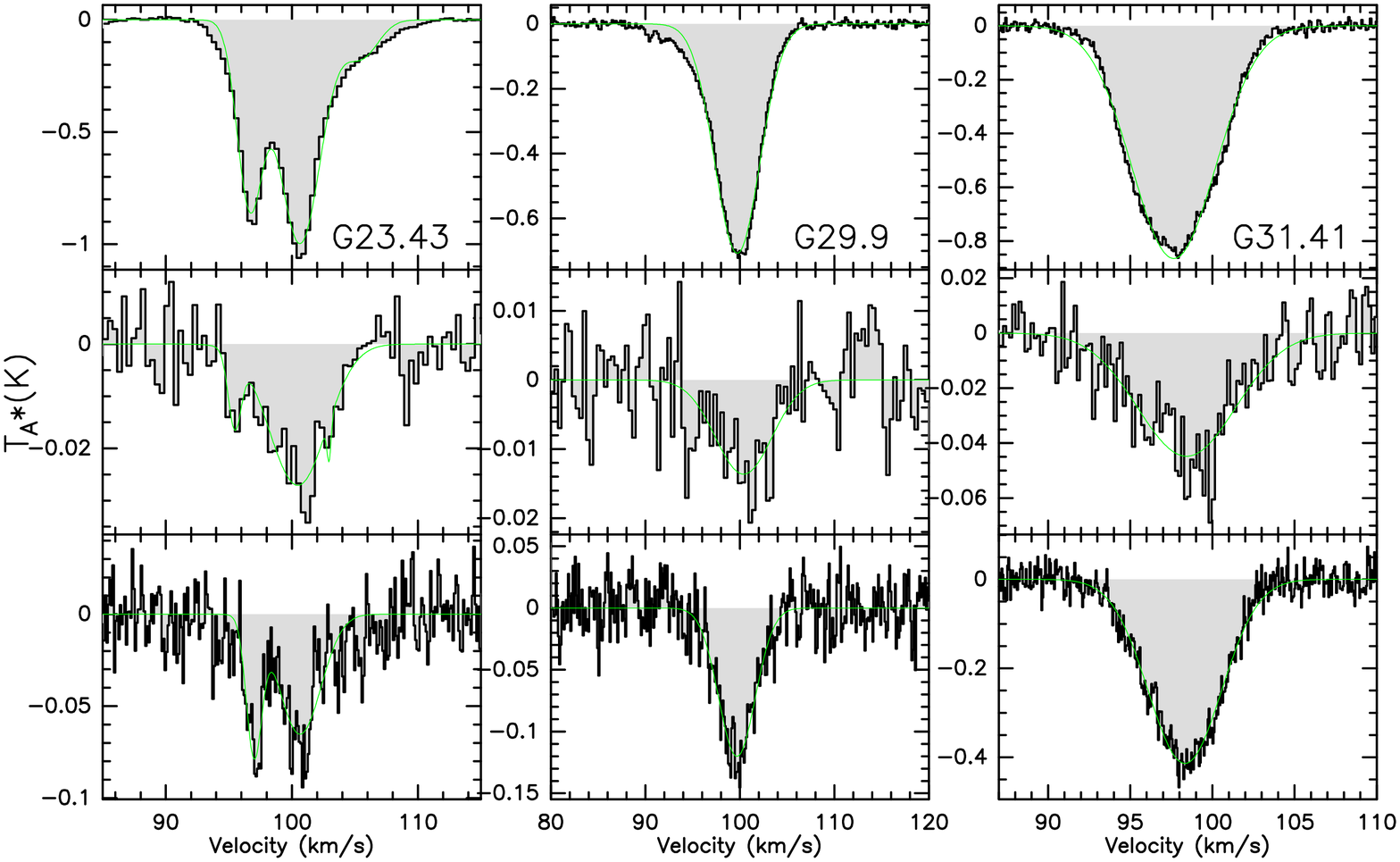}
    \includegraphics[width=233pt]{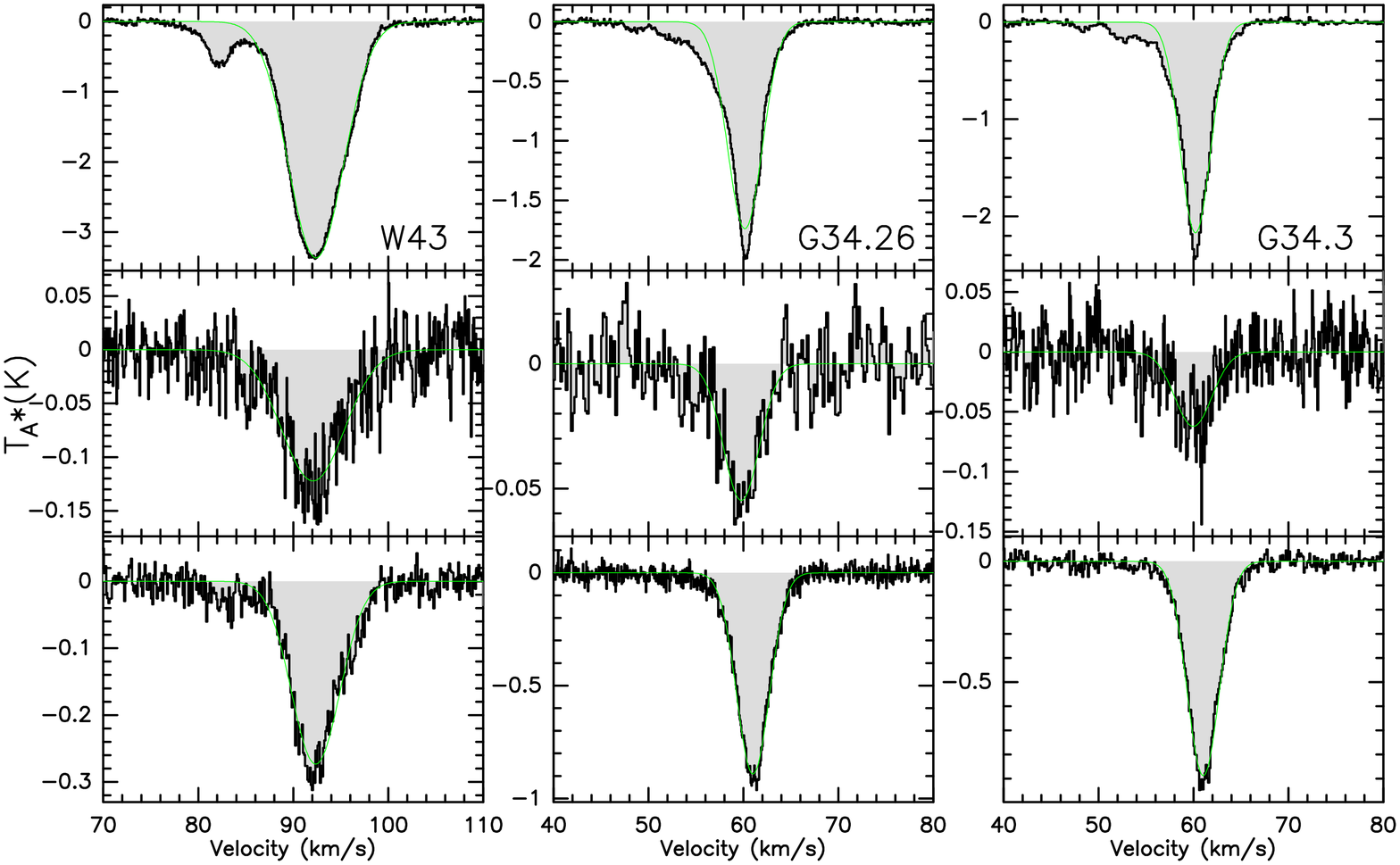}
  \includegraphics[width=233pt]{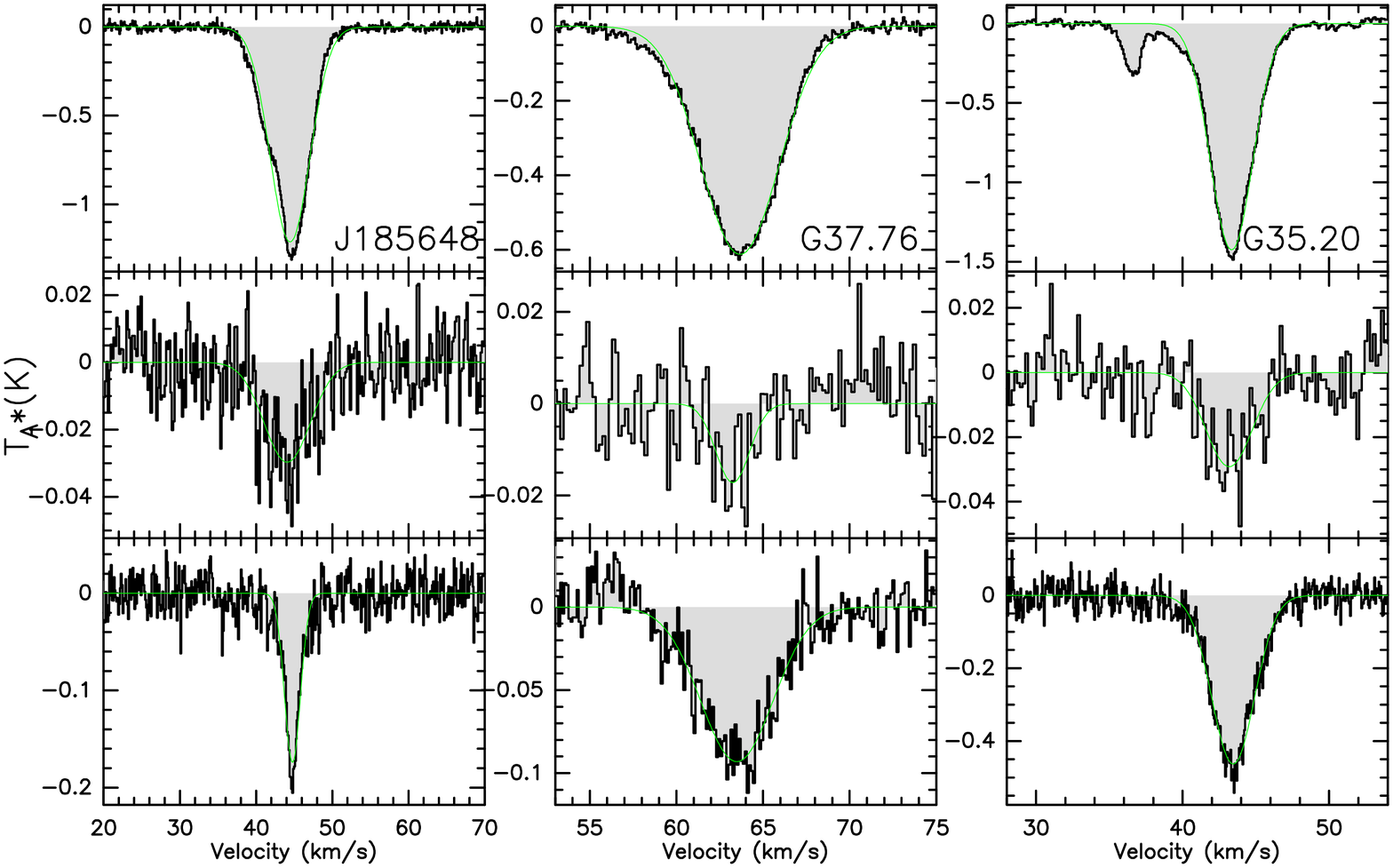}
    \includegraphics[width=233pt]{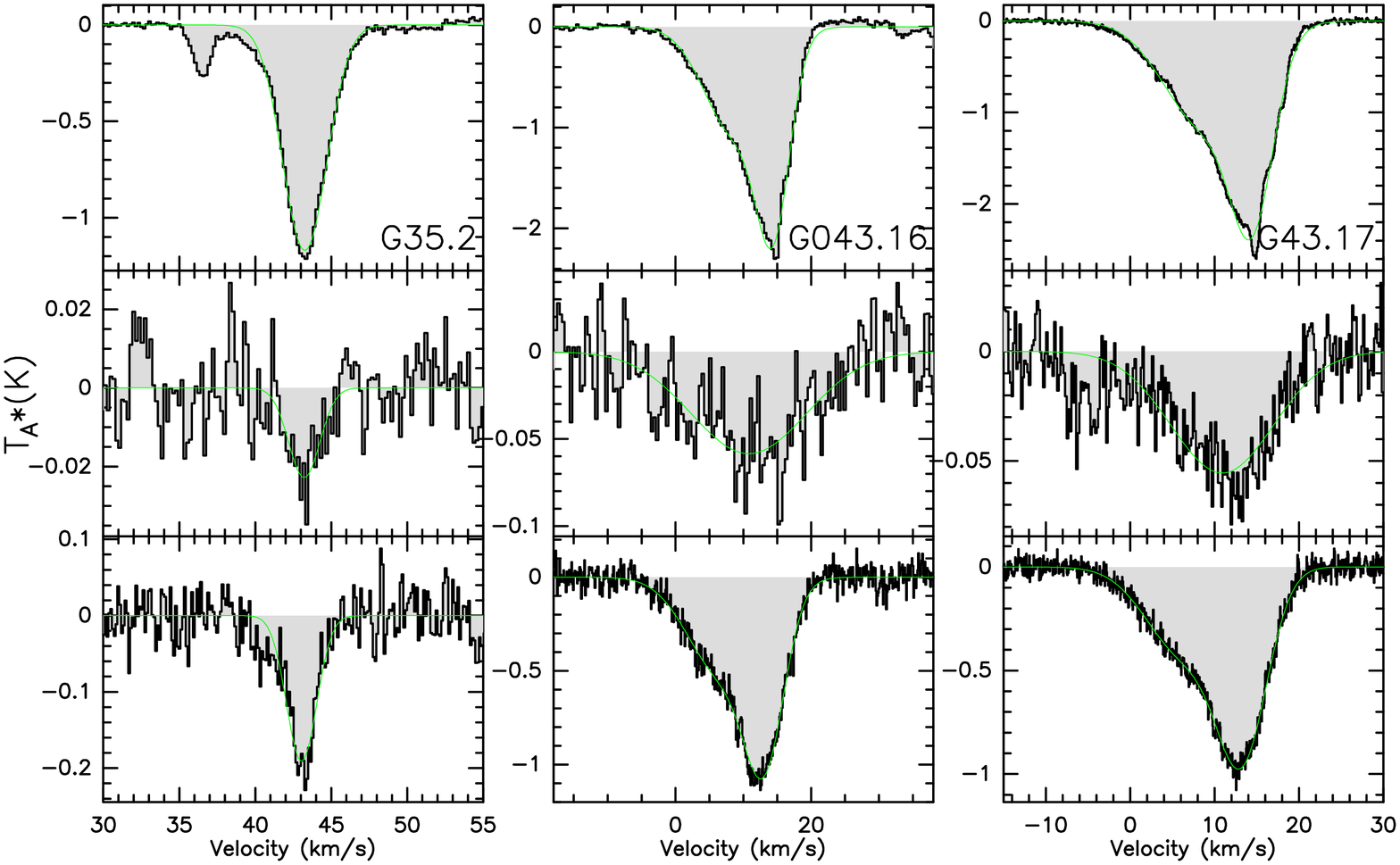}
  \includegraphics[width=233pt]{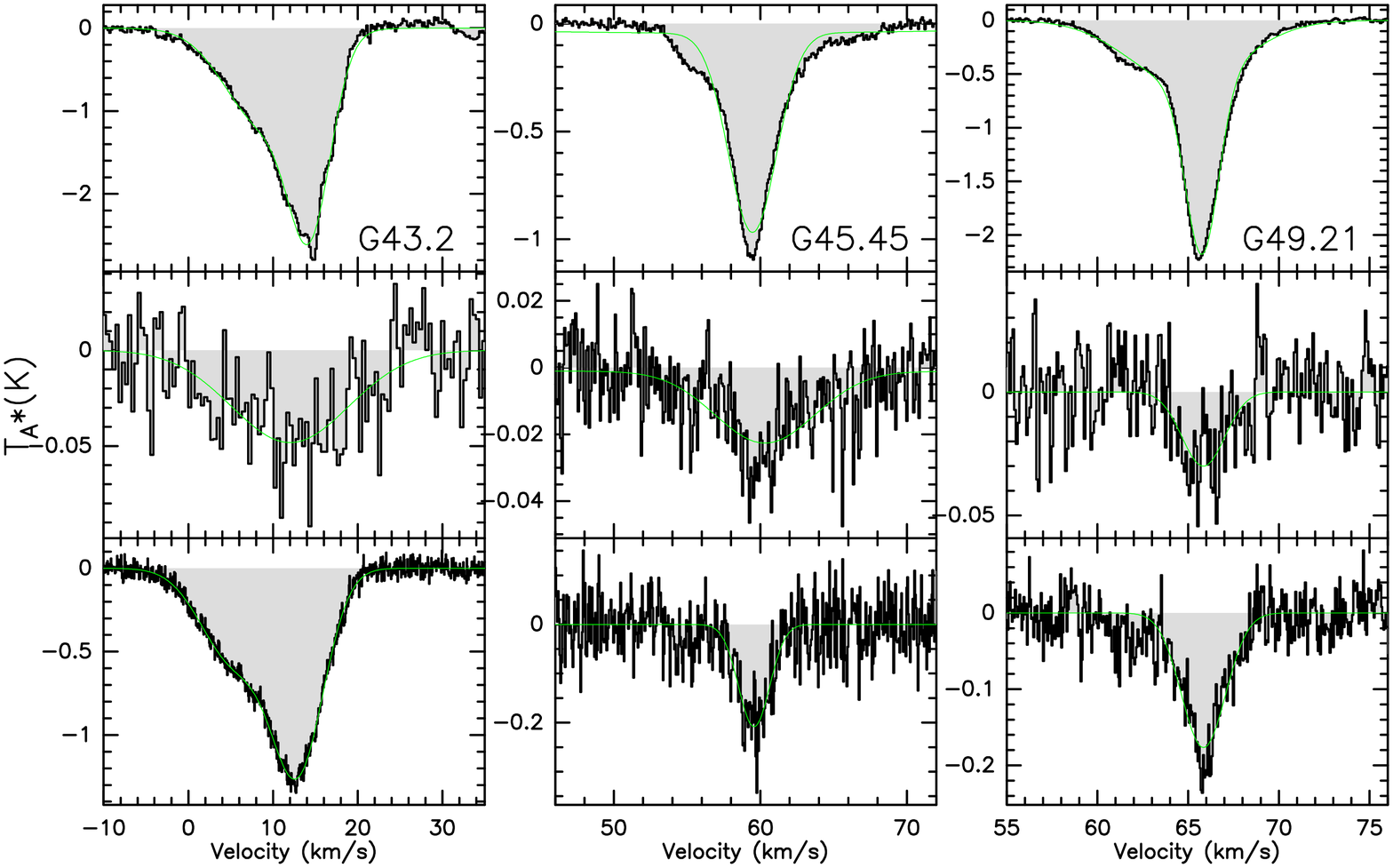}
  \end{figure*}
 \begin{figure*}[h]
\center
  \includegraphics[width=233pt]{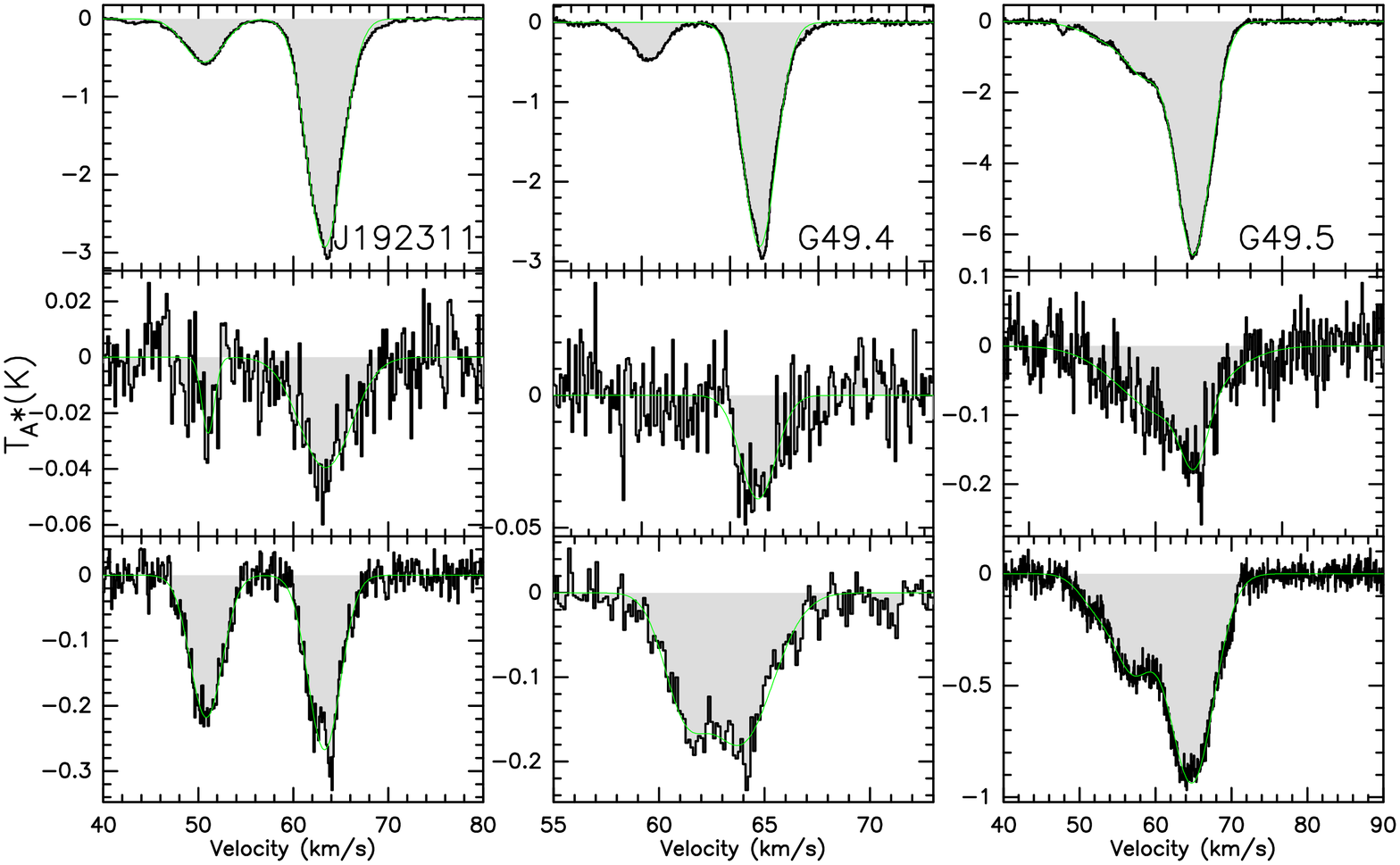}
  \includegraphics[width=233pt]{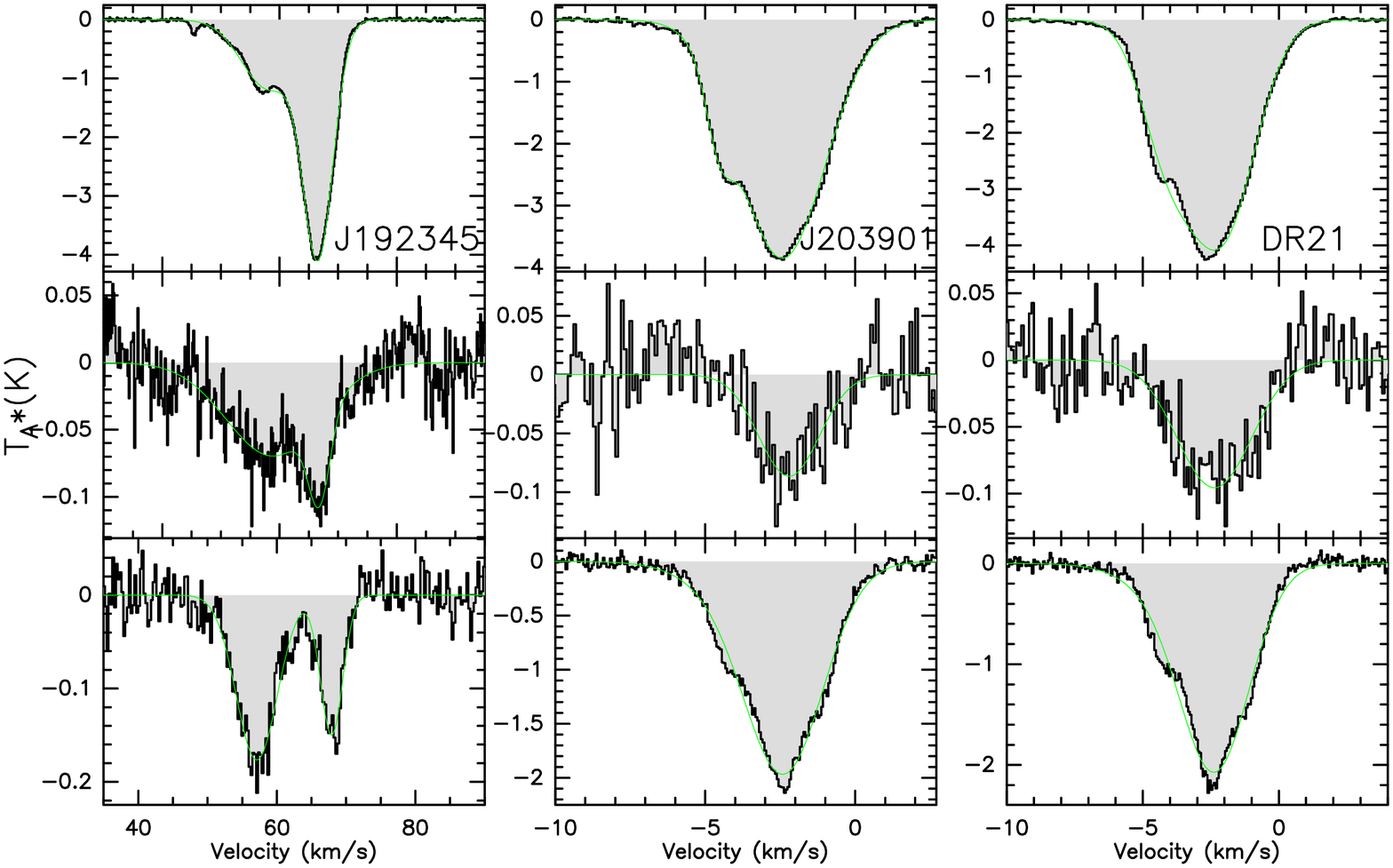}
  \includegraphics[width=165pt]{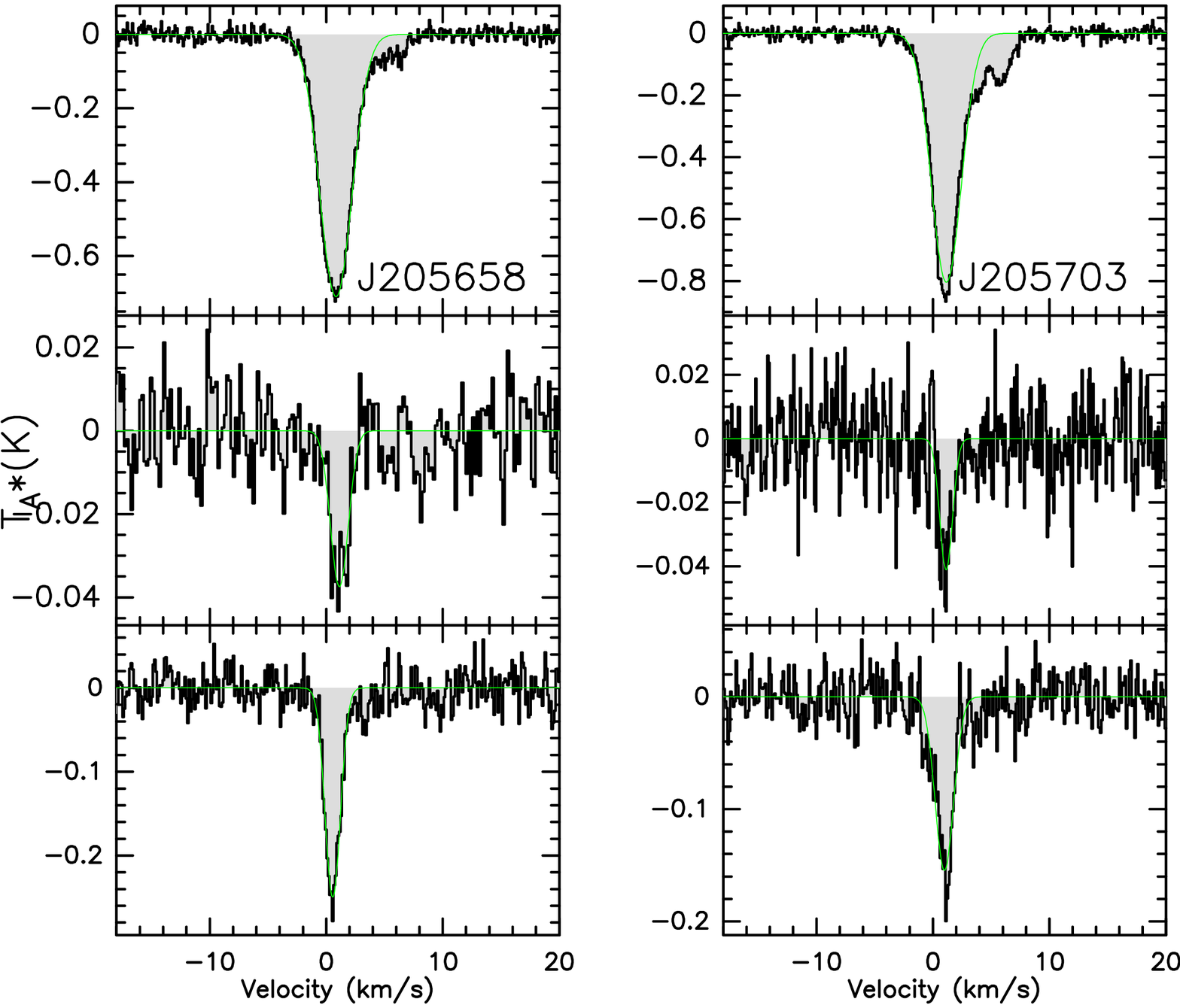}
  \caption{The $1_{10}-1_{11}$ and $2_{11}-2_{12}$ spectra of 38 sources, after subtracting baselines and applying Hanning smoothing. The upper profiles are the $1_{10}-1_{11}$ transitions of H$_2$$^{12}$CO; the middle profiles are the $1_{10}-1_{11}$ transitions of H$_2$$^{13}$CO and the lower profiles are the $2_{11}-2_{12}$ transitions of H$_2$$^{12}$CO. Antenna temperature scales are presented on the left hand side of the profiles.}
  \label{fig1}
\end{figure*}

\begin{figure*}[h]
\center
  \includegraphics[width=200pt]{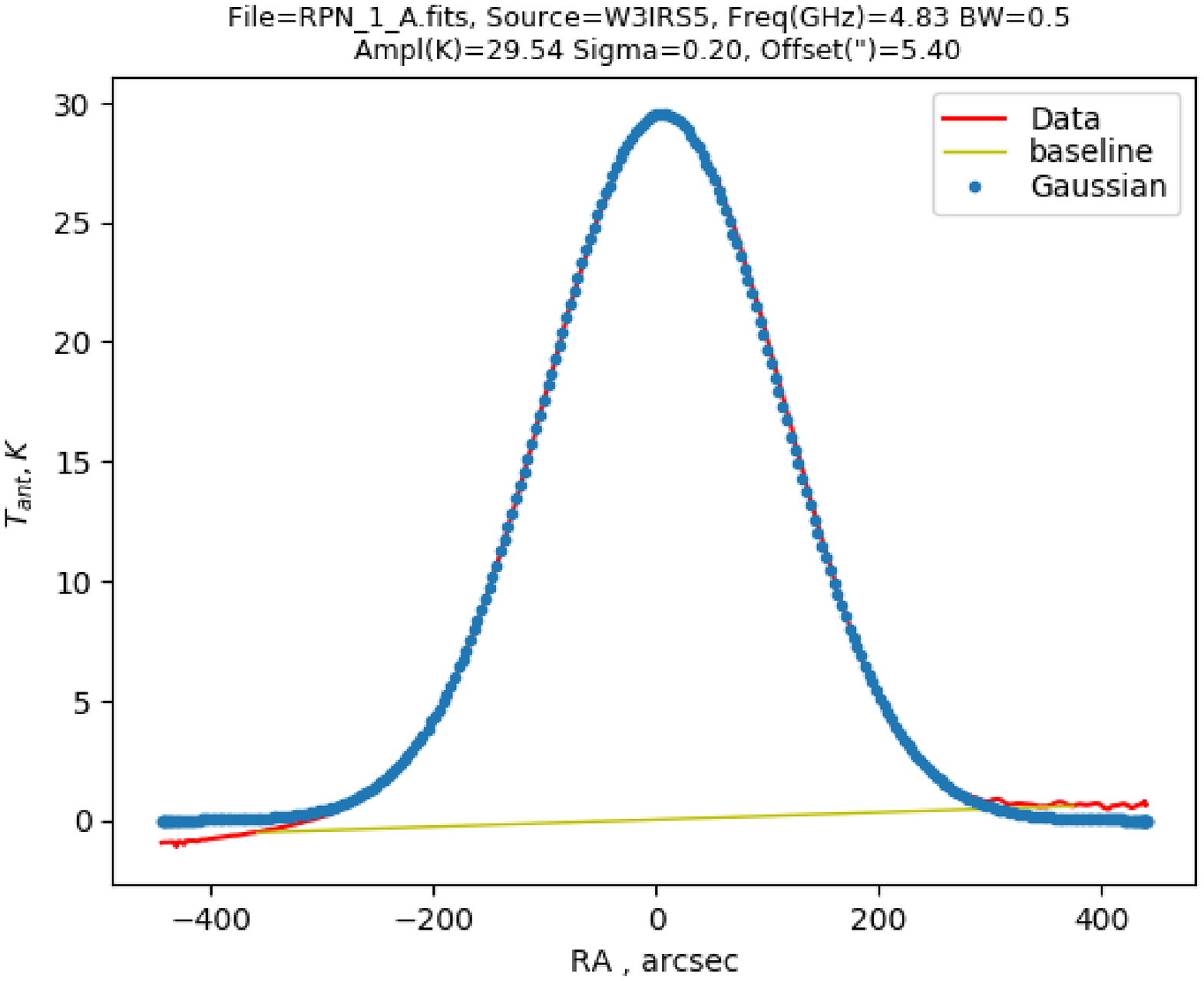}
  \includegraphics[width=200pt]{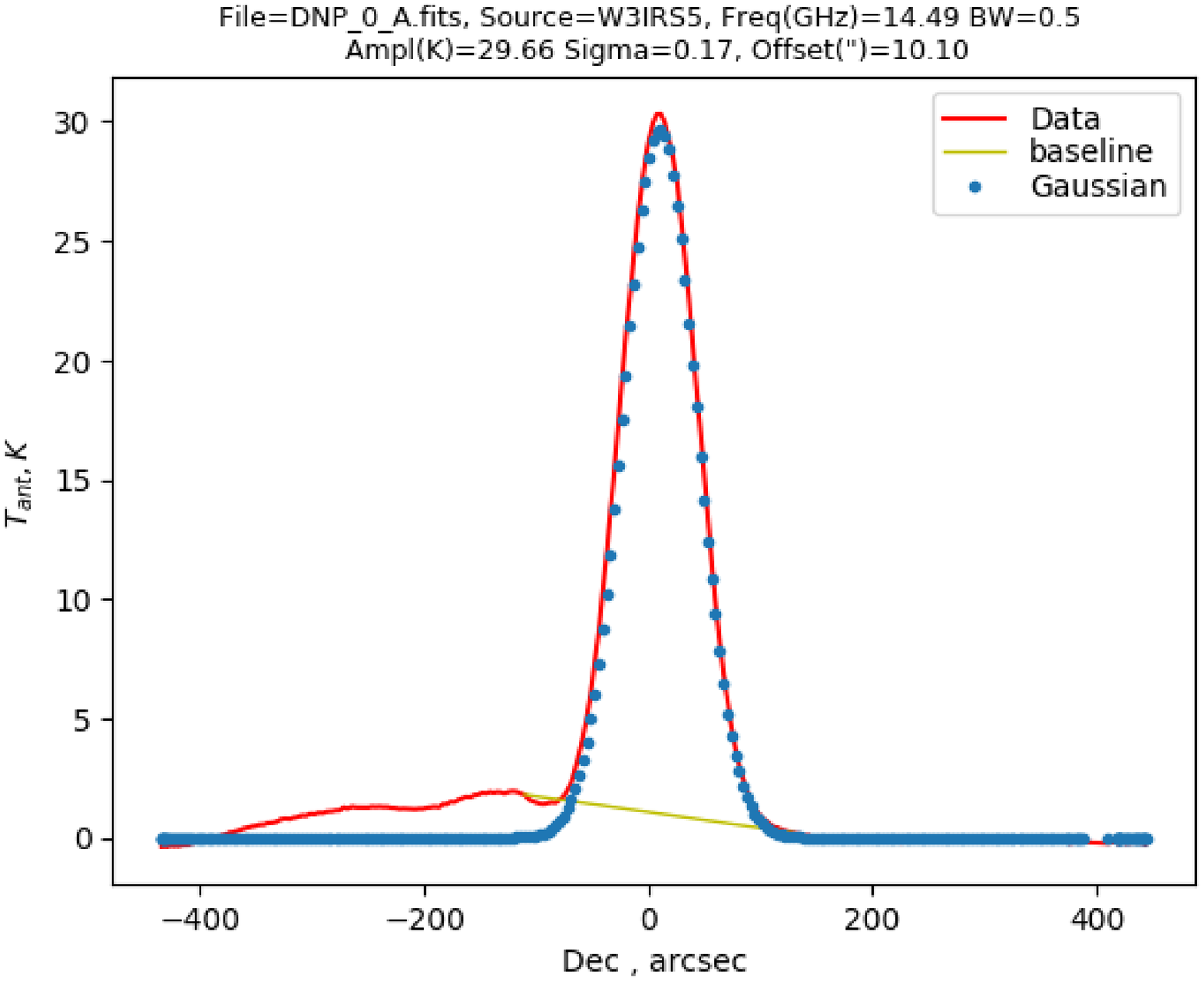}
  \caption{Continuum cross scan of W3IRS5 at 4.8 GHz(left) and 14.5 GHz (right). Red: observed cross scans. Blue: Gaussian fits. Yellowish: First order baseline fits to the measured data.}
  \label{fig2}
\end{figure*}

The continuum data calibration was obtained with the following procedure: First, baselines were subtracted, being defined by the lengths of the continuum scans avoiding the source itself at $\pm$ 1.5$\times$HPBW (half power beam width), relative to the center of the scan. Second, Gaussian profiles were fitted to obtain the position offset between the real position of the source and the center of the cross scans. Then, the obtained amplitude was corrected for the pointing error, adopting a two-dimensional Gaussian intensity profile, using the formula:
\begin{equation}
T_A^{*'} = T_A^*/Pointing,
\end{equation}
where $Pointing = exp(-4 \times ln2 \times (offset/HPBW)^2)$. Finally, the antenna temperature has been corrected for the elevation-dependent gain, defined by the parabolic equation:
\begin{equation}
T = T_A^{*'} / Gain,
\end{equation}
where $Gain = A \times Elevation^2 + B \times Elevation + C$. A, B, C are the coefficients of a 2nd order polynomial fit, obtained from "$T/T_{max} - Elevation$" plots of well-known stable calibrators (e.g. 3C286, 3C123 and 3C48). For Ku-band observations the last step of the data reduction was omitted because of an absence of a gain curve equation for this band, which provides up to $10\%$ of uncertainty. The resulting measured amplitude of each source is an average from all its scans, in units of antenna temperature in Kelvin (K). The measurement uncertainty is formed by the statistical errors of the Gaussian fits. Figure \ref{fig2} shows one example characterizing our sample of observed sources. The line parameters and continuum temperatures are listed in Table \ref{Table1}.

Then we used these parameters to calculate optical depths (\citealt{1985A&A...149..195G})
\begin{equation}
\tau = - \ln(1+T_{L}/(T_{C}+T_{BB}-T_{ex})),
\end{equation}
where $T_L$ is the observed line temperature, $T_C$ is the continuum temperature, $T_{BB}$ is the 2.7 K background radiation, and $T_{ex}$ is the excitation temperature. We have used the RADEX non-LTE model\footnote{http://var.sron.nl/radex/radex.php\label{Radex}} (\citealt{Van der Tak2007}) to provide excitation temperatures for our sources (see details in Section 3.2). The apparent maximum optical depths $\tau$, which are the optical depths at the velocities with the most negative (absorption) line temperatures, in the $2_{11}-2_{12}$ transition of H$_2$$^{12}$CO are listed in Table \ref{Table1}. In Table \ref{Table2}, the apparent maximum optical depths of the $1_{10}-1_{11}$ transitions are listed. Finally, we can get with the velocity integrated optical depth the column density, following Wilson et al.(\citeyear{1976A&A....51..303W}):
\begin{equation}
N_1[cm^{-2}] = 1.26 \times 10^{13}T_{ex}\int\tau(V)dV.
\end{equation}
The radial velocity $V$ is in km s$^{-1}$ and the excitation temperature $T_{ex}$ in K. The numerical coefficient is valid for H$_2$$^{12}$CO; the coefficient for H$_2$$^{13}$CO is 1.32 $\times$ 10$^{13}$. The column densities in the 1$_{11}$ level divided by the a priori unknown quantity $T_{ex}$ for our sources are listed in Columns 6 and 7 of Table \ref{Table2}. Used velocity ranges are given in Columns 4 and 5.

We derived the H$_2$$^{12}$CO/H$_2$$^{13}$CO isotope ratios in two different ways: (1) from Gaussian least square fits and (2) using planimetry to derive the ratio of their column densities (Col.(8) of Table \ref{Table2}; see details in Wilson et al. \citeyear{1976A&A....51..303W}). Since the ratios obtained by these two methods differ by only 10\% within the permissible margin of error, we average these two ratios to get the final H$_2$$^{12}$CO/H$_2$$^{13}$CO ratios. The average ratios corrected for telescope gain (see details in Section \ref{sec:obser}) for our sample are listed in Col.(4) of Table \ref{Table3}.

\begin{deluxetable}{c|cc|cc|cc|cc}
%% Use 8pt
\tabletypesize{\scriptsize}
%% This is the title of the table.
\tablecaption{H$_2$$^{12}$CO$\diagup$H$_2$$^{13}$CO isotope ratios obtained with the planimetry method}

%\tablenum{2}

\tablehead{
\colhead{Source} & \multicolumn{2}{c}{Apparent } & \multicolumn{2}{c}{Velocity range} & \multicolumn{2}{c}{\multirow{3}{*}{ $ \displaystyle\int_{V_1}^{V_2}\frac{N_1dV}{T_{ex}} $}} & \multirow{3}{*}{$\displaystyle R = \frac{col. (6)}{col. (7)}$} & \colhead{Distance of} \\
\colhead{ } &\multicolumn{2}{c}{maximum} & \multicolumn{2}{c}{used for integrated} & \colhead{ } & \colhead{ } & \colhead{ } & \colhead{cloud from } \\
\colhead{ } &\multicolumn{2}{c}{optical depth} & \multicolumn{2}{c}{optical depth} & \colhead{ } & \colhead{ } & \colhead{ } & \colhead{ Galactic center} \\
\colhead{ } &\colhead{H$_2$$^{12}$CO} & \colhead{H$_2$$^{13}$CO} & \colhead{V$_1$} & \colhead{V$_2$} & \colhead{H$_2$$^{12}$CO} & \colhead{H$_2$$^{13}$CO} & \colhead{ } & \colhead{} \\
\colhead{ } &\colhead{ } & \colhead{ } & \multicolumn{2}{c}{(km s$^{-1}$)} & \colhead{(10$^{13}$cm$^{-2}$K$^{-1}$)} & \colhead{(10$^{11}$cm$^{-2}$K$^{-1}$)} & \colhead{ } & \colhead{(kpc)} \\
}
\colnumbers
%% All data must appear between the \startdata and \enddata commands
\startdata
\label{Table2}
 W3IRS5  &  0.088  &  0.0018  &  -44  &  -34  &  0.353 $\pm$ 0.002  &  0.464 $\pm$ 0.082  &  76 $\pm$ 14  &  10.3 \\
Mol12  &  0.464   &  0.0221  &  -1  &  8  &  1.401 $\pm$ 0.008  &  4.698 $\pm$ 0.839  &  30 $\pm$ 5    &  9.71\tablenotemark{a} \\
NGC2024  &  0.205  &  0.0024  &  7  &  15  &  0.465 $\pm$ 0.001  &  0.778 $\pm$ 0.100  &  60 $\pm$ 7.9  &  7.22 \\
 M-0.13  &  0.431  &  0.0615  &  -25  &  71  &  17.47 $\pm$ 0.021  &  229.8 $\pm$ 1.521 &  7.6 $\pm$ 0.06  &  0.06  \\
SgrA  &  0.132  &  0.0166  &  4  &  91  &  5.01 $\pm$ 0.004  &  6.964 $\pm$ 0.401  &  9.1 $\pm$ 0.05  &  0.17  \\
 Sgr B2  &  0.116  &  0.0043  &  -13  &  23  &  1.83 $\pm$ 0.025  &  1.829 $\pm$ 0.427  &  100 $\pm$ 25 &  7.81 \\
 &  0.933  &  0.108  &  37  &  103  &  25.48 $\pm$ 0.002 &  311.97 $\pm$ 0.469  &  8.2 $\pm$ 0.01  &  0.38\tablenotemark{a} \\
 M+1.6  &  0.943   &  0.106  &  30  &  85  &  17.8 $\pm$ 0.296  &  204.3 $\pm$ 22.71 &   8.7 $\pm$ 0.97   &  0.77 \\
G10.16-0.35  &  0.063  &  0.0105  &  -8  &  4  &  1.32 $\pm$ 0.005  &  5.54 $\pm$ 0.646  &  24 $\pm$ 3  &  7.44 \\
  &  0.281  &  0.0135  &  4  &  23  &  3.48 $\pm$ 0.006  &  10.56 $\pm$ 0.693  &  33 $\pm$ 2  &  6.24 \\
W31  &  0.164  &  0.0111  &  -13  &  7  &  1.42 $\pm$ 0.001 &  10.35 $\pm$ 0.058  &  14 $\pm$ 0.1  &  3.38\tablenotemark{a} \\
  &  0.160  &  0.006  &  24  &  36  &  0.902 $\pm$ 0.001  &  1.568 $\pm$ 0.058  &  58 $\pm$ 2  &  5.0  \\
G11.93-0.61  &  0.45  &  0.022  &  32  &  43  &  2.08 $\pm$ 0.021  &  6.144 $\pm$ 1.282  &  34$\pm$ 7  &  4.88\tablenotemark{a} \\
W33  &  0.344  &  0.0135  &  25  &  43  &  2.26 $\pm$ 0.002  &  9.94 $\pm$ 0.140  &  23 $\pm$ 1  &  5.32\tablenotemark{a} \\
G13.88+0.28  &  0.17  &  0.0094  &  43  &  54  &  0.77 $\pm$ 0.004  &  1.87 $\pm$ 0.353  &  41 $\pm$ 8  &  4.41\tablenotemark{a} \\
G12.91-0.26  &  0.506  &  0.0253  &  28  &  43  &  3.61 $\pm$ 0.015 &  9.49 $\pm$ 0.748  &  38 $\pm$ 3  &  5.69\tablenotemark{a} \\
G19.62-0.23  &  0.124  &  0.008  &  32  &  50  &  1.17 $\pm$ 0.007  &  5.3 $\pm$ 0.779 &  22 $\pm$ 3 &  5.23 \\
G023.44-00.18  &  0.05  &  0.0049  &  70  &  87  &  0.32 $\pm$ 0.004  &  5.1 $\pm$ 0.623  &  6 $\pm$ 1  &  2.48 \\
  &  0.155  &  0.0065  &  90  &  115  &  1.31 $\pm$ 0.005  &  4.35 $\pm$ 0.473  &  30 $\pm$ 3  &  3.59\tablenotemark{a} \\
G23.43-0.21  &  0.174  &  0.005  &  90  &  113  &  1.4 $\pm$ 0.005 &  3.83 $\pm$ 0.445  &  37 $\pm$ 4  &  3.9 \\
G29.9-0.0  &  0.088  &  0.0042  &  86  &  108  &  0.666 $\pm$ 0.003 &  1.3 $\pm$ 0.257 &  51 $\pm$ 10  &  4.38 \\
G31.41+0.31  &  0.654  &  0.0478  &  88  &  107  &  4.71 $\pm$ 0.050  &  24.3 $\pm$ 1.934  &  19 $\pm$ 2  &  4.51 \\
W43  &  0.131  &  0.0059  &  84  &  103  &  1.16 $\pm$ 0.002  &  4.12 $\pm$ 0.167 &  28 $\pm$ 1  &  4.58  \\
G34.26+0.15  &  0.655  &  0.018  &  47  &  68  &  3.24 $\pm$ 0.011  &  9.27 $\pm$ 0.958  &  35 $\pm$ 4  &  5.58 \\
G34.3+0.1  &  0.251  &  0.0087  &  53  &  68  &  1.22 $\pm$ 0.005  &  3.93 $\pm$ 0.497  &  31 $\pm$ 4  &  5.58 \\
J185648.26  &  0.45  &  0.0153  &  35  &  54  &  3.18 $\pm$ 0.015  &  8.55 $\pm$ 1.010 &  37 $\pm$ 5  &  4.8 \\
G37.76-0.20  &  0.153  &  0.0076  &  55  &  71  &  1.05 $\pm$ 0.004  &  0.89 $\pm$ 0.415 &  118 $\pm$ 55  &  5.62 \\
G35.20-1.74  &  0.104  &  0.0034  &  38  &  49  &  0.46 $\pm$ 0.002  &  1.07 $\pm$ 0.145  &  43 $\pm$ 6  &  5.77\tablenotemark{a} \\
G35.2-1.8  &  0.1  &  0.0029  &  38  &  48  &  0.45 $\pm$ 0.046  &  0.92 $\pm$ 0.159  &  49 $\pm$ 14  &  6.16 \\
G043.16+00.01  &  0.066  &  0.0028  &  -8  &  23  &  0.78 $\pm$ 0.006  &  4.56 $\pm$ 0.293  &  17 $\pm$ 1  &  7.6\tablenotemark{a} \\
G43.17+0.00  &  0.069  &  0.0026  &  -8  &  24  &  0.83 $\pm$ 0.001  &  3.15 $\pm$ 0.107  &  26 $\pm$ 1  &  7.56 \\
G43.2+0.0  &  0.09  &  0.0216  &  -8  &  24  &  1.04 $\pm$ 0.009  &  6.04 $\pm$ 0.299  &  17 $\pm$ 1  &  7.55 \\
G45.45+0.06  &  0.128  &  0.0065  &  52  &  70  &  0.75 $\pm$ 0.003  &  3.02 $\pm$ 0.318  &  25 $\pm$ 3  &  6.39\tablenotemark{a} \\
G49.21-0.35  &  0.224  &  0.0049  &  58  &  72  &  0.98 $\pm$ 0.002  &  0.78 $\pm$ 0.240  &  126 $\pm$ 39  &  6.15  \\
J192311.17  &  0.042  &  0.0033  &  44  &  56  &  0.25 $\pm$ 0.002  &  0.31 $\pm$ 0.155 &  81 $\pm$ 41  &  6.31 \\
  &  0.246  &  0.0056  &  57  &  73  &  1.35 $\pm$ 0.002  &  2.19 $\pm$ 0.155  &  62 $\pm$ 4  &  6.17 \\
G49.4-0.3  &  0.218  &  0.0037  &  56  &  73  &  1.49 $\pm$ 0.003  &  1.58 $\pm$ 0.121  &  94 $\pm$ 7  &  6.17 \\
G49.5-0.4  &  0.128  &  0.0054  &  47  &  76  &  1.4 $\pm$ 0.002  &  5.09 $\pm$ 0.235  &  28 $\pm$ 1  &  6.18 \\
J192345.73  &  0.117  &  0.0032  &  47  &  76  &  1.36 $\pm$ 0.002  &  4.84 $\pm$ 0.266 &  28 $\pm$ 2  &  6.18\\
J203901.04  &  0.221  &  0.0066  &  -8  &  2  &  1.05 $\pm$ 0.002  &  1.44 $\pm$ 0.165  &  73 $\pm$ 8  &  8.14 \\
DR21  &  0.297  &  0.0077  &  -8  &  2  &  1.38 $\pm$ 0.002  &  2.92 $\pm$ 0.167  &  47 $\pm$ 3  &  8.05\tablenotemark{a} \\
J205658.56  &  0.935   &  0.051  &  -5  &  9  &  3.67 $\pm$ 0.025   &  12.0 $\pm$ 1.729 &  31 $\pm$ 5  &  7.38 \\
J205703.98  &   1.604  &  0.051  &  -4  &  5  &  5.10 $\pm$ 0.033   &  11.7 $\pm$ 1.915 & 44 $\pm$ 7 &  7.37 \\
\enddata
\tablenotetext{a}{From Reid et al. (2014).}
%% Include any \tablenotetext{key}{text}, \tablerefs{ref list},
%% or \tablecomments{text} between the \enddata and
%% \end{deluxetable} commands

\tablecomments {Column (1): source name; Columns (2) and (3): the apparent optical depths of H$_2$$^{12}$CO and H$_2$$^{13}$CO $1_{10}-1_{11}$ respectively; Columns (4) and (5): the velocity range used for integrated optical depth; Columns (6) and (7): the values of the column densities in the 1$_{11}$ level divided by $T_{ex}$ for H$_2$$^{12}$CO and H$_2$$^{13}$CO respectively; Column (8): the H$_2$$^{12}$CO$\diagup$H$_2$$^{13}$CO isotope ratios obtained with the planimetry method; Column (9): the distance from the Galactic center.}

%% No \tablerefs indicated
\end{deluxetable}

\subsection{Beam Size Effect} \label{subsec:beam}

The beam sizes are $\sim$4 arcmin at C-band and $\sim$1.3 arcmin at Ku-band, respectively. The temperature difference between the microwave background (2.73 K) and the excitation temperature of the collisionally cooled (\citealt{1975ApJ...196..433}) H$_2$CO lines is about 1.5 K, as derived from our non-LTE calculations. For a cloud fully filling the beam, and being optically moderately thin, e.g. $\tau = 0.3$, this would result in about 0.5 K across that part of the beam not covered by a background H\uppercase\expandafter{\romannumeral2} region. Assuming that the H\uppercase\expandafter{\romannumeral2} region has a diameter of 6 arcsec (e.g., \citealt{2009ApJ...689..1422}) and a radiation temperature of 6000 K (e.g., \citealt{1970A&A...4..357}), we obtain for T$_{non-H\uppercase\expandafter{\romannumeral2}-region(H_2CO)}$/T$_{H\uppercase\expandafter{\romannumeral2}-region(H_2CO)}$ values of about 0.01 and 0.1 for a 1.3 arcmin beam and 4 arcmin beam, respectively. The result indicates that in this case the tiny spot in front of the H\uppercase\expandafter{\romannumeral2} region dominates the H$_2$CO absorption budget and the difference in beam sizes can be neglected. If the respective clouds are not filling the entire 4 arcmin beam, the comparatively small area covered by the continuum emission becomes even more dominant and the difference in beam size can also be neglected.

\subsection{Photon Trapping Corrections } \label{subsec:correct}
\begin{figure*}[h]
\center
  \includegraphics[width=300pt]{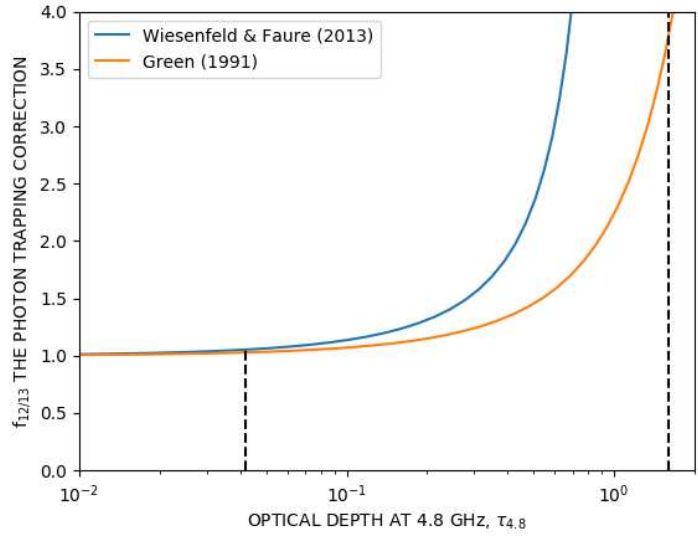}
  \caption{The comparison of $f_{12/13}$ calculated with the new collision rates (blue line; \citealt{Wiesenfeld2013}) and the collision rates of Green (\citeyear{1991ApJS...76..979}; orange line) for a kinetic temperature of 30 K and a molecular hydrogen density of 10$^{5}$ cm$^{-3}$. The black dotted lines indicate the range of optical depth of our sources at 4.8 GHz.}
  \label{fig3}
\end{figure*}

\begin{figure*}[h]
\center
  \includegraphics[width=300pt]{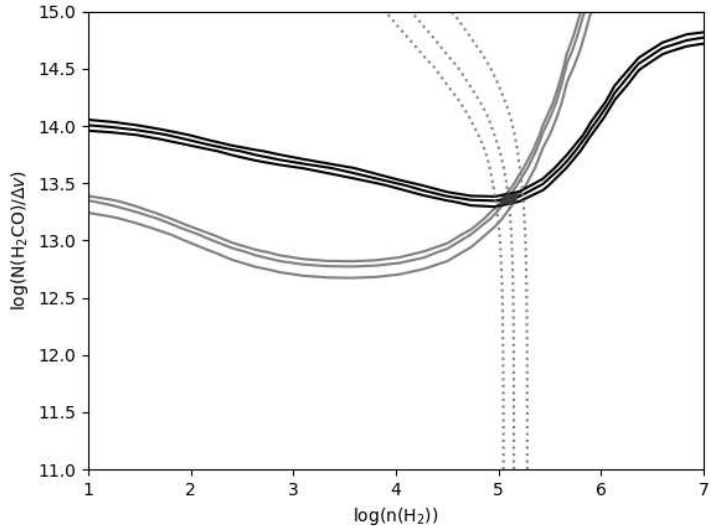}
  \caption{Determination of the H$_2$ density for one of our sample sources, W3-IRS5. The black and gray solid curves reproduce the 1$_{10}$-1$_{11}$ and 2$_{11}$-2$_{12}$ optical depths, respectively. The almost vertical gray dotted curves delineate the ratio of these two optical depths. For each of these parameters, the central line refers to the measured value, while the pair of curves around it shows the standard deviations (rms), which were calculated with the rms noise obtained from Gaussian-fitting of the 1$_{10}$-1$_{11}$ and 2$_{11}$-2$_{12}$ transitions. The small shaded region where the curves overlap shows the allowed parameter space for which the physical parameters are derived.}
  \label{fig4}
\end{figure*}

\begin{figure*}[h]
\center
  \includegraphics[width=300pt]{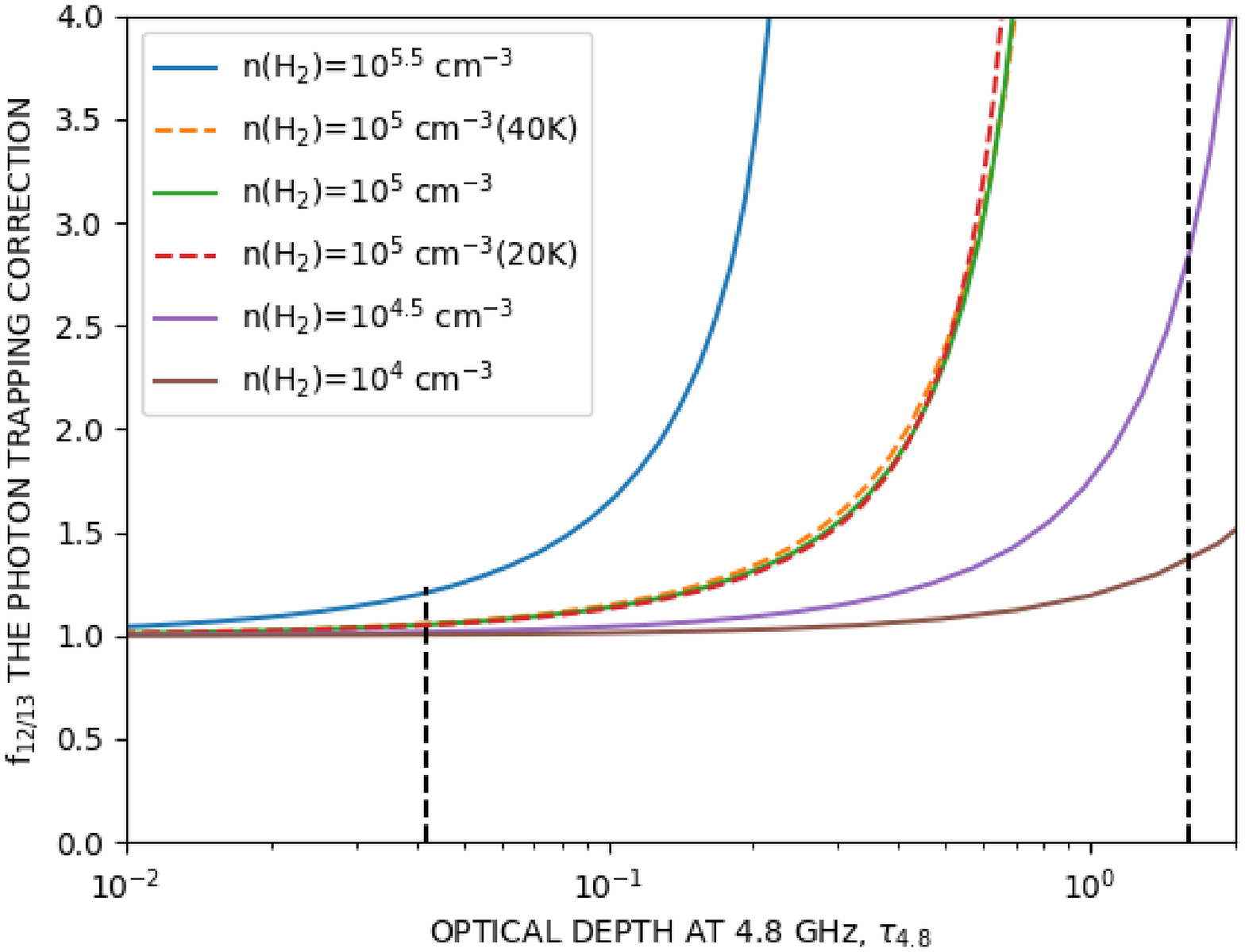}
  \caption{The full lines are curves showing the theoretical dependence of the correction factor for photon trapping $f_{12/13}$ on the 4.8 GHz optical depth of H$_2$$^{12}$CO for a kinetic temperature of 30 K, and for molecular hydrogen densities varying between 10$^4$ cm$^{-3}$ and 10$^{5.5}$ cm$^{-3}$. The dashed lines show the correction factor of two temperatures (20K, 40K) for n(H$_2$) = 10$^5$ cm$^{-3}$. The black dotted lines indicate the range of optical depth of our sources at 4.8 GHz.}
  \label{fig5}
\end{figure*}

Photon trapping in the millimeter H$_2$CO rotation lines, connecting the J = 1 and 2 K-doublets (i.e. the 2$_{11}$ - 1$_{10}$ and 2$_{12}$ - 1$_{11}$ transitions), has also to be considered. In case the two millimeter lines (2$_{11}$ - 1$_{10}$ and 2$_{12}$ - 1$_{11}$) are not entirely optically thin, their excitation temperature values rise and are higher than in the optically thin case, represented by H$_2$$^{13}$CO. Thus, more population ends up in the H$_2$CO J = 2 doublet than in the entirely optically thin case represented by H$_2$$^{13}$CO. The J = 1 H$_2$CO doublet then gets a little depopulated and thus the C-band H$_2$CO/H$_2$$^{13}$CO line intensity ratio becomes smaller than if both H$_2$CO and H$_2$$^{13}$CO lines were all optically thin. Here, the RADEX non-LTE model\textsuperscript{\ref{Radex}} was used to correct the ratios for photon trapping. As already mentioned in Sect. \ref{sec:intro}, the collision rates of H$_2$CO are taken from \citet{Wiesenfeld2013}, which are calculated for H$_2$CO in collision with H$_2$ and the
high accuracy potential energy surface of Troscompt et al. (\citeyear{2009aA&A...493..687}). These are significantly different from the scaled rates of H$_2$CO in collision with He (\citealt{1991ApJS...76..979}; see below). Correction factors $f_{12/13}$ calculated by the new collision rates tend to be larger than those derived by the old collision rates, especially for larger optical depths at 4.8 GHz, as shown in Figure \ref{fig3}. The correction factor $f_{12/13}$ is defined as in Henkel et al. (\citeyear{1980A&A...82..41}):
\begin{equation}
f_{12/13} = 50 \tau^{0}_{4.8}/\tau^{*}_{4.8},
\end{equation}
where the star indicates the RADEX model and the superscript zero refers to the model with the same H$_2$ density and temperature but with a formaldehyde column density which is a factor 50 times lower. The results are listed in Table \ref{Table3}. We ran the RADEX offline code to independently estimate the H$_2$ density for our sources, and find in good agreement that it ranges from 10$^{4.7}$ cm$^{-3}$ to 10$^{5.3}$ cm$^{-3}$. An example of this fitting process (see details in \citealt{2011ApJ...736..149}) is shown in Figure \ref{fig4}. The H$_2$ density of the H{\uppercase\expandafter{\romannumeral2}} regions is $\thicksim$ 10$^5$ cm$^{-3}$ (\citealt{1980A&A...82..41}; \citealt{2011ApJ...736..149}; \citealt{2017A&A...598..30}). For the 6 sources for which we only derived the upper limit of the 2$_{11}$-2$_{12}$ optical depths due to the non-detection of continuum at Ku-band, we assume that the H$_2$ density in these 6 sources is also $\thicksim$ 10$^5$ cm$^{-3}$. Thus we adopt n(H$_2$) = 10$^5$ cm$^{-3}$ for all of our sources. In Figure \ref{fig5}, we show the dependence of $f_{12/13}$ on $\tau_{4.8}$ for various molecular hydrogen densities and temperatures. As expected, the corrections to the measured isotope ratios become more important with larger optical depth. As for the kinetic temperatures, molecular clouds near ultracompact H\uppercase\expandafter{\romannumeral2} (UCH\uppercase\expandafter{\romannumeral2}) regions have a temperature of around 40 K, which is warmer than the molecular environment of more evolved H\uppercase\expandafter{\romannumeral2} regions (\citealt{Rivera2010ApJ}; \citealt{2011ApJ...736..149}). Therefore, we chose the kinetic temperature range of 20K-40K to analyze the effect of temperature on $f_{12/13}$. The results show that there is little difference for $f_{12/13}$ in the kinetic temperature range of 20K-40K, as can be seen in Figure \ref{fig5}. Thus we adopt 30 K as the kinetic temperature for our sources.
\subsection{Hyperfine Structure}  \label{subsec:hyperfine}

Both the $1_{10}-1_{11}$ H$_2$$^{12}$CO and H$_2$$^{13}$CO lines are split by hyperfine interactions (\citealt{1971ApJ...169..429T}). The H$_2$$^{12}$CO line consists of six hyperfine components extended over 30 kHz (\citealt{1971J.Mol...38..130}; \citealt{1972Phys.Chem...1..1011}; \citealt{1979MNRAS...188..331}), corresponding to a velocity range between -1.13 km s$^{-1}$ and +0.71 km s$^{-1}$ relative to the line center (4829.66 MHz). For line widths of order 1 km s$^{-1}$, the line profile can be characterized by two components. 89\% of the intensity go into features between -0.07 km s$^{-1}$ and +0.71 km s$^{-1}$, 11\% into the feature at -1.13 km s$^{-1}$. This implies that for wider lines, 100\% of the intensity can be observed. While for very narrow lines from dark clouds, the -1.13 km s$^{-1}$ feature may be seen separately (e.g., \citealt{1981A&A...99..270}). However, this did not occur in our sources. The hfs slightly broadens the H$_2$$^{13}$CO lines with respect to the H$_2$$^{12}$CO lines (\citealt{1974ApJ...189..217}). This should be taken into account. For 21 hyperfine components in the H$_2$$^{13}$CO $1_{10}-1_{11}$ transition, 77\% go into components between -1.15 km s$^{-1}$ and +1.15 km s$^{-1}$ relative to the line center (4593.089 MHz), while 11.5\% go into -8.5 and -6.0 km s$^{-1}$ and another 11.5\% into +5.7 to +8.1 km s$^{-1}$ features. Thus, in case of weakly detected H$_2$$^{13}$CO lines, we obtain only 77\% of the total intensity for most of our sources. However, in case of really strong H$_2$$^{13}$CO lines, we can also see the outer features near -7 and +7 km s$^{-1}$. Based on the different S/N ratio of each source, we derived the factors of hyperfine splitting ($f_{hfs}$) individually for each source. The corrections for hyperfine splitting, also accounting for the fact that H$_2$CO and H$_2$$^{13}$CO are measured at slightly different frequencies ($(\nu_{13}/\nu_{12})\cdot f_{hfs}$), are listed in Column 5 of Table \ref{Table3}.

With corrections due to photon trapping ($f_{12/13}$), hyperfine splitting ($f_{hfs}$) and line frequencies, the corrected isotope ratios are derived from the measured isotope ratio R for our 38 sources (\citealt{1985A&A...149..195G}; \citealt{1985A&A...143..148H}):
\begin{equation}
[H_2^{12}CO]/[H_2^{13}CO] = (\nu_{13}/\nu_{12})\cdot f_{hfs}\cdot f_{12/13}\cdot R.
\end{equation}

For comparison, previously published ratios are also collected and are included in Table \ref{Table4}. There are 15 sources from H$_2$CO/H$_2$$^{13}$CO measurements (Henkel et al. \citeyear{1980A&A...82..41},\citeyear{1982A&A...109..344},\citeyear{1983A&A...127..388},\citeyear{1985A&A...143..148H}), 18 from CN/$^{13}$CN measurements (\citealt{2002ApJ...578..211}; \citealt{2005ApJ...634...1126}) and 9 from C$^{18}$O/$^{13}$C$^{18}$O measurements (e.g., \citealt{1990ApJ...357..477}; \citealt{1996A&AS...119..439}; \citealt{1998ApJ...494...L107}). With respect to these previous measurements, our sample of 38 sources and 43 velocity components covering a wide range of Galactocentric distances provides a significant sample for studying the radial gradient of the isotope ratio $^{12}$C/$^{13}$C.

\subsection{Distances} \label{subsec:distance}
Therefore, we also need Galactocentric distances for our sources that are more accurate than those used in previous less extended surveys. The trigonometric parallax is a very accurate method to measure the distance of sources, which can directly and geometrically determine source distances from the Sun (Reid et al. \citeyear{2009ApJ...700..137}; \citeyear{2014ApJ...783..130}). Based on trigonometric parallax data, Reid et al. (\citeyear{2014ApJ...783..130}) provide an accurate method (see details in \citealt{2009ApJ...700..137}) for estimating revised kinematic distances with improved Galactic constants and rotation curve. Thus we derived the heliocentric distance for 12 of our sources (marked in Table \ref{Table2}) from their trigonometric parallax data (\citealt{2014ApJ...783..130}). For the other 26 sources without trigonometric parallax data, we estimated their heliocentric distance using the Revised Kinematic Distance Calculator\footnote{\url{http://bessel.vlbi-astrometry.org/revised_kd_2014}}, also based on the results of the parallax measurements. This also includes 7 sources, which are not part of our sample but which were included in previous measurements. A more accurate distance was adopted here (listed in Table \ref{Table4}), including 5 sources (W3-OH, W51M, Orion A, Orion Bar and NGC7538) with trigonometric parallax distances (\citealt{2007A&A...474..515}; \citealt{2014ApJ...783..130}) and 2 sources (S156, WB 391) with new heliocentric distances from the Revised Kinematic Distance calculator. The latter were obtained with a solar rotational velocity of V$_0$ = 240 km s$^{-1}$, assuming the derivative of the rotation curve beyond R = R$_0$ (for the solar galactocentric value, we adopted 8.125 kpc, \citealt{2018A&A...615..L15}) to be dV/dR = 0.2 km s$^{-1}$ kpc$^{-1}$ and the solar motion with respect to the LSR to be (U$_\odot$, V$_\odot$, W$_\odot$) = (10.7, 15.6, 8.9) km s$^{-1}$ (\citealt{2014ApJ...783..130}). Then, we can determine the Galactocentric distance from the heliocentric distance (\citealt{2009ApJ...699...1153}):
\begin{equation}
r = \sqrt{( R_0 cos(l) - d )^2 + R_0^2 sin^2(l)},
\end{equation}
where $l$ is the Galatic longitude of the source and $d$ is the kinematic distance.

The new Galactocentric distances for our 38 detected sources and 7 sources not in our sample are listed in Table \ref{Table2} and Table \ref{Table4}, respectively.

\begin{deluxetable}{ccccccc}
\tablecaption{The $^{12}$C/$^{13}$C isotope ratios from corrected H$_2$$^{12}$CO/H$_2$$^{13}$CO ratios}
%\tablenum{3}
\tablehead{\colhead{Source} & \multicolumn{2}{c}{Excitation temperature} & \colhead{ Average ratios } & \colhead{Corrections }& \colhead{Corrections} & \colhead{Corrected} \\
\colhead{} & \colhead{4.8 GHz} & \colhead{14.5 GHz} & \colhead{with corrections due }& \colhead{for Line Frequencies} & \colhead{for Photon} & \colhead{ Ratios}\\
\colhead{} & \colhead{ } & \colhead{ } & \colhead{to telescope gain }& \colhead{and Hyperfine Effects} & \colhead{ Trapping} & \colhead{} \\
\colhead{} & \colhead{(K)} & \colhead{(K)} & \colhead{} & \colhead{$(\nu_{13}/\nu_{12})\cdot f_{hfs}$} & \colhead{$f_{12/13}$ } & \colhead{}}
\colnumbers
\startdata
\label{Table3}
 W3IRS5 & 1.21  & 1.56  & 84 $\pm$ 15 &0.67 & 1.06  &         60$\pm$ 11       \\
Mol12 &  1.85 & 2.03 & 37 $\pm$ 6 &0.68 & 1.89 &              47 $\pm$ 8       \\
NGC2024 & 1.31  & 1.62  & 65 $\pm$ 8 &0.68 & 1.18  &          52 $\pm$ 6       \\
 M-0.13 & 1.45  & 1.71  & 8.59 $\pm$ 0.06 &0.95 & 1.36  &     11.10 $\pm$ 0.08 \\
SgrA & 1.22  & 1.57  & 10.45 $\pm$ 0.11&0.95  & 1.08  &       10.71 $\pm$ 0.11 \\
Sgr B2 & 1.20  & 1.56  & 96 $\pm$ 18  &0.76& 1.05  &          77 $\pm$ 15      \\
 & 1.46  & 1.72  & 8.78 $\pm$ 0.02 &0.95 & 1.38  &            11.48 $\pm$ 0.03 \\
 M+1.6 & 1.26 & 1.59 & 9.45 $\pm$ 0.63&0.95  & 1.13 &         10.15 $\pm$ 0.68 \\
G10.16-0.35 & 1.33  & 1.63& 30 $\pm$ 5 &0.66 & 1.21 &         24 $\pm$ 4       \\
 & 1.35  & 1.64  & 45 $\pm$ 3 &0.75 & 1.23  &                 42 $\pm$ 3       \\
W31 & 1.17  & 1.55  & 21$\pm$ 5\tablenotemark{*} &0.76 & 1.02 &               16 $\pm$ 4       \\
 & 1.17  & 1.55  & 64 $\pm$ 7 &0.71 & 1.02  &                 47 $\pm$ 5       \\
G11.93-0.61 & 1.69  & 1.90  & 36 $\pm$ 6 &0.74 & 1.68 &       45 $\pm$ 7        \\
W33 & 1.28  & 1.60  & 27 $\pm$ 1 &0.74 & 1.15  &              23 $\pm$ 1        \\
G13.88+0.28 & 1.31  & 1.62  & 43 $\pm$ 7 &0.68 & 1.18 &       35 $\pm$ 6        \\
G12.91-0.26 & 1.65  & 1.87  & 41 $\pm$ 3 &0.74 & 1.63  &      49 $\pm$ 3        \\
G19.62-0.23 & 1.28  & 1.60  & 34 $\pm$ 4 &0.73 & 1.15   &     28 $\pm$ 3        \\
G023.44-00.18 & 1.20  & 1.56  & 23 $\pm$ 17\tablenotemark{*} &0.74 & 1.05 &      18 $\pm$ 13        \\
 & 1.28  & 1.60  & 34 $\pm$ 3 &0.76 & 1.15  &                 30 $\pm$ 3       \\
G23.43-0.21 & 1.28  & 1.60  & 44 $\pm$ 5 &0.76 & 1.15 &       39 $\pm$ 4        \\
G29.9-0.0 & 1.23  & 1.57  & 54 $\pm$ 11 &0.76 & 1.09   &      44 $\pm$ 9        \\
G31.41+0.31 & 1.95  & 2.13  & 21 $\pm$ 2 &0.76 & 2.04 &       32 $\pm$ 2        \\
W43 & 1.25  & 1.58  & 31 $\pm$ 1 &0.77 & 1.11  &              26 $\pm$ 1        \\
G34.26+0.15 & 1.66  & 1.87  & 38 $\pm$ 3 &0.73 & 1.64  &      46 $\pm$ 4        \\
G34.3+0.1 & 1.34  & 1.64  & 35 $\pm$ 4 &0.74 & 1.22  &        31$\pm$ 4        \\
J185648.26 & 1.55  & 1.79  & 43 $\pm$ 4 &0.75 & 1.49  &       48 $\pm$ 5        \\
G37.76-0.20 & 1.30  & 1.62  & 42 $\pm$ 28\tablenotemark{*} &0.69 & 1.18  &    34 $\pm$ 23       \\
G35.20-1.74 & 1.23  & 1.58  & 53 $\pm$ 8 &0.74 & 1.09  &      43 $\pm$ 6       \\
G35.2-1.8 & 1.23  & 1.58  & 72 $\pm$ 15 &0.68 & 1.09  &       53 $\pm$ 11       \\
G043.16+00.01 & 1.20  & 1.56  & 39 $\pm$ 14\tablenotemark{*} &0.95 & 1.05  &    39 $\pm$ 14       \\
G43.17+0.00 & 1.20  & 1.56  & 45$\pm$ 14\tablenotemark{*} &0.84 & 1.05  &       40 $\pm$ 13       \\
G43.2+0.0 & 1.21  & 1.57  & 89 $\pm$ 57\tablenotemark{*} &0.77 & 1.07  &        72$\pm$ 46        \\
G45.45+0.06 & 1.26  & 1.59  & 36 $\pm$ 5 &0.74 & 1.12  &      30 $\pm$ 4       \\
G49.21-0.35 & 1.32  & 1.63  & 119 $\pm$ 29 &0.69 & 1.20   &   99 $\pm$ 24       \\
J192311.17 & 1.19  & 1.55  & 26 $\pm$ 13\tablenotemark{*} &0.67 & 1.04   &     18 $\pm$ 9       \\
 & 1.33  & 1.63  & 66 $\pm$ 5 &0.75 & 1.20 &                  59 $\pm$ 4        \\
G49.4-0.3 & 1.34  & 1.64  & 95 $\pm$ 8 &0.74 & 1.22  &        86 $\pm$ 7         \\
G49.5-0.4 & 1.25  & 1.58  & 50 $\pm$ 18\tablenotemark{*} &0.81 & 1.11  &        45 $\pm$ 16         \\
J192345.73 & 1.24  & 1.58  & 22 $\pm$ 9\tablenotemark{*} &0.81 & 1.10  &       19 $\pm$ 8         \\
J203901.04 & 1.31  & 1.62  & 76 $\pm$ 8 &0.71 & 1.18  &       64 $\pm$ 7         \\
DR21 & 1.36  & 1.65  & 55 $\pm$ 3 &0.71& 1.24  &              48 $\pm$ 3         \\
J205658.56 & 1.67 & 1.88 & 38 $\pm$ 5 &0.67 & 1.65 &          42$\pm$ 5         \\
J205703.98 & 1.71 & 1.91 & 23 $\pm$ 7\tablenotemark{*} &0.66 & 1.70 &          26 $\pm$ 8         \\
\enddata
\tablenotetext{*}{For those sources with poor H$_2$$^{13}$CO Signal-to-Noise ratios, the abundance ratios are derived from the peak value, instead of the integrated intensity.}
\tablecomments {Column (1): source name; Columns (2) and (3): the excitation temperatures of H$_2$$^{12}$CO $1_{10}-1_{11}$ and 2$_{11}$-2$_{12}$ lines derived from the RADEX model, respectively; Column (4): H$_2$CO/H$_2$$^{13}$CO line intensity ratios (averages from values obtained by Gaussian fits and the planimetry method) corrected for differences in telescope gain; Column (5): the corrections for line frequencies and hyperfine effects; Column (6): the correction factor $f_{12/13}$ obtained from RADEX; Column (7): the corrected H$_2$$^{12}$CO$\diagup$H$_2$$^{13}$CO isotope ratios.}
\end{deluxetable}

\begin{figure*}
\center
  \includegraphics[width=500pt]{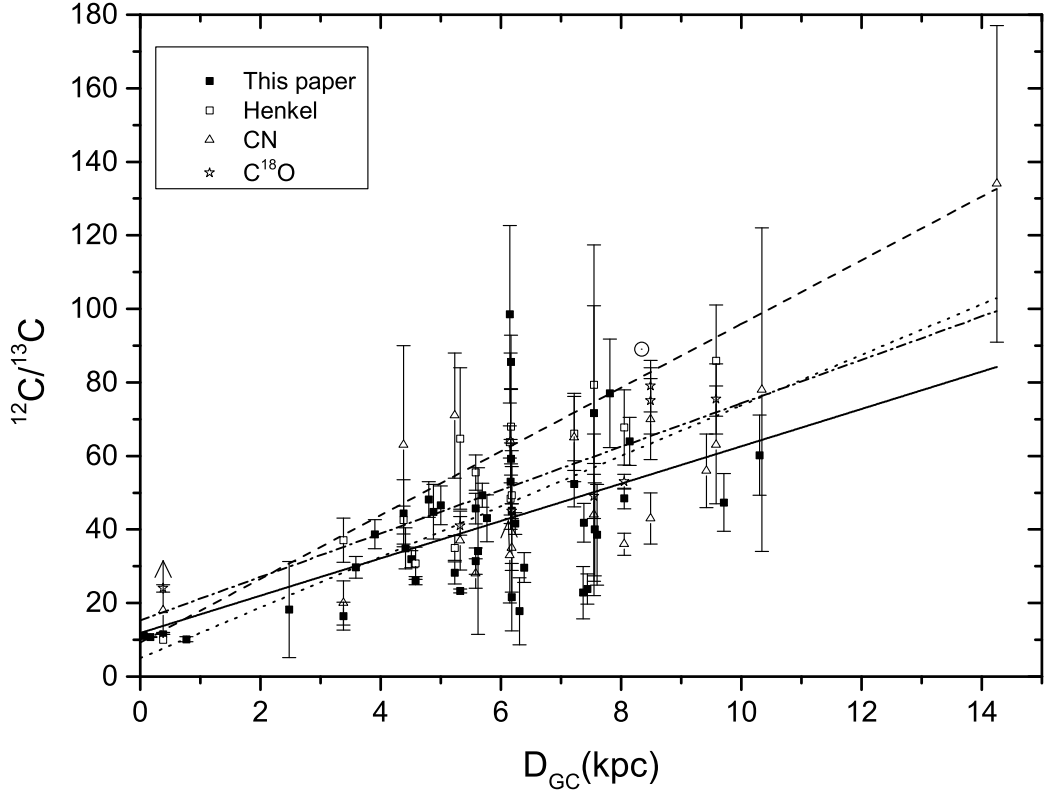}
  \caption{The $^{12}$C/$^{13}$C isotope ratios from H$_2$CO/H$_2$$^{13}$CO, CN/$^{13}$CN, and C$^{18}$O/$^{13}$C$^{18}$O plotted as functions of the distance from the Galactic center. The filled black squares are the results derived in the current work from H$_2$CO and the result of a first order polynomial fit is plotted as a solid line. The open squares are values derived with H$_2$CO from Henkel et al.(1980, 1982, 1983, 1985), using state of the art distances, and are fitted by a dashed line. The empty triangles and stars are values from CN (Savage et al. 2002, Milam et al. 2005) and C$^{18}$O (Langer \& Penzias 1990, Wouterloot \& Brand 1996, Keene et al. 1998), respectively, also using modified distances. The dotted line presents the linear fit from CN, and the dash-dotted line is the fit from C$^{18}$O. The symbol $\odot$ indicates the $^{12}$C/$^{13}$C isotope ratio of the Sun. All of the values are also presented in Table \ref{Table4}. }
  \label{fig6}
\end{figure*}

\begin{figure*}
\center
  \includegraphics[width=400pt]{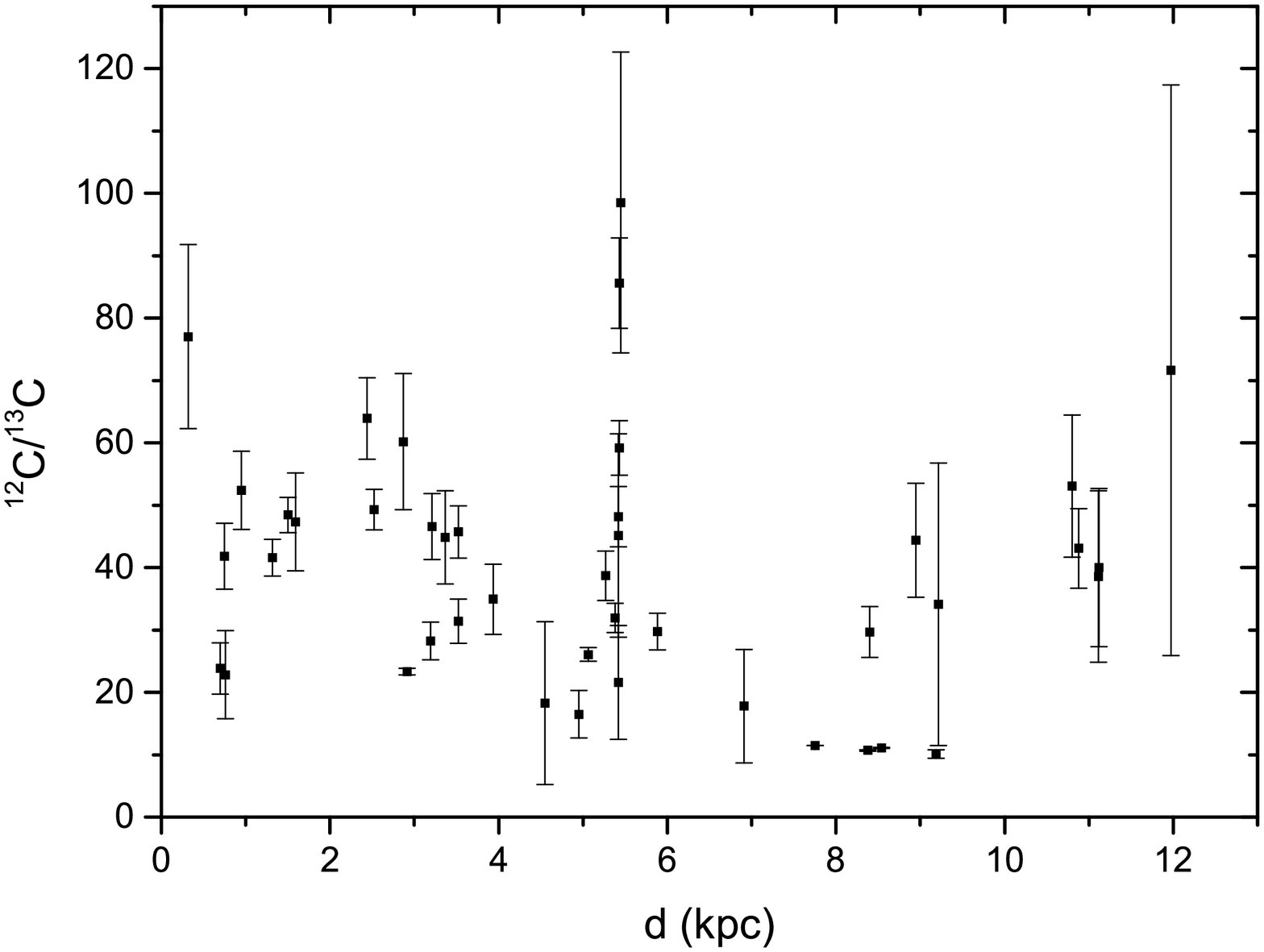}
  \caption{The $^{12}$C/$^{13}$C isotope ratios from our H$_2$CO/H$_2$$^{13}$CO measurements plotted as functions of the distance from the Sun.}
  \label{fig7}
\end{figure*}

\section{Discussion} \label{sec:discussions}
With the accurate Galactocentric distance and the corrected H$_2$$^{12}$CO/H$_2$$^{13}$CO ratios, we can study the variation of the $^{12}$C/$^{13}$C isotope ratios as a function of Galactocentric distance. We show our results as filled black squares in Figure \ref{fig6}. Previous results of the ratio are also plotted, but against the new distance values (see details in Sect. \ref{subsec:distance}), and new linear fitting lines are also presented in Figure \ref{fig6}. A comparison shows that using the new distances really affects the fitting results (see details in Table \ref{Table5}). E.g., for the gradient results from H$_2$$^{12}$CO/H$_2$$^{13}$CO in Henkel et al. (\citeyear{1980A&A...82..41},\citeyear{1982A&A...109..344},\citeyear{1983A&A...127..388},\citeyear{1985A&A...143..148H}), the slope/intercept becomes (8.66 $\pm$ 1.64)/(9.30 $\pm$ 10.37) from (7.60 $\pm$ 1.79)/(18.05 $\pm$ 10.88). The fitting for CN/$^{13}$CN in Milam et al. (\citeyear{2005ApJ...634...1126}) becomes (6.87 $\pm$ 1.46)/(5.06 $\pm$ 11.07) from (6.01 $\pm$ 1.19)/(12.28 $\pm$ 9.33). In addition we give in Table \ref{Table5} the ratios that would be derived using the old distances and the collision rates of Green et al. (\citeyear{1991ApJS...76..979}). It is clear that the $^{12}$C/$^{13}$C gradient in the case of the collision rates taken from \citet{Wiesenfeld2013} and new distances are closer to the gradient derived from CN and C$^{18}$O. In order not to bias our results towards low values with small error bars, the gradient was calculated from an unweighted least-squares fit of our corrected H$_2$$^{12}$CO/H$_2$$^{13}$CO ratios with the collision rates taken from \citet{Wiesenfeld2013} and the new, more reliable distances:
\begin{equation}
^{12}C/^{13}C=(5.08\pm1.10)D_{GC}+(11.86\pm6.60).
\end{equation}
This gradient is flatter than the gradient derived from previous H$_2$CO measurements (Henkel et al. \citeyear{1980A&A...82..41},\citeyear{1982A&A...109..344}) but agrees very well with those from CN and C$^{18}$O measurements (\citealt{2005ApJ...634...1126}).

Although our results are more reliable (due to our bigger sample, the new collision rates, more accurate distances and presumably a more realistic determination of the photon trapping effect), some other uncertainties (observational bias due to different distances, beam sizes, excitation temperatures, isotope selective photodissociation and chemical fractionation) should be mentioned.

\begin{enumerate}

\item Observational bias due to distance effects

The $^{12}$C/$^{13}$C isotope ratios from our H$_2$CO/H$_2$$^{13}$CO measurements plotted as functions of the distance from the Sun are shown in Figure \ref{fig7}. No apparent gradient can be found, which suggests that any observational bias due to distance related effects is not significant for the $^{12}$C/$^{13}$C gradient as a function of Galactocentric distance.

\item Excitation temperature

We assumed that the excitation temperature of the $1_{10}-1_{11}$ transition for H$_2$CO equals that for H$_2$$^{13}$CO. It is generally accepted that these lines are affected by a collisional pumping mechanism (\citealt{1975ApJ...196..433}; \citealt{1976A&A....51..303W}), which should produce nearly the same excitation temperature for the two species. Hence any differences in the excitation between the two species should be negligible.

\item Isotope selective photodissociation

Isotope selective photodissociation can occur in a molecular cloud affected by high UV radiation, which increases the $^{12}$C/$^{13}$C isotope ratio (\citealt{2010A&A...523..A51}). This effect is particularly pronounced in photon-dominated regions (PDRs). The UV photons can photodissociate the rarer isotopologues and affect less the main isotopologues (the more abundant molecules) due to their more efficient self-shielding, which would lead to higher $^{12}$C/$^{13}$C isotope ratios. However, isotope selective photodissociation hardly affects the $^{12}$C/$^{13}$C ratio from high density tracers like CN (\citealt{2005ApJ...634...1126}). And since the $^{12}$C/$^{13}$C ratios derived from H$_2$CO, C$^{18}$O and CN are quite similar in the average (Fig. \ref{fig6}), isotope selective photodissociation should not be a dominant effect.

\item Chemical fractionation

In order to investigate the effects of chemical fractionation, we must understand the formation of formaldehyde (H$_2$CO). In classic gas-phase reaction networks, H$_2$CO is formed from neutral-neutral reactions between CH$_3$ and atomic oxygen, where CH$_3$ is built from carbon ions reacting with molecular hydrogen (e.g., \citealt{2012LPCS...43..1611}). Alternatively, formaldehyde is efficiently formed on the surface of icy dust grains via the hydrogenation of CO (e.g., \citealt{1997ApJ...205..L165}; \citealt{2002ApJ...571..L173}). Subsequently, it may be released into the gas phase by photoevaporation or shocks and then behave, in the first case, like a tracer of photon dominated regions (PDRs), and in the second like a shock tracer. The relative importance of the two main pathways of H$_2$CO formation, gas-phase chemistry or dust grain mantle evaporation, is poorly constrained.

Small differences in the zero point energy between reactants and products of isotopically distinct species may cause fractionation. Due to the charge exchange reaction of CO with $^{13}$C$^+$ (\citealt{1976ApJ...205..L165}), gas phase CO should have a tendency to be enriched in $^{13}$CO. Other molecules, like perhaps H$_2$CO, formed in the gas phase through different chemical pathways, should then be depleted in the $^{13}$C bearing isotopologue (e.g., \citealt{1984ApJ...277..581}). However, if most of the formaldehyde originates alternatively from dust grain mantles, the situation is different. In this case, H$_2$CO formed by the hydrogenation of CO, would be similarly enriched in $^{13}$C as CO.

As a result, the carbon isotope ratios obtained from formaldehyde could be similar or larger than that from C$^{18}$O and could provide us with a useful hint for the predominant H$_2$CO formation scenario in massive star forming regions. In previous studies (see, e.g., \citealt{2005ApJ...634...1126}), larger values were found, suggesting a predominantly gas-phase origin of formaldehyde. In this study, however, we find slightly smaller values, consistent with formaldehyde formation on dust grains and subsequent release into the interstellar medium. Within the range of uncertainty, it could even contradict both grain mantle and gas phase formation scenarios outlined above. First results from four sources, hinting at this latter possibility, i.e. suggesting significantly $^{13}$C enhanced formaldehyde as compared to CO and methanol (CH$_3$OH), were already published by Wirstr\"{o}m et al. \citeyear{2012LPCS...43..1611}. Constructing more extensive fractionation networks including both gas-phase, dust grain and gas-dust interactions might provide further insight into this interesting and quite basic astrochemical puzzle.

\end{enumerate}

\begin{deluxetable}{cccccc}
\tablecaption{Comparison of $^{12}$C/$^{13}$C ratios from H$_2$CO, CN, and CO}
%\tablenum{4}
\tablehead{\colhead{Source} & \colhead{D$_{GC}$} & \multicolumn{2}{c}{H$_2$CO} & \colhead{CN\tablenotemark{b}} & \colhead{CO\tablenotemark{c}} \\
\colhead{} & \colhead{(kpc)} & \colhead{This paper} & \colhead{Previous results\tablenotemark{a}} & \colhead{} & \colhead{} }
\renewcommand\arraystretch{1.1}
%% All data must appear between the \startdata and \enddata commands
\startdata
\label{Table4}
W3IRS5 & 10.3 &    60$\pm$ 11           & $\cdots$ & $\cdots$ & $\cdots$ \\
Mol12 & 9.71 &     47 $\pm$ 8           & $\cdots$ & $\cdots$ & $\cdots$ \\
NGC2024 & 7.22 &   52 $\pm$ 6           & 72 $\pm$ 11 & 65 $\pm$ 12 & $\cdots$ \\
M-0.13 & 0.06 &    11.10 $\pm$ 0.08   & $\cdots$ & $\cdots$ & $\cdots$ \\
Sgr A & 0.17 &     10.71 $\pm$ 0.11   & $\cdots$ & $\cdots$ & $\cdots$ \\
Sgr B2 & 7.81 &    77 $\pm$ 15       & $\cdots$ & $\cdots$ & $\cdots$ \\
 & 0.38 &          11.48 $\pm$ 0.03   & 10& $\geq$18$_{-6}$ &24 $\pm$ 1 \\
M+1.6 & 0.77 &     10.15 $\pm$ 0.68   & $\cdots$ & $\cdots$ & $\cdots$ \\
G10.16-0.35 &7.44& 24 $\pm$ 4         & $\cdots$ & $\cdots$ & $\cdots$ \\
 & 6.24 &          42 $\pm$ 3        & $\cdots$ & $\cdots$ & $\cdots$ \\
W31 & 3.38 &       16 $\pm$ 4        & 37 $\pm$ 6 & 20 $\pm$ 6 & $\cdots$ \\
 & 5.0 &           47 $\pm$ 5         & $\cdots$ & $\cdots$ & $\cdots$ \\
G11.93-0.61 &4.88& 45 $\pm$ 7       & $\cdots$ & $\cdots$ & $\cdots$ \\
W33 & 5.32 &       23 $\pm$ 1       & 74 $\pm$ 22 & 37 $\pm$ 8 & 39 $\pm$ 1, 43 $\pm$ 4\tablenotemark{d} \\
G13.88+0.28 &4.41& 35 $\pm$ 6       & $\cdots$ & $\cdots$ & $\cdots$ \\
G12.91-0.26 &5.69& 49 $\pm$ 3       & $\cdots$ & $\cdots$ & $\cdots$ \\
G19.62-0.23 &5.23& 28 $\pm$ 3       & 41 $\pm$ 4 & 71 $\pm$ 17 & $\cdots$ \\
G023.44-00.18&2.48&18 $\pm$ 13        & $\cdots$ & $\cdots$ & $\cdots$ \\
 & 3.59 &          30 $\pm$ 3        & $\cdots$ & $\cdots$ & $\cdots$ \\
G23.43-0.21 &3.9&  39 $\pm$ 4        & $\cdots$ & $\cdots$ & $\cdots$ \\
G29.9-0.0 & 4.38 & 44 $\pm$ 9        & 45 $\pm$ 5 & 63 $\pm$ 27 & $\cdots$ \\
G31.41+0.31&4.51 & 32 $\pm$ 2        & $\cdots$ & $\cdots$ & $\cdots$ \\
W43 & 4.58 &       26 $\pm$ 1        & 42 $\pm$ 6 & $\cdots$ & $\cdots$ \\
G34.26+0.15 &5.58& 46 $\pm$ 4        & $\cdots$ & $\cdots$ & $\cdots$ \\
G34.3+0.1 &5.58 &  31$\pm$ 4         & 58 $\pm$ 5 & 28 $\pm$ 4 & $\cdots$ \\
J185648.26 & 4.8 & 48 $\pm$ 5        & $\cdots$ & $\cdots$ & $\cdots$ \\
G37.76-0.20 &5.62& 34 $\pm$ 23        & $\cdots$ & $\cdots$ & $\cdots$ \\
G35.20-1.74 &5.77& 43 $\pm$ 6        & $\cdots$ & $\cdots$ & $\cdots$ \\
G35.2-1.8 &6.16 &  53 $\pm$ 11        & 55 $\pm$ 5 & 64 $\pm$ 24 & $\cdots$ \\
G043.16+00.01&7.6& 39 $\pm$ 14        & $\cdots$ & $\cdots$ & $\cdots$ \\
G43.17+0.00 &7.56& 40 $\pm$ 13        & $\cdots$ & $\cdots$ & $\cdots$ \\
G43.2+0.0 & 7.55 & 72$\pm$ 46         & 74 $\pm$ 20 & 44 $\pm$ 22 & 49 $\pm$ 6 \\
G45.45+0.06 &6.39& 30 $\pm$ 4        & $\cdots$ & $\cdots$ & $\cdots$ \\
G49.21-0.35 &6.15& 99 $\pm$ 24        & 58 $\pm$ 4 & $\geq$33$_{-13}$ & $\cdots$ \\
J192311.17 & 6.31& 18 $\pm$ 9        & $\cdots$ & $\cdots$ & $\cdots$ \\
 & 6.17 &          59 $\pm$ 4        & $\cdots$ & $\cdots$ & $\cdots$ \\
G49.4-0.3& 6.17 &  86 $\pm$ 7        & 81 $\pm$ 12 & $\cdots$ & $\cdots$ \\
G49.5-0.4 & 6.18 & 45 $\pm$ 16        & 70 $\pm$ 11 & $\cdots$ & $\cdots$ \\
J192345.73 & 6.18& 19 $\pm$ 8        & $\cdots$ & $\cdots$ & $\cdots$ \\
J203901.04 & 8.14& 64 $\pm$ 7        & $\cdots$ & $\cdots$ & $\cdots$ \\
DR21 & 8.05 &      48 $\pm$ 3        & 73 $\pm$ 11 & 36 $\pm$ 3 & 53 $\pm$ 2 \\
J205658.56 &7.38&  42$\pm$ 5         & $\cdots$ & $\cdots$ & $\cdots$ \\
J205703.98 & 7.37& 26 $\pm$ 8        & $\cdots$ & $\cdots$ & $\cdots$ \\
W3-OH & 9.58 & $\cdots$ & 91 $\pm$ 16 & 63 $\pm$ 6 & 66 $\pm$ 4, 85 $\pm$ 15\tablenotemark{d} \\
W51M & 6.18 & $\cdots$ & $\cdots$ & 35 $\pm$ 12 & 45 $\pm$ 2 \\
Orion A & 8.49 & $\cdots$ & $\cdots$ & 43 $\pm$ 7 & 79 $\pm$ 7 \\
Orion Bar & 8.49 & $\cdots$ & $\cdots$ & 70 $\pm$ 11 & 75 $\pm$ 9\tablenotemark{e} \\
NGC7538 & 9.42 & $\cdots$ & $\cdots$ & 56 $\pm$ 10 & 56 $\pm$ 10 \\
S156& 10.34 & $\cdots$ & $\cdots$ & 78 $\pm$ 44 & $\cdots$ \\
WB89 391 & 14.25 & $\cdots$ & $\cdots$ & 134 $\pm$ 43 & $\cdots$ \\
\enddata
\tablenotetext{a}{From Henkel et al. (1980, 1982 ,1983, 1985).}
\tablenotetext{b}{From Milam et al. (2005), Savage et al. (2002).}
\tablenotetext{c}{From Langer \& Penzias (1990).}
\tablenotetext{d}{From Wouterloot \& Brand (1996).}
\tablenotetext{e}{From Keene et al. (1998).}

%% No \tablerefs indicated
\end{deluxetable}

\begin{deluxetable}{ccc|cc}
\tablecaption{Measurements of the $^{12}$C/$^{13}$C gradient.}
%\tablenum{5}
\tablehead{ \colhead{} &\multicolumn{2}{c}{Previous fitting results\tablenotemark{a}} & \multicolumn{2}{c}{This work\tablenotemark{b}} \\
 \colhead{} & \colhead{slope} & \colhead{intercept} & \colhead{slope} & \colhead{intercept} }
%\colnumbers
\startdata
\label{Table5}
CN\tablenotemark{c} &6.01 $\pm$ 1.19 & 12.28 $\pm$ 9.33 &6.87 $\pm$ 1.46&5.06 $\pm$ 11.07\\
C$^{18}$O\tablenotemark{d} &5.41 $\pm$ 1.07 & 19.03 $\pm$ 7.90 &5.90 $\pm$ 1.32&15.30 $\pm$ 9.65\\
H$_2$CO\tablenotemark{e}& 7.60 $\pm$ 1.79 & 18.05 $\pm$ 10.88& 8.66 $\pm$ 1.64 &9.30 $\pm$ 10.37\\
This work (H$_2$CO) &4.94 $\pm$ 0.96 & 9.37 $\pm$ 6.05 & 5.08 $\pm$ 1.10&11.86 $\pm$ 6.60\\
\enddata

\tablecomments{The $^a$ and $^b$ represent the fitting results for the old and new distances, respectively. For this work, $^a$ and $^b$ also represent the corrected results using the collision rates of H$_2$CO taken from Green et al. (1991) and Wiesenfeld \& Faure (2013), respectively.}
\tablenotetext{c}{From Milam et al. (2005).}
\tablenotetext{d}{From Langer \& Penzias (1990), Wouterloot \& Brand (1996) and Keene et al. (1998).}
\tablenotetext{e}{From Henkel et al. (1980, 1982 ,1983, 1985).}
\end{deluxetable}

\section{Summary} \label{sec:summary}
With the Tianma Radio Telescope (TMRT), we performed observations of the $1_{10}-1_{11}$ and $2_{11}-2_{12}$ lines of H$_2$CO and the $1_{10}-1_{11}$ line of H$_2$$^{13}$CO toward a big sample of 112 Galactic molecular lines-of-sight.  All three lines are detected toward 38 targets (43 radial velocity components), with a detection rate of $\sim$34\%. For these 38 sources, their continuum at C- ($\sim$ 5 GHz) and Ku-band (14.5 GHz) were also observed and detected, at C-band in all sources and, at Ku-band, in 32 objects. Spectral line and continuum data for those 38 sources were analyzed. Our main results are :
\begin{enumerate}
\item
Based on spectral line parameters and continuum temperatures, we obtained column densities, optical depths and the molecular abundance ratios.
\item
We used the RADEX non-LTE model for the radiative transfer and took the new collision rates from \citet{Wiesenfeld2013} to determine the photon trapping effect in the mm-wave lines connecting the J = 1 and 2 K-doublets of ortho-H$_2$CO.
\item
We took reliable distance values from trigonometric parallax measurements and the Revised Kinematic Distance Calculator (\citealt{2014ApJ...783..130}) for our sources.  Thus the $^{12}$C/$^{13}$C gradient along the Galactocentric distance is confirmed from a linear fit to the $^{12}$C/$^{13}$C data resulting in (5.08$\pm$1.10)D$_{GC}$+(11.86$\pm$6.60), with a correlation coefficient of 0.58. Measurements of more sources, especially those with large Galactocentric distance, are needed to further improve the statistical significance.
\item
The gradient determined by us is flatter than that obtained from previous studies of formaldehyde, but is consistent, within the uncertainties, with those obtained from CN and C$^{18}$O. While the previous results may suggest an H$_2$CO formation mechanism dominated by gas-phase reactions, our new result tends to support a formation on dust grain mantles followed by evaporation.

\end{enumerate}
\acknowledgments
This work is supported by the Natural Science Foundation of China (No. 11473007, 11590782) and the Guangzhou Education Bureau (No. 1201410593). We would like to thank for the assistance of the TMRT operators during the observations. Y.T.Y. wishes to thank J.Z.Wang and X.Chen for their help with the observations and comments.
\software{GILDAS/CLASS (Pety 2005, GILDAS team 2013), RADEX (Van der Tak et al. 2007)}

\appendix
\section{}
\label{A}
\begin{longdeluxetable}{lcccccc}
%% Use 8pt
\tabletypesize{\scriptsize}
\tablewidth{400pt}

%% This is the title of the table.
\tablecaption{H$_2$$^{12}$CO and H$_2$$^{13}$CO $1_{10}-1_{11}$ observations of the entire sample}

%\tablenum{6}

\tablehead{
\colhead{Sources} &  \colhead{R.A.(J2000)} & \colhead{Dec.(J2000)} & \colhead{Line}  &\colhead{Integration time}& \colhead{rms}& \colhead{$\Delta V$}\\
\colhead{} &  \colhead{($h\quad m\quad s$)} & \colhead{($\arcdeg\quad \arcmin\quad \arcsec$)}& \colhead{}&\colhead{(minutes)}& \colhead{(mK)}& \colhead{(km s$^{-1}$)}\\}
\tablecolumns{6}
\colnumbers
\startdata
\label{Table6}
G121.29+00.65\tablenotemark{a} & 00:36:47.35 & 63:29:02.2 & H$_2$$^{12}$CO (C)& 90 &11.1&0.089  \\
  &   &   & H$_2$$^{13}$CO (C)&82 &10.8 &0.093 \\
IRAS00338\tablenotemark{a}  & 00:36:47.50 & 63:29:02.0  & H$_2$$^{12}$CO (C)& 39&17.0 & 0.089 \\
  &   &   & H$_2$$^{13}$CO (C)& 39 & 17.8 & 0.093\\
G122.01-07.08\tablenotemark{a}  & 00:44:58.40 & 55:46:47.6  & H$_2$$^{12}$CO (C)&90 & 11.4 & 0.089\\
  &   &   & H$_2$$^{13}$CO (C)& 90 &11.0 & 0.093\\
G123.06-06.30\tablenotemark{a}   & 00:52:24.20 & 56:33:43.2 & H$_2$$^{12}$CO (C)& 30& 19.4 & 0.089\\
  &   &   & H$_2$$^{13}$CO (C)&30 & 20.6 & 0.093\\
G134.62-02.19\tablenotemark{b}  & 02:22:51.71 & 58:35:11.4 & H$_2$$^{12}$CO (C)& 42& 15.3 & 0.089\\
  &   &   & H$_2$$^{13}$CO (C)& 42 & 15.3 & 0.093\\
W3IRS5 & 02:25:40.80 & 62:05:53.00 & H$_2$$^{12}$CO (C)& 42.9 & 48.2 & 0.178\\
  &   &   & H$_2$$^{13}$CO (C)& 102 & 15.2 & 0.374\\
    &   &   & H$_2$$^{12}$CO (Ku)&40 &67.4 &0.059 \\
W3\tablenotemark{a}    & 02:25:44.19 & 62:06:00.9 & H$_2$$^{12}$CO (C)& 276 & 17.2 & 0.089\\
  &   &   & H$_2$$^{13}$CO (C)& 438 & 19.1 & 0.093\\
W3-OH \tablenotemark{a}  & 02:27:03.82 & 61:52:25.2  & H$_2$$^{12}$CO (C)& 88 & 12.2 & 0.089\\
  &   &   & H$_2$$^{13}$CO (C)& 88 & 12.3 & 0.093\\
G133.947\tablenotemark{a}      & 02:27:16.50 & 61:52:24.5 & H$_2$$^{12}$CO (C)& 60 & 18.8 & 0.089\\
  &   &   & H$_2$$^{13}$CO (C)& 42.9 & 20.1 & 0.093\\
G135.27+02.79\tablenotemark{b}     & 02:43:28.57 & 62:57:08.4 & H$_2$$^{12}$CO (C)& 30 & 23.6 & 0.089\\
  &   &   & H$_2$$^{13}$CO (C)& 30 & 23.6 & 0.093\\
WB443\tablenotemark{b}   & 02:47:15.43 & 60:30:49.0  & H$_2$$^{12}$CO (C)& 30 & 20.9 & 0.089\\
  &   &   & H$_2$$^{13}$CO (C)& 30 & 20.9 &0.093\\
WB448\tablenotemark{b} & 02:50:08.6 & 61:59:52.3& H$_2$$^{12}$CO (C)& 30 & 24.4 & 0.089\\
  &   &   & H$_2$$^{13}$CO (C)& 30 & 24.4 & 0.093\\
G138.295\tablenotemark{b}      & 03:00:10.08 & 02:12:09.1 & H$_2$$^{12}$CO (C)& 39 & 22.2 & 0.089\\
  &   &   & H$_2$$^{13}$CO (C)& 39 & 22.2 & 0.093\\
AFGL490\tablenotemark{a}         & 03:27:38.80 & 58:47:00.0 & H$_2$$^{12}$CO (C)& 42.9 & 17.6 & 0.089\\
  &   &   & H$_2$$^{13}$CO (C)& 42.9 & 35.2 & 0.093\\
G160.14+03.15\tablenotemark{b} & 05:01:40.24 & 47:07:19.0 & H$_2$$^{12}$CO (C)& 28 & 23.0 & 0.089\\
  &   &   & H$_2$$^{13}$CO (C)& 28 & 23.0 & 0.093\\
G168.06+00.82\tablenotemark{b}    & 05:17:13.74 & 39:22:19.9  & H$_2$$^{12}$CO (C)& 28 & 21.4 & 0.089\\
  &   &   & H$_2$$^{13}$CO (C)& 28 & 21.4 & 0.093\\
ORI\tablenotemark{b}     & 05:35:17.46 & -05:23 15.7 & H$_2$$^{12}$CO (C)& 39 & 128 & 0.089\\
  &   &   & H$_2$$^{13}$CO (C)& 39 & 128 & 0.093\\
G176.51+00.20\tablenotemark{a}    & 05:37:52.14 & 32:00:03.9 & H$_2$$^{12}$CO (C)& 30 & 19.6 & 0.089\\
  &   &   & H$_2$$^{13}$CO (C)& 30 & 18.9 & 0.093\\
IRAS05358\tablenotemark{a}    & 05:39:13.00 & 35:45:49.0 & H$_2$$^{12}$CO (C)& 42.9 & 15.3 & 0.089\\
  &   &   & H$_2$$^{13}$CO (C)& 42.9 & 15.2 & 0.093\\
G182.67-03.26\tablenotemark{b}          & 05:39:28.42 & 24:56:32.1  & H$_2$$^{12}$CO (C)& 30 & 19.9 & 0.089\\
  &   &   & H$_2$$^{13}$CO (C)& 30 & 19.9 & 0.093\\
G183.72-03.66\tablenotemark{a}          & 05:40:24.23 & 23:50:54.7 & H$_2$$^{12}$CO (C)& 20 & 24.6 & 0.089\\
  &   &   & H$_2$$^{13}$CO (C)& 20 & 21.2 & 0.093\\
Mol12 & 05:40:24.40 & 23:50:54 & H$_2$$^{12}$CO (C)& 432 & 5.5 & 0.089\\
  &   &   & H$_2$$^{13}$CO (C)& 402 & 4.3 & 0.093\\
      &   &   & H$_2$$^{12}$CO (Ku)&40 &31.5 &0.059 \\
NGC2024 & 05:41:45.50 & -01:54:46.7 & H$_2$$^{12}$CO (C)& 120 & 23.1 & 0.089\\
  &   &   & H$_2$$^{13}$CO (C)& 102 & 15.1 & 0.187\\
      &   &   & H$_2$$^{12}$CO (Ku)&40 &43.5 &0.059 \\
G192.16-03.81\tablenotemark{a}   & 05:58:13.53 & 16:31:58.9  & H$_2$$^{12}$CO (C)& 28 & 14.6 & 0.178\\
  &   &   & H$_2$$^{13}$CO (C)& 28 & 20.3 & 0.093\\
MonR2\tablenotemark{b} & 06:07:52.43 & -01:06:50.8 & H$_2$$^{12}$CO (C)& 29.2 & 19.4 & 0.089\\
  &   &   & H$_2$$^{13}$CO (C)& 29.2 & 19.4 & 0.093\\
G188.94+00.88\tablenotemark{a} & 06:08:53.35 & 21:38:28.7 & H$_2$$^{12}$CO (C)& 28 & 20.1 & 0.089\\
  &   &   & H$_2$$^{13}$CO (C)& 28 & 18.5 & 0.093\\
G188.79+01.03\tablenotemark{a} & 06:09:06.97 & 21:50:41.4  & H$_2$$^{12}$CO (C)& 30 & 16.5 & 0.089\\
  &   &   & H$_2$$^{13}$CO (C)& 30 & 17.8 & 0.093\\
H192.584\tablenotemark{b}   & 06:10:56.51 & -01:06:53.3 & H$_2$$^{12}$CO (C)& 39 & 21.9 & 0.089\\
  &   &   & H$_2$$^{13}$CO (C)& 39 & 21.9 & 0.093\\
G192.60-00.04\tablenotemark{a} & 06:12:54.02 & 17:59:23.3 & H$_2$$^{12}$CO (C)& 38 & 16.3 & 0.089\\
  &   &   & H$_2$$^{13}$CO (C)& 38 & 18.2 & 0.093\\
G196.45-01.67\tablenotemark{a} & 06:14:37.08 & 13:49:36.7 & H$_2$$^{12}$CO (C)& 60 & 14.4 & 0.089\\
  &   &   & H$_2$$^{13}$CO (C)& 60 & 15.2 & 0.093\\
NGC2264-1\tablenotemark{a} & 06:41:09.80 & 09:29:32:0& H$_2$$^{12}$CO (C)& 39 & 18.5 & 0.089\\
  &   &   & H$_2$$^{13}$CO (C)& 39 & 16.9 & 0.093\\
G211.59+01.05\tablenotemark{b} & 06:52:45.32 & 01:40:23.1 & H$_2$$^{12}$CO (C)& 38 & 17.9 & 0.089\\
  &   &   & H$_2$$^{13}$CO (C)& 38 & 17.9 & 0.093\\
G239.35-05.06\tablenotemark{b} & 07:22:58.33 & -25:46:03.1  & H$_2$$^{12}$CO (C)& 34 & 17.1 & 0.089\\
  &   &   & H$_2$$^{13}$CO (C)& 34 & 17.1 & 0.093\\
G229.57+00.15\tablenotemark{a}  & 07:23:01.84 & -14:41:32.8& H$_2$$^{12}$CO (C)& 60 & 11.6 & 0.089\\
  &   &   & H$_2$$^{13}$CO (C)& 60 & 13.8 & 0.093\\
G232.62+00.99\tablenotemark{a} & 07:32:09.78 & -16:58:12.8 & H$_2$$^{12}$CO (C)& 82 & 12.5 & 0.089\\
  &   &   & H$_2$$^{13}$CO (C)& 108 & 10.5 & 0.093\\
G236.81+01.98\tablenotemark{a}  & 07:44:28.24 & -20:08:30.2 & H$_2$$^{12}$CO (C)& 28 & 19.8 & 0.089\\
  &   &   & H$_2$$^{13}$CO (C)& 28 & 18.5 & 0.093\\
G240.31+00.07\tablenotemark{b} & 07:44:51.92 & -24:07:41.5 & H$_2$$^{12}$CO (C)& 28 & 20.5 & 0.089\\
  &   &   & H$_2$$^{13}$CO (C)& 28 & 20.5 & 0.093\\
IRC+10216\tablenotemark{b}       & 09.47:57.49 & 13:16:47.8 & H$_2$$^{12}$CO (C)& 102 & 11.2 & 0.089\\
  &   &   & H$_2$$^{13}$CO (C)& 102 & 11.2 & 0.093\\
M-0.13 & 17:45:37.37 & -29:05:19.74 & H$_2$$^{12}$CO (C)&33.2  & 41.2 & 0.089\\
  &   &   & H$_2$$^{13}$CO (C)& 39 & 16.5 & 0.84\\
      &   &   & H$_2$$^{12}$CO (Ku)&8 &42.5 &0.059 \\
Sgr A & 17:45:42.63 & -29:00:32.39 & H$_2$$^{12}$CO (C)& 24 & 99.2 & 0.089\\
  &   &   & H$_2$$^{13}$CO (C)& 24 & 39.6 & 0.747\\
    &   &   & H$_2$$^{12}$CO (Ku)&16 &78.5 &0.059 \\
Sgr B2 & 17:47:20.00 & -28:22:40.0 & H$_2$$^{12}$CO (C)& 32 & 186 & 0.178\\
  &   &   & H$_2$$^{13}$CO (C)& 32 & 24.1 & 0.093\\
      &   &   & H$_2$$^{12}$CO (Ku)&36 &65.6 &0.059 \\
M+1.6 & 17:49:22.90 & -27:34:02.11 & H$_2$$^{12}$CO (C)& 39 & 23.0 & 0.089\\
  &   &   & H$_2$$^{13}$CO (C)& 39 & 9.9 & 0.467\\
      &   &   & H$_2$$^{12}$CO (Ku)&8 &42.3 &0.059 \\
G19.6-0.2  & 18 27:38.05 & -11:56:41.4 & H$_2$$^{12}$CO (C)& 27.3 & 24.0 & 0.089\\
  &   &   & H$_2$$^{13}$CO (C)& 39 & 24.3 & 0.093\\
G005.88-00.39\tablenotemark{a}      & 18:00:30.31 & -24:04:04.5 & H$_2$$^{12}$CO (C)& 28 & 15.2 & 0.178\\
  &   &   & H$_2$$^{13}$CO (C)& 28 & 15.4 & 0.093\\
G009.62+00.19\tablenotemark{a}          & 18:06:14.66 & -20:31:31.7& H$_2$$^{12}$CO (C)& 22 & 16.3 & 0.178\\
  &   &   & H$_2$$^{13}$CO (C)& 22 & 25.8 & 0.093\\
G010.47+00.02\tablenotemark{a}  & 18:08:38.23 & -19:51:50.3 & H$_2$$^{12}$CO (C)& 66 & 12.5 & 0.089\\
  &   &   & H$_2$$^{13}$CO (C)& 66 & 14.8 & 0.093\\
G10.16-0.35 & 18:09:24.60 & -20:19:29.0 & H$_2$$^{12}$CO (C)& 180 & 8.5 & 0.178\\
  &   &   & H$_2$$^{13}$CO (C)& 180 & 7.7 & 0.187\\
      &   &   & H$_2$$^{12}$CO (Ku)&8 &48.2 &0.059 \\
W31 & 18:10:28.55 & -19:55:48.6 & H$_2$$^{12}$CO (C)& 114 & 16.7 & 0.178\\
  &   &   & H$_2$$^{13}$CO (C)& 108 & 6.3 & 0.374\\
     &   &   & H$_2$$^{12}$CO (Ku)&8 &42.6 &0.059 \\
G011.91-00.61\tablenotemark{a} & 18:13:58.12 & -18:54:20.3  & H$_2$$^{12}$CO (C)& 28 & 21.8 & 0.089\\
  &   &   & H$_2$$^{13}$CO (C)& 28 & 20.0 & 0.093\\
G11.93-0.61 & 18:14:01.00 & -18:53:23.0 & H$_2$$^{12}$CO (C)& 80 & 13.7 & 0.089\\
  &   &   & H$_2$$^{13}$CO (C)& 80 & 9.5 & 0.187\\
     &   &   & H$_2$$^{12}$CO (Ku)&8 &41.0 &0.059 \\
W33 & 18:14:13.98 & -17:55:50.17 & H$_2$$^{12}$CO (C)& 30&34.3 & 0.089\\
  &   &   & H$_2$$^{13}$CO (C)& 60 & 26.1 & 0.093\\
     &   &   & H$_2$$^{12}$CO (Ku)&20 &68.9 &0.059 \\
G13.88+0.28 & 18:14:35.20 & -16:45:21.0 & H$_2$$^{12}$CO (C)& 80 & 15.3 & 0.089\\
  &   &   & H$_2$$^{13}$CO (C)& 76 & 12.7 & 0.093\\
     &   &   & H$_2$$^{12}$CO (Ku)&8 &37.7 &0.059 \\
G12.91-0.26 & 18:14:39.00 & -17:52:03.0 & H$_2$$^{12}$CO (C)& 78 & 15.1 & 0.089\\
  &   &   & H$_2$$^{13}$CO (C)& 80 & 10.5 & 0.093\\
     &   &   & H$_2$$^{12}$CO (Ku)&8 &37.9 &0.059 \\
G011.49-01.48\tablenotemark{a} & 18:16:22.13 & -19:41:27.2 & H$_2$$^{12}$CO (C)& 28 & 15.3 & 0.178\\
  &   &   & H$_2$$^{13}$CO (C)& 28 & 19.8 & 0.093\\
M17\tablenotemark{a}   & 18:20:26.14 & -16.11.24.0  & H$_2$$^{12}$CO (C)& 96 & 49.6 & 0.089\\
  &   &   & H$_2$$^{13}$CO (C)& 180 & 55.4 & 0.093\\
J182708.27\tablenotemark{b}    & 18:27:08.27 & -10.46.09.9 & H$_2$$^{12}$CO (C)& 60 & 14.4 & 0.089\\
  &   &   & H$_2$$^{13}$CO (C)& 60 & 14.4 & 0.093\\
G19.62-0.23 & 18:27:38.08 & -11:56:35.5 & H$_2$$^{12}$CO (C)& 132 & 9.9 & 0.178\\
  &   &   & H$_2$$^{13}$CO (C)& 132 & 8.9 & 0.187\\
     &   &   & H$_2$$^{12}$CO (Ku)&8 &41.5 &0.059 \\
G023.44-00.18 & 18:34:39.19 & -08:31:25.4 & H$_2$$^{12}$CO (C)& 198 & 8.2 & 0.178\\
  &   &   & H$_2$$^{13}$CO (C)& 210 & 8.0 & 0.374\\
     &   &   & H$_2$$^{12}$CO (Ku)&8 &48.8 &0.059 \\
G23.43-0.21 & 18:34:43.60 & -08:32:25.0 & H$_2$$^{12}$CO (C)& 138 & 6.3 & 0.355\\
  &   &   & H$_2$$^{13}$CO (C)& 132 & 5.7 & 0.374\\
     &   &   & H$_2$$^{12}$CO (Ku)&16 &15.8 &0.059 \\
G029.95-00.01\tablenotemark{a} & 18:46:03.74 & -02:39:22.3  & H$_2$$^{12}$CO (C)& 26 & 15.7 & 0.178\\
  &   &   & H$_2$$^{13}$CO (C)& 2 & 30.1 & 0.093\\
G29.9-0.0 & 18:46:08.10 & -02:41:40.20 & H$_2$$^{12}$CO (C)& 120 & 7.8 & 0.178\\
  &   &   & H$_2$$^{13}$CO (C)& 132 & 6.1 & 0.374\\
     &   &   & H$_2$$^{12}$CO (Ku)&8 &38.4 &0.059 \\
G31.41+0.31 & 18:47:34.60 & -01:12:43.0 & H$_2$$^{12}$CO (C)& 72 & 13.4 & 0.089\\
  &   &   & H$_2$$^{13}$CO (C)& 72 & 9.1 & 0.188\\
     &   &   & H$_2$$^{12}$CO (Ku)&8 &38.3 &0.059 \\
W43 & 18:47:36.68 & -01:59:01.85 & H$_2$$^{12}$CO (C)& 30 & 29.2 & 0.089\\
  &   &   & H$_2$$^{13}$CO (C)& 60 & 24.8 & 0.093\\
     &   &   & H$_2$$^{12}$CO (Ku)&8 &38.9 &0.059 \\
J184741.61\tablenotemark{a}     & 18:47:41.61 & -01.35.05.3& H$_2$$^{12}$CO (C)& 56 & 8.4 & 0.089\\
  &   &   & H$_2$$^{13}$CO (C)& 56 & 14.8 & 0.093\\
J184754.69\tablenotemark{a}   & 18:47:54.70 & -01.34.57.0& H$_2$$^{12}$CO (C)& 60 & 11.7 & 0.178\\
  &   &   & H$_2$$^{13}$CO (C)& 60 & 16.2 & 0.093\\
G031.28+00.06\tablenotemark{a} & 18:48:12.39 & -01:26:30.7 & H$_2$$^{12}$CO (C)& 62 & 9.3 & 0.178\\
  &   &   & H$_2$$^{13}$CO (C)& 62 & 13.7 & 0.093\\
G34.26+0.15 & 18:53:18.54 & 01:14:58 & H$_2$$^{12}$CO (C)& 80 & 15.1 & 0.089\\
  &   &   & H$_2$$^{13}$CO (C)& 80 & 12.6 & 0.188\\
     &   &   & H$_2$$^{12}$CO (Ku)&8 &42.8 &0.059 \\
J185319.73\tablenotemark{b}      & 18:53:19.73 & 00:29:59.8 & H$_2$$^{12}$CO (C)& 30 & 19.6 & 0.089\\
  &   &   & H$_2$$^{13}$CO (C)& 30 & 19.6 & 0.093\\
G34.3+0.1 & 18:53:20.21 & 01:14:31.74 & H$_2$$^{12}$CO (C)& 33.2 & 19.2 & 0.178\\
  &   &   & H$_2$$^{13}$CO (C)& 39 & 22.1 & 0.093\\
     &   &   & H$_2$$^{12}$CO (Ku)&8 &46.0 &0.059 \\
J185648.26 & 18:56:48.27 & 01:18:47 & H$_2$$^{12}$CO (C)& 72 & 16.3 & 0.089\\
  &   &   & H$_2$$^{13}$CO (C)& 102 & 9.6 & 0.187\\
     &   &   & H$_2$$^{12}$CO (Ku)&8 &35.3 &0.059 \\
G37.76-0.20 & 19:00:59.30 & 04:12:06 & H$_2$$^{12}$CO (C)& 120 & 9.8 & 0.089\\
  &   &   & H$_2$$^{13}$CO (C)& 120 & 7.7 & 0.187\\
     &   &   & H$_2$$^{12}$CO (Ku)&16 &28.2 &0.059 \\
G35.20-1.74 & 19:01:47.00 & 01:13:08 & H$_2$$^{12}$CO (C)& 186 & 18.2 & 0.089\\
  &   &   & H$_2$$^{13}$CO (C)& 186 & 9.1 & 0.187\\
     &   &   & H$_2$$^{12}$CO (Ku)&8 &45.2 &0.059 \\
G35.2-1.8 & 19:01:47.76 & 01:12:51 & H$_2$$^{12}$CO (C)& 79.1 & 17.0 & 0.178\\
  &   &   & H$_2$$^{13}$CO (C)& 83 & 9.5 & 0.187\\
     &   &   & H$_2$$^{12}$CO (Ku)&20 &38.7 &0.059 \\
G49.2-0.3\tablenotemark{a}  & 19.23:01.21 & 14:16:40.4 & H$_2$$^{12}$CO (C)& 60 & 21.4 & 0.089\\
  &   &   & H$_2$$^{13}$CO (C)& 60 & 26.8 & 0.093\\
G043.16+00.01 & 19:10:13.41 & 09:06:12.8 & H$_2$$^{12}$CO (C)& 26 & 45.4 & 0.355\\
  &   &   & H$_2$$^{13}$CO (C)& 26 & 20.4 & 0.374\\
     &   &   & H$_2$$^{12}$CO (Ku)&4 &129 &0.059 \\
G43.17+0.00 & 19:10:15.30 & 09:06:17 & H$_2$$^{12}$CO (C)& 150 & 16.0 & 0.089\\
  &   &   & H$_2$$^{13}$CO (C)& 150 & 12.5 & 0.187\\
     &   &   & H$_2$$^{12}$CO (Ku)&8 &55.7 &0.059 \\
G43.2+0.0 & 19:10:15.71 & 09:06:05.48 & H$_2$$^{12}$CO (C)& 39 & 64.5 & 0.178\\
  &   &   & H$_2$$^{13}$CO (C)& 39 & 18.7 & 0.374\\
     &   &   & H$_2$$^{12}$CO (Ku)&12 &57.3 &0.059 \\
W49\tablenotemark{a}     & 19:10:17.61 & 09:05:58.6 & H$_2$$^{12}$CO (C)& 30 & 35.5 & 0.089\\
  &   &   & H$_2$$^{13}$CO (C)& 60 & 25.9 & 0.093\\
G043.79-00.12\tablenotemark{a}     & 19:11:53.99 & 09:35:50.3  & H$_2$$^{12}$CO (C)& 26 & 14.9 & 0.178\\
  &   &   & H$_2$$^{13}$CO (C)& 26 & 21.6 & 0.093\\
G045.07+00.13\tablenotemark{b} & 19:13:22.04 & 10:50:53.3 & H$_2$$^{12}$CO (C)& 18 & 23.5 & 0.089\\
  &   &   & H$_2$$^{13}$CO (C)& 18 & 23.5 & 0.093\\
G45.45+0.06 & 19:14:21.30 & 11:09:13 & H$_2$$^{12}$CO (C)& 132 & 13.3 & 0.089\\
  &   &   & H$_2$$^{13}$CO (C)& 138 & 10.5 & 0.093\\
     &   &   & H$_2$$^{12}$CO (Ku)&4 &25.3 &0.059 \\
G043.89-00.78\tablenotemark{a}           & 19:14:26.39 & 09:22:36.5& H$_2$$^{12}$CO (C)& 58 & 14.8 & 0.089\\
  &   &   & H$_2$$^{13}$CO (C)& 58 & 13.7 & 0.093\\
G49.21-0.35 & 19:23:02.10 & 14:16:40 & H$_2$$^{12}$CO (C)& 120 & 14.5 & 0.089\\
  &   &   & H$_2$$^{13}$CO (C)& 120 & 15.8 & 0.093\\
     &   &   & H$_2$$^{12}$CO (Ku)&16 &39.9 &0.059 \\
J192311.17 & 19:23:11.17 & 14:26:33 & H$_2$$^{12}$CO (C)& 132 & 10.7 & 0.089\\
  &   &   & H$_2$$^{13}$CO (C)& 138 & 10.5 & 0.187\\
     &   &   & H$_2$$^{12}$CO (Ku)&8 &40.1 &0.059 \\
G49.4-0.3 & 19:23:14.01 & 14:27:07 & H$_2$$^{12}$CO (C)& 84 & 21.8 & 0.089\\
  &   &   & H$_2$$^{13}$CO (C)& 108 & 12.2 & 0.187\\
     &   &   & H$_2$$^{12}$CO (Ku)&8 &38.2 &0.059 \\
G052.10+01.04\tablenotemark{b} & 19:23:37.32 & 17:29:10.5 & H$_2$$^{12}$CO (C)& 16 & 34.1 & 0.089\\
  &   &   & H$_2$$^{13}$CO (C)& 16 & 34.1 & 0.093\\
G49.5-0.4 & 19:23:43.96 & 14:30:33 & H$_2$$^{12}$CO (C)& 40 & 41.1 & 0.089\\
  &   &   & H$_2$$^{13}$CO (C)& 40 & 37.2 & 0.187\\
     &   &   & H$_2$$^{12}$CO (Ku)&8 &54.4 &0.059 \\
J192345.73 & 19:23:45.74 & 14:28:45 & H$_2$$^{12}$CO (C)& 78 & 18.3 & 0.178\\
  &   &   & H$_2$$^{13}$CO (C)& 114 & 23.3 & 0.187\\
     &   &   & H$_2$$^{12}$CO (Ku)&4 &63.5 &0.059 \\
G61.48+0.09\tablenotemark{a}     & 19:46:47.33 & 25:12:45.6 & H$_2$$^{12}$CO (C)& 10 & 26.9 & 0.178\\
  &   &   & H$_2$$^{13}$CO (C)& 20 & 51.3 & 0.093\\
G70.33+1.59\tablenotemark{a}     & 20:01:54.50 & 33:34.15.0& H$_2$$^{12}$CO (C)& 60 & 17.5 & 0.089\\
  &   &   & H$_2$$^{13}$CO (C)& 60 & 19.7 & 0.093\\
G069.54-00.97\tablenotemark{a}  & 20:10:09.07 & 31:31:36.0  & H$_2$$^{12}$CO (C)& 28 & 23.1 & 0.089\\
  &   &   & H$_2$$^{13}$CO (C)& 28 & 17.7 & 0.093\\
G075.29+01.32\tablenotemark{b} & 20:16:16.01 & 37:35:45.8 & H$_2$$^{12}$CO (C)& 12 & 36.8 & 0.089\\
  &   &   & H$_2$$^{13}$CO (C)& 12 & 36.8 & 0.093\\
G074.03-01.71\tablenotemark{a}   & 20:25:07.11 & 34:49:57.6& H$_2$$^{12}$CO (C)& 78 & 10.3 & 0.089\\
  &   &   & H$_2$$^{13}$CO (C)& 88 & 12.0 & 0.093\\
G079.87+01.17\tablenotemark{b} & 20:30:29.14 & 41:15:53.6 & H$_2$$^{12}$CO (C)& 46 & 14.7 & 0.089\\
  &   &   & H$_2$$^{13}$CO (C)& 46 & 14.7 & 0.093\\
W75N\tablenotemark{a}           & 20:38:36.93 & 42:37:37.5& H$_2$$^{12}$CO (C)& 3 & 48.5 & 0.178\\
  &   &   & H$_2$$^{13}$CO (C)& 5.9 & 51.6 & 0.093\\
J203901.04 & 20:39:01.00 & 42:19:53.0 & H$_2$$^{12}$CO (C)& 52 & 21.6 & 0.089\\
  &   &   & H$_2$$^{13}$CO (C)& 58 & 33.2 & 0.093\\
     &   &   & H$_2$$^{12}$CO (Ku)&8 &62.3 &0.059 \\
DR21 & 20:39:01.30 & 42:19:32.86 & H$_2$$^{12}$CO (C)& 108 & 17.8 & 0.089\\
  &   &   & H$_2$$^{13}$CO (C)& 108 & 22.2 & 0.093\\
     &   &   & H$_2$$^{12}$CO (Ku)&8 &64.9 &0.059 \\
J205658.56 & 20:56:58.57 & 43:43:10.0 & H$_2$$^{12}$CO (C)& 64 & 15.4 & 0.089\\
  &   &   & H$_2$$^{13}$CO (C)& 144 & 9.3 & 0.187\\
     &   &   & H$_2$$^{12}$CO (Ku)&4 &46.9 &0.059 \\
J205703.98 & 20:57:03.98 & 43:37:46.5 & H$_2$$^{12}$CO (C)& 102 & 14.0 & 0.089\\
  &   &   & H$_2$$^{13}$CO (C)& 174 & 12.0 & 0.093\\
     &   &   & H$_2$$^{12}$CO (Ku)&8 &31.5 &0.059 \\
WB018\tablenotemark{b}  & 20:58:23.00 & 48:32:48.0 & H$_2$$^{12}$CO (C)& 52 & 18.2 & 0.089\\
  &   &   & H$_2$$^{13}$CO (C)& 52 & 18.2 & 0.093\\
WB021\tablenotemark{b} & 21:01:34.94 & 48:55:01.0 & H$_2$$^{12}$CO (C)& 30 & 18.6 & 0.089\\
  &   &   & H$_2$$^{13}$CO (C)& 30 & 18.6 & 0.093\\
WB024\tablenotemark{b} & 21:02:21.79 & 50:48:34.6 & H$_2$$^{12}$CO (C)& 28 & 24.0 & 0.089\\
  &   &   & H$_2$$^{13}$CO (C)& 28 & 24.0 & 0.093\\
WB026\tablenotemark{a}  & 21:02:48.50 & 59:30:48.0  & H$_2$$^{12}$CO (C)& 64 & 13.6 & 0.089\\
  &   &   & H$_2$$^{13}$CO (C)& 64 & 15.7 & 0.093\\
WB082\tablenotemark{b} & 21:27:33.00 & 56:05:09.0 & H$_2$$^{12}$CO (C)& 30 & 19.2 & 0.089\\
  &   &   & H$_2$$^{13}$CO (C)& 30 & 19.2 & 0.093\\
WB091\tablenotemark{a} & 21:32:10.39 & 55:52:42.3& H$_2$$^{12}$CO (C)& 28 & 15.7 & 0.178\\
  &   &   & H$_2$$^{13}$CO (C)& 28 & 20.5 & 0.093\\
G105.41+09.8\tablenotemark{a} & 21:43:06.48 & 66:06:55.3 & H$_2$$^{12}$CO (C)& 52 & 13.6 & 0.089\\
  &   &   & H$_2$$^{13}$CO (C)& 56 & 16.2 & 0.093\\
WB140\tablenotemark{b} & 21:56:15.09 & 57:50:41.9 & H$_2$$^{12}$CO (C)& 28 & 24.3 & 0.089\\
  &   &   & H$_2$$^{13}$CO (C)& 28 & 24.3 & 0.093\\
WB160\tablenotemark{b} & 22:15:09.51 & 58:49:06.0 & H$_2$$^{12}$CO (C)& 34 & 18.7 & 0.089\\
  &   &   & H$_2$$^{13}$CO (C)& 34 & 18.7 & 0.093\\
G100.37-03.57\tablenotemark{b} & 22:16:10.37 & 52:21:34.1 & H$_2$$^{12}$CO (C)& 44 & 15.6 & 0.089\\
  &   &   & H$_2$$^{13}$CO (C)& 44 & 15.6 & 0.093\\
G107.29+05.63\tablenotemark{b}      & 22:21:26.73 & 63:51:37.9 & H$_2$$^{12}$CO (C)& 20 & 26.4 & 0.089\\
  &   &   & H$_2$$^{13}$CO (C)& 20 & 26.4 & 0.093\\
CASA\tablenotemark{a}   & 23.23:02.61 & 58:48:46.1 & H$_2$$^{12}$CO (C)& 108 & 103 & 0.089\\
  &   &   & H$_2$$^{13}$CO (C)& 126 & 97.7 & 0.093\\
NGC7538\tablenotemark{a}      & 23:13:45.40 & 61:28:15.0 & H$_2$$^{12}$CO (C)& 126 & 12.6 & 0.089\\
  &   &   & H$_2$$^{13}$CO (C)& 132 & 12.7 & 0.093\\
\enddata

%% Include any \tablenotetext{key}{text}, \tablerefs{ref list},
%% or \tablecomments{text} between the \enddata and
%% \end{deluxetable} commands
\tablenotetext{a}{The sources only detected in the $1_{10}-1_{11}$ lines of H$_2$$^{12}$CO.}
\tablenotetext{b}{The sources not detected in both the $1_{10}-1_{11}$ lines of H$_2$$^{12}$CO and H$_2$$^{13}$CO.}
\tablecomments {Column (1): source name; Columns (2) and (3): Equatorial coordinates of sources; Column (4): molecular species, (C) and (Ku) indicate the C- and Ku-band, respectively; Column (5): integration time (ON-source + OFF-source); Column (6): rms noise obtained from Gaussian-fitting; Column (7): corresponding channel width.}

\end{longdeluxetable}

\newpage
\section{}
\label{B}
\begin{figure*}[h]
\center
  \includegraphics[width=233pt]{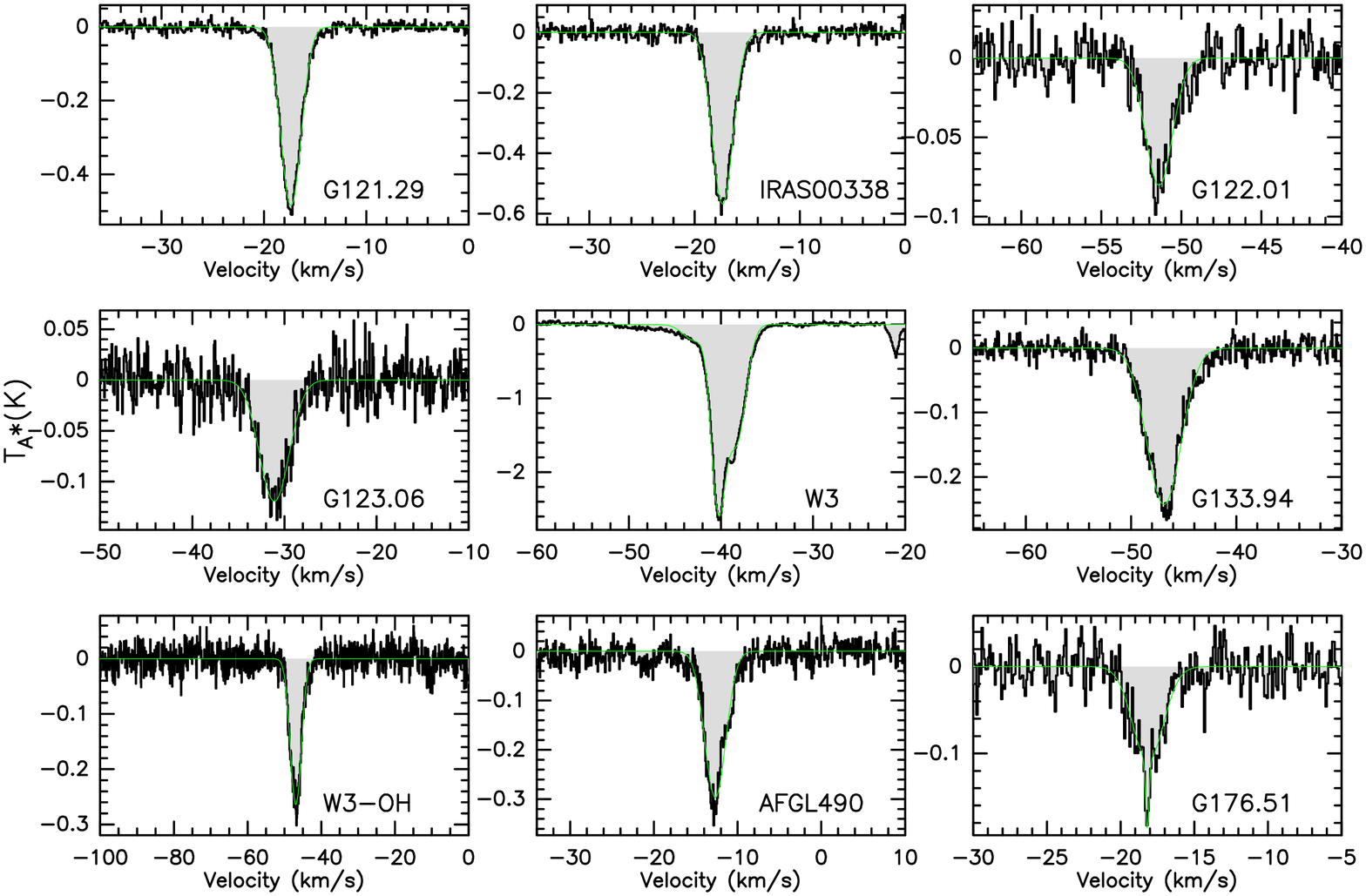}
  \includegraphics[width=233pt]{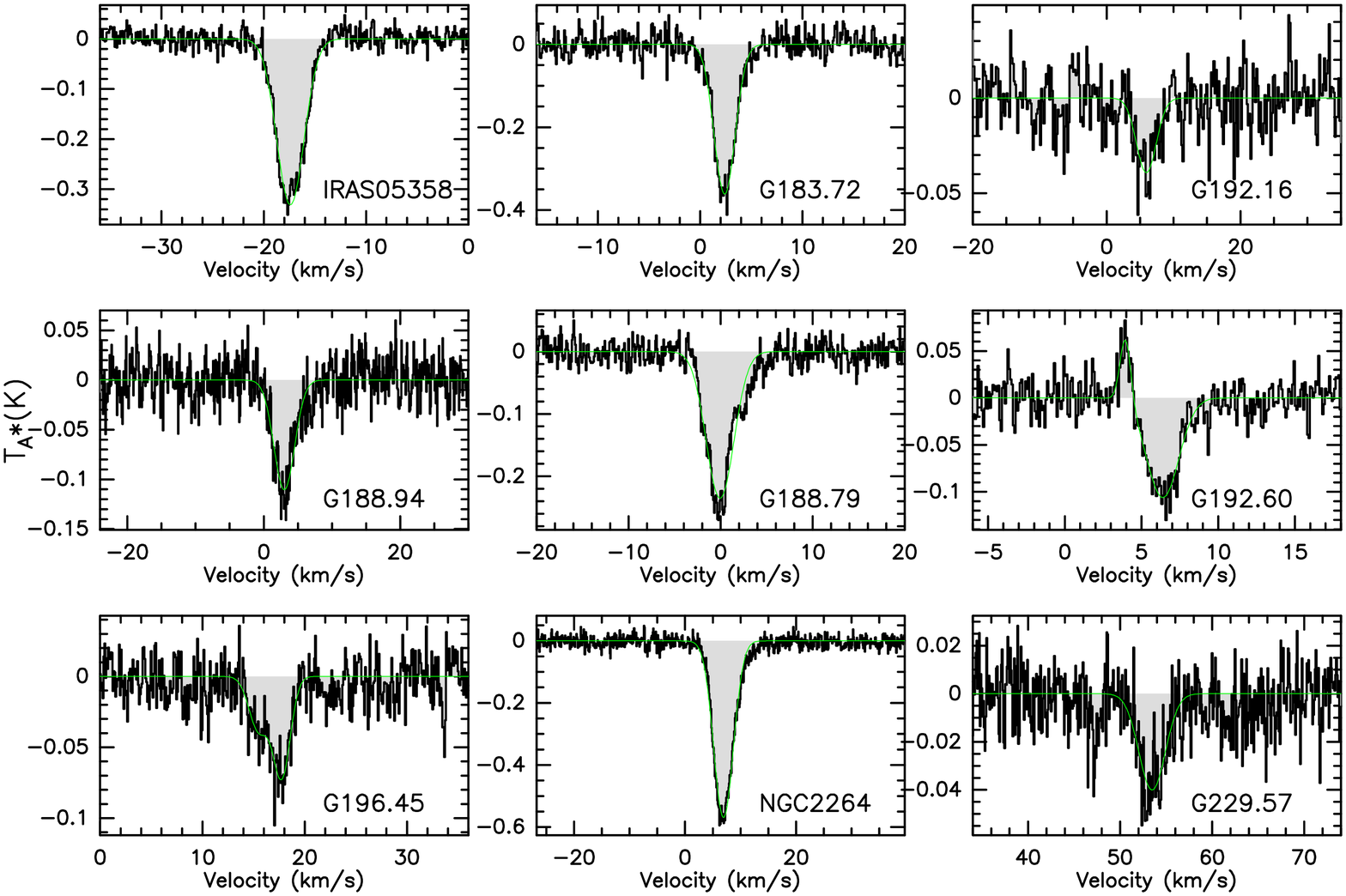}
  \includegraphics[width=233pt]{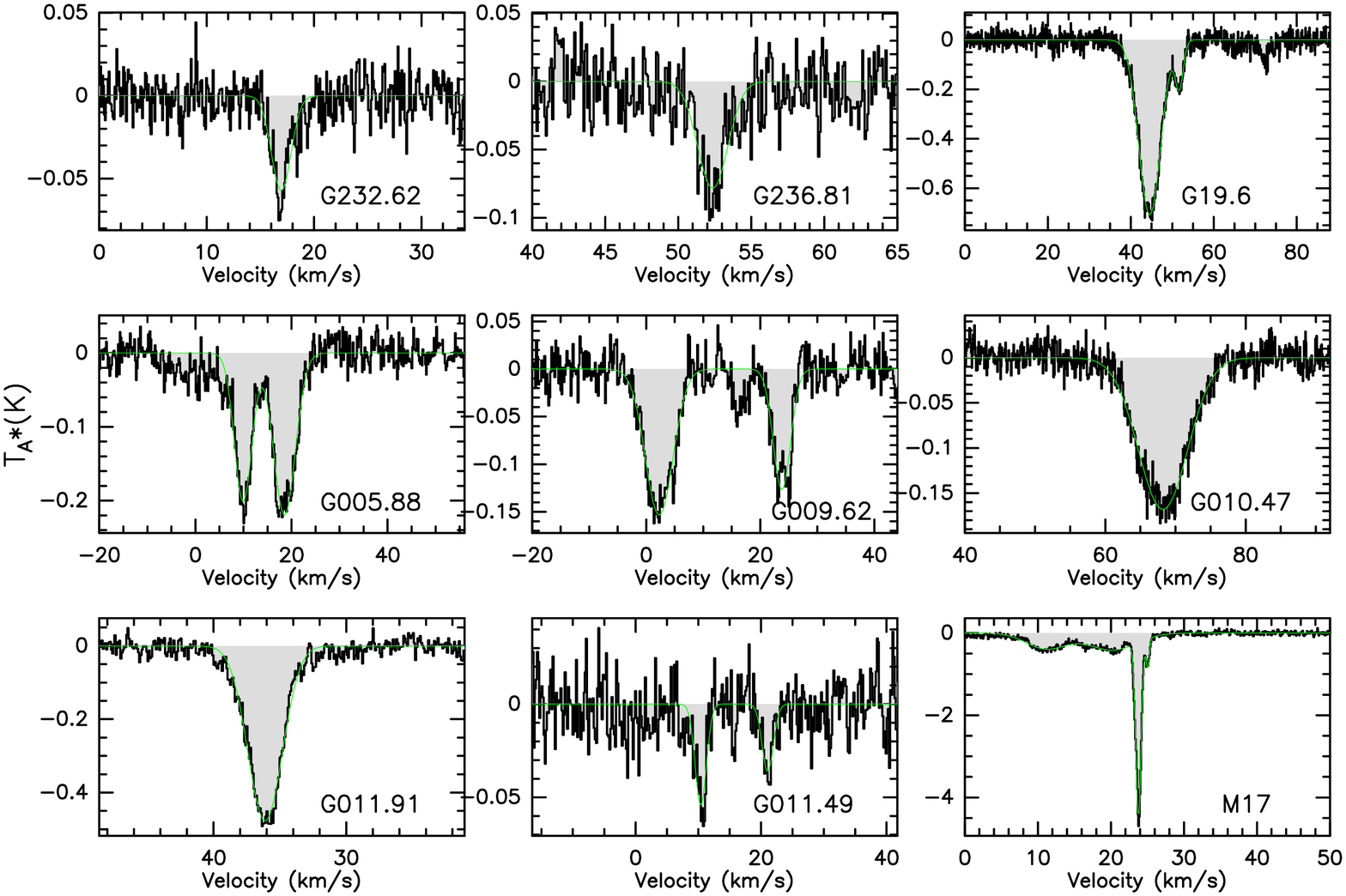}
  \includegraphics[width=233pt]{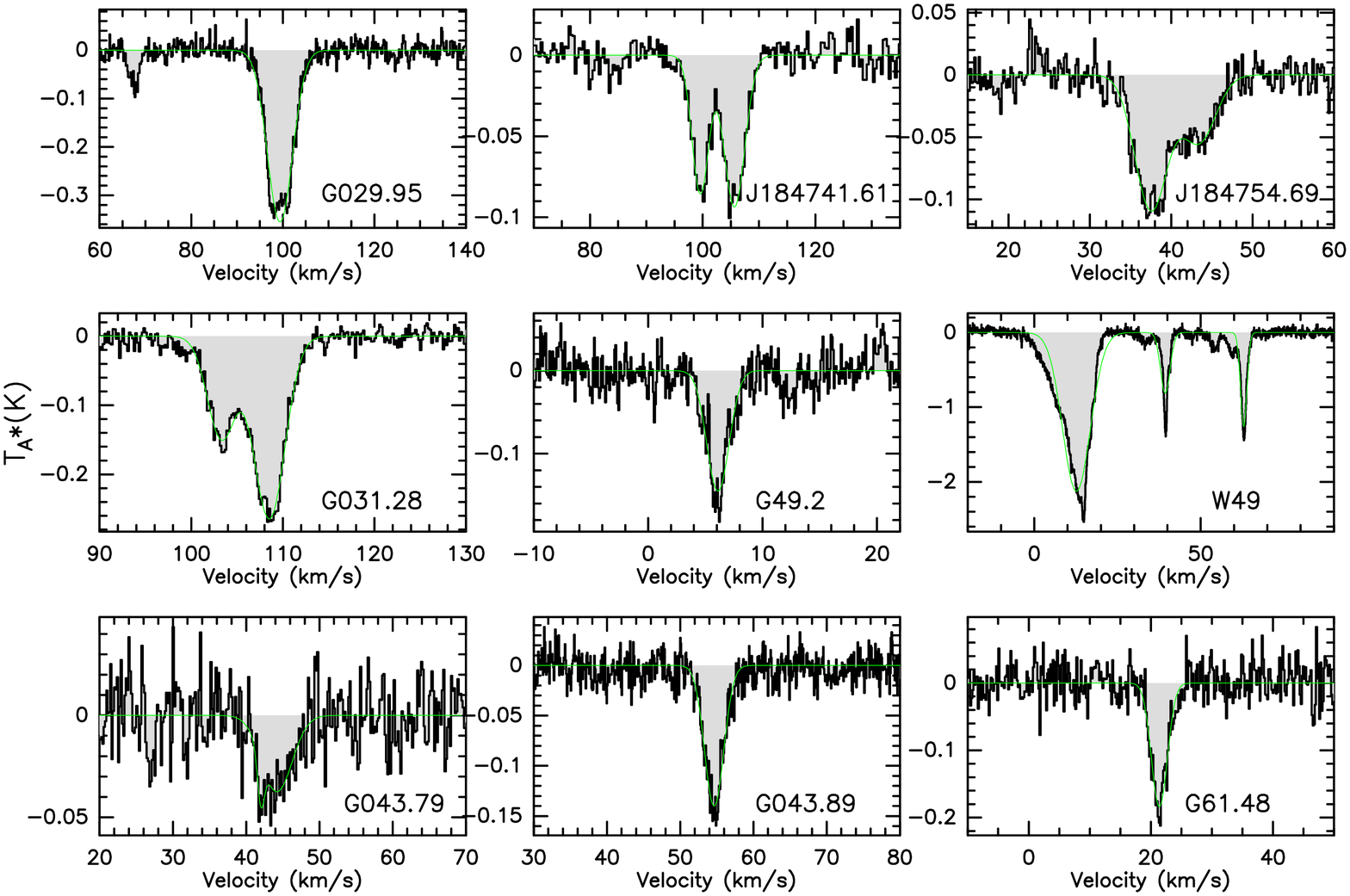}
  \includegraphics[width=233pt]{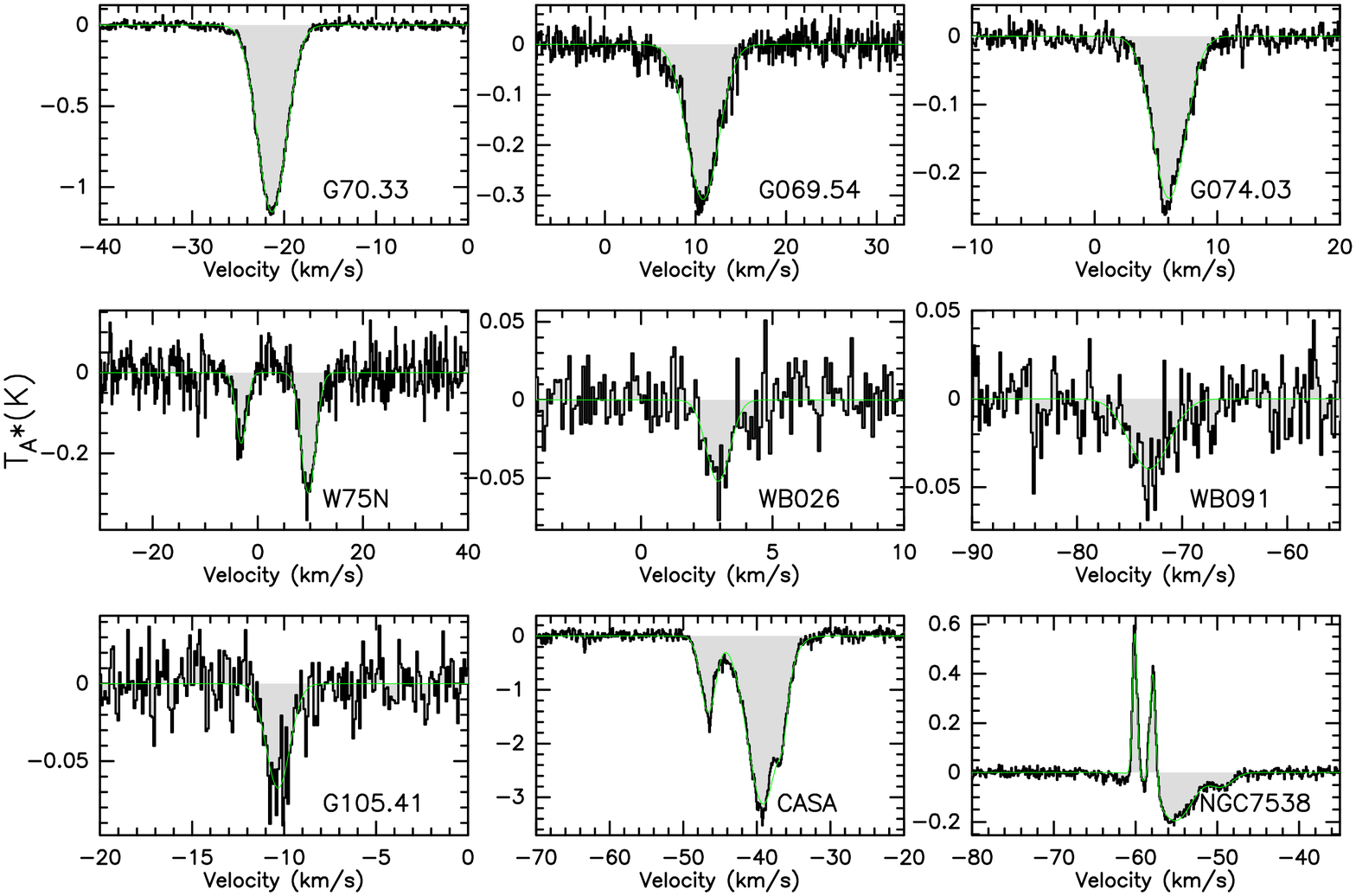}
  \caption{Spectra of 46 sources, only detected in the $1_{10}-1_{11}$ lines of H$_2$$^{12}$CO, after subtracting baselines and applying Hanning smoothing. Antenna temperature scales can be found on the left side of the profiles.}
  \label{fig8}
\end{figure*}

\newpage
\section{}
\label{C}
\begin{longdeluxetable}{ccccc}
%% Use 8pt
\tabletypesize{\small}

%% This is the title of the table.
\tablecaption{Gaussian-fitting results for 46 sources with only the H$_2$$^{12}$CO $1_{10}-1_{11}$ line detected}

%\tablenum{7}
\tablehead{
\colhead{Source}& \colhead{$\int{T_{mb}dv}$} & \colhead{Velocity} & \colhead{Width} & \colhead{Tpeak} \\
\colhead{} & \colhead{(K km s$^{-1}$)} & \colhead{(km s$^{-1}$)} & \colhead{(km s$^{-1}$)} & \colhead{(K)} }
\colnumbers
%% All data must appear between the \startdata and \enddata commands
\startdata
\label{Table7}
G121.29+00.65 & -1.217(0.008)& -17.357(0.007) & 2.356(0.018) & -0.485 \\
IRAS00338  & -1.429(0.012)& -17.338(0.010)& 2.366(0.024)&  -0.567 \\
G122.01-07.08   & -0.168(0.007) & -51.397(0.041)& 1.978(0.107) &     -0.080\\
G123.06-06.30   & -0.485(0.017)& -31.090(0.065) & 3.814(0.155)& -0.119\\
W3     & -4.627(0.331)& -40.345(0.042)& 1.982(0.068) & -2.193\\
       & -4.081(0.331) & -38.307(0.093)& 2.55(0.142)& -1.504\\
G133.94+01.06    & -0.991(0.012) & -46.865(0.022) & 2.832(0.055) & -0.243\\
W3-OH       & -0.984(0.035)& -46.049(0.055)& 3.242(0.146) & -0.286\\
G133.947       &  -1.009(0.017)&   -46.882(0.029) & 3.586(0.188) & -0.264 \\
AFGL490          & -0.894(0.016) & -12.676(0.025) &  2.874(0.064) & -0.292  \\
G176.51+00.20    & -0.302(0.014)& -18.141(0.051)& 2.231(0.125) & -0.127 \\
IRAS05358     &  -1.124(0.013) & -17.393(0.018)& 3.178(0.040)& -0.332 \\
G183.72-03.66          & -0.943(0.018) & 2.403(0.022)& 2.448(0.056) & -0.362 \\
G192.16-03.81     & -0.157(0.016)& 5.938(0.186)& 3.667(0.445) & -0.040 \\
G188.94+00.88  & -0.385(0.019) &  2.967(0.078) &  3.324(0.205) & -0.109\\
G188.79+01.03  & -0.888(0.016)& -0.052(0.032)& 3.559(0.084)& -0.234 \\
G192.60-00.04 & -0.274(0.013)& 6.464(0.054)& 2.239(0.112) & -0.115\\
G196.45-01.67  & -0.211(0.012)& 17.158(0.092)& 3.230(0.206)& -0.061\\
NGC2264-1 & -2.538(0.017)&     6.928(0.013) & 4.189(0.032)& -0.569\\
G229.57+00.15  & -0.143(0.010) & 53.488(0.120) & 3.362(0.291)&  -0.040\\
G232.62+00.99  & -0.145(0.009)& 16.957(0.070) & 2.365(0.202)& -0.058 \\
G236.81+01.98  & -0.183(0.015) & 52.339(0.079) & 2.190(0.247)& -0.078\\
G19.6-0.2  & -4.689(0.037)& 44.56(0.023)& 6.231(0.061) & -0.707\\
            & -0.452(0.023)&  51.692(0.054)&  2.285(0.089)& -0.186\\
G005.88-00.39      & -0.924(0.031) & 9.936(0.064)& 4.327(0.193)& -0.201 \\
                & -1.244(0.031) & 18.521(0.064)& 5.332(0.149) & -0.219\\
G009.62+00.19          & -0.935(0.028)& 2.274(0.088)& 5.736(0.189)& -0.153 \\
                      &-0.484(0.022) & 23.783(0.084) & 3.587(0.179)& 0.127 \\
G010.47+00.02 &-1.563(0.019)& 68.161(0.052)& 8.586(0.120) & -0.171\\
G011.91-00.61  & -1.552(0.018)& 36.113(0.017)& 3.061(0.043) & -0.476\\
G011.49-01.48  & -0.122(0.015)& 10.389(0.132)& 2.121(0.358)& -0.054 \\
M17   &  -2.822(0.047)& 11.977(0.089)& 7.259(0.089)& -0.365\\
       & -3.493(0.047)& 21.342(0.089)&  7.499(0.089)& -0.438\\
      &  -4.066(0.047)& 23.768(0.089)& 0.876(0.089)& -4.360\\
       & -0.476(0.050) & 25.031(0.089) & 0.592(0.089)& -0.755\\
G029.95-00.01  &-2.50(0.025)& 99.403(0.044)& 6.571(0.075)& -0.357\\
J184741.61    &  -0.318(0.016)& 99.584(0.086)& 3.512(0.214)& -0.085 \\
               & -0.435(0.017)& 105.64(0.082)& 4.368(0.202) & -0.094\\
J184754.69   & -0.542(0.037) &  37.541(0.149)& 4.699(0.293) & -0.108 \\
             &-0.283(0.037) & 43.377(0.33)& 4.914(0.633)& -0.054\\
G031.28+00.06  & -0.576(0.022)& 103.317(0.058)& 3.657(0.162) & -0.148\\
             & -1.138(0.021) & 108.533(0.036)& 4.051(0.089) &-0.264\\
G49.2-0.3  & -0.354(0.015)&  6.098(0.047)& 2.298(0.123) & -0.145\\
W49      & -22.935(0.139) & 12.668(0.028) & 10.132(0.077)& -2.127\\
        & -2.449(0.064)&  39.470(0.023)& 1.956(0.064)& -1.176\\
         &  -2.957(0.067)& 62.935(0.022)& 2.115(0.065) & -1.314\\
G043.79-00.12       & -0.222(0.022)& 43.776(0.259)& 5.112(0.540)& -0.041 \\
G043.89-00.78          & -0.462(0.012)& 54.643(0.036)& 3.097(0.090)& -0.140\\
G61.48+0.09     & -0.622(0.034) & 21.406(0.084)& 3.151(0.209) & -0.185\\
G70.33+1.59    & -4.557(0.015)& -21.306(0.006) & 3.714(0.014) & -1.153\\
G069.54-00.97  &  -1.360(0.021)& 10.824(0.029)& 4.153( 0.078)&  -0.308\\
G074.03-01.71  & -0.787(0.010)& 6.068(0.018)& 3.125(0.045)& -0.237\\
W75N           & -0.443(0.052)& -3.143(0.128)& 2.372(0.354)& -0.175\\
              & -1.033(0.057)& 9.627(0.088)&  3.251(0.211) & -0.298\\
WB026   & -0.068(0.009)& 2.952(0.072)& 1.207(0.214)& -0.053\\
WB091 & -0.161(0.021)& -73.164(0.220)& 3.766(0.689)& -0.040\\
G105.41+09.8 & -0.115(0.009) & -10.322(0.061)& 1.575(0.137) & -0.068\\
CASA   & -3.656(0.058)& -46.597(0.011)& 2.441(0.049)& -1.407\\
       & -16.523(0.169)& -39.265(0.023)& 4.971(0.057)& -3.122\\
       & -1.335(0.12) & -36.377(0.025) & 1.591(0.083)& -0.788\\
NGC7538       & -1.253(0.034)& -55.518(0.064)& 6.289(0.192)& -0.187\\
              & -0.123(0.011)& -49.113(0.085)& 2.189(0.191)& -0.053\\
\enddata

%% Include any \tablenotetext{key}{text}, \tablerefs{ref list},
%% or \tablecomments{text} between the \enddata and
%% \end{deluxetable} commands
\tablecomments {Column (1): source name; Columns (2), (3), (4) and (5): the area, position, width and peak antenna temperature of Gaussian-fitting results respectively.}

%% No \tablerefs indicated

\end{longdeluxetable}

\end{document}